\documentclass[fleqn,usenatbib]{mnras}
\pdfoutput=1

\usepackage{amsmath} 
\usepackage{amssymb}

\usepackage[T1]{fontenc}
\usepackage{ae,aecompl}
\usepackage{newtxtext,newtxmath}

\usepackage{graphicx}

\usepackage{color}

\usepackage{subcaption}
\usepackage{mathtools}

\usepackage{mathrsfs}

\defcitealias{Lopez-Sanjuan2014}{LS15}
\defcitealias{Henriques2014}{H15}
\defcitealias{Duncan2015b}{D15b}

\def\mstar{\mathcal{M}_{*}}
\def\msun{\text{M}_\odot}
\def\eazy{{\tt{EAZY}}}

\title[A consistent study of major mergers at $z < 3.5$]{A consistent measure of the merger histories of massive galaxies using close-pair statistics I: Major mergers at $z < 3.5$}

\author[Carl J. Mundy et al.]{Carl J. Mundy,$^{1}$\thanks{\texttt{carl.j.mundy@gmail.com}}
Christopher J. Conselice,$^{1}$\thanks{\texttt{conselice@nottingham.ac.uk}}
Kenneth J. Duncan$^{1,2}$\thanks{\texttt{duncan@strw.leidenuniv.nl}}
Omar Almaini,$^{1}$
\newauthor Boris H\"au\ss ler,$^{3,4,5}$
and William G. Hartley$^{1,6,7}$
\\
$^{1}$School of Physics and Astronomy, University of Nottingham, Nottingham NG7 2RD, UK\\
$^{2}$Leiden Observatory, Leiden University, PO Box 9513, 2300RA Leiden, The Netherlands\\
$^{3}$University of Oxford, Denys Wilkinson Building, Keble Road, Oxford, Oxon OX1 3RH, UK\\
$^{4}$University of Hertfordshire, Hatfield, Hertfordshire AL10 9AB, UK\\
$^{5}$European Southern Observatory, Alonso de Cordova 3107, Vitacura, Casilla 19001, Santiago, Chile\\
$^{6}$ETH Zürich, Institut für Astronomie, Wolfgang-Pauli-Str. 27, 8093, Zürich, Schweiz\\
$^{7}$Department of Physics \& Astronomy, University College London, Gower Street, London, WC1E 6BT, UK}

\date{\today}

\pagerange{\pageref{firstpage}--\pageref{lastpage}} \pubyear{2017} 

\begin{document} 

\maketitle

\label{firstpage}

\begin{abstract}
We use a large sample of $\sim 350,000$ galaxies constructed by combining the UKIDSS UDS, VIDEO/CFHT-LS, UltraVISTA/COSMOS and GAMA survey regions to probe the major (1:4 stellar mass ratio) merging histories of massive galaxies ($>10^{10}\ \mathrm{M}_\odot$) at $0.005 < z < 3.5$. We use a method adapted from that presented in \citet{Lopez-Sanjuan2014}, using the full photometric redshift probability distributions, to measure pair \textit{fractions} of flux-limited, stellar mass selected galaxy samples using close-pair statistics. The pair fraction is found to weakly evolve as $\propto (1+z)^{0.8}$ with no dependence on stellar mass. We subsequently derive major merger \textit{rates} for galaxies at $> 10^{10}\ \mathrm{M}_\odot$ and at a constant number density of $n > 10^{-4}$ Mpc$^{-3}$, and find rates a factor of 2--3 smaller than previous works, although this depends strongly on the assumed merger timescale and likelihood of a close-pair merging. Galaxies undergo approximately 0.5 major mergers at $z < 3.5$, accruing an additional 1--4 $\times 10^{10}\ \mathrm{M}_\odot$ in the process. On average this represents an increase in stellar mass of 20--30\% (40--70\%) for constant stellar mass (constant number density) samples. Major merger accretion rate densities of $\sim 2 \times 10^{-4}$ $\mathrm{M}_\odot$ yr$^{-1}$ Mpc$^{-3}$ are found for number density selected samples, indicating that direct progenitors of local massive ($>10^{11}\mathrm{M}_\odot$) galaxies have experienced a steady supply of stellar mass via major mergers throughout their evolution. While pair fractions are found to agree with those predicted by the \citet{Henriques2014} semi-analytic model, the Illustris hydrodynamical simulation fails to quantitatively reproduce derived merger rates. Furthermore, we find major mergers become a comparable source of stellar mass growth compared to star-formation at $z < 1$, but is 10--100 times smaller than the SFR density at higher redshifts.
\end{abstract}


\section{Introduction}
\label{sec:intro}

The hierarchical growth of matter in the Universe naturally emerges from cold dark matter (CDM) dominated paradigms whereby systems observed today are produced through the repeated merging of smaller systems across cosmic time. While such models make clear predictions on the evolution of dark matter halos \citep[e.g., ][]{Jenkins1998, Maller2006}, the consequences for galaxy formation and evolution are not trivial to infer. Observing galaxies in the process of merging therefore represents a probe of these models and of galaxy formation and evolution, and allows constraints to be placed on evolutionary models of massive galaxies as well as cosmology and the nature of dark matter \citep[e.g.,][]{Bertone2009,Conselice2014b}.

Both major and minor galaxy mergers have been observationally and theoretically implicated in various aspects of galaxy formation and evolution. Mergers were first employed to explain the observed morphological transformations of galaxies over time. For example, galaxy mergers are most likely an important process in the evolution of massive elliptical galaxies \citep{Toomre1972, Barnes1996, Bell2006}. Furthermore, massive quiescent galaxies selected at fixed stellar mass are observed to be a factor of 3--6 times smaller at $z \sim 2$ than in the local Universe \citep{Daddi2005, Trujillo2007, Buitrago2008}, while massive galaxies have increased their stellar mass by a factor of 2--3 over the same time period \citep{Ilbert2010, Ilbert2013, vanderwel2014, Mortlock2015, Ownsworth2016}. Major mergers have been invoked as a possible mechanism responsible for this drastic evolution, and their role has been increasingly constrained over time \citep[e.g.,][]{Conselice2003, Conselice2006, Bluck2012, Lopez-Sanjuan2011, Lopez-Sanjuan2012, Lopez-Sanjuan2013, Man2012,Man2014}, albeit with merger histories often derived from relatively small samples, especially at high redshift. While some works suggest major mergers do play a significant role in the evolution of massive galaxies, other studies exclude major mergers as the main driver and instead suggest that minor mergers are responsible, at least at high redshift \citep[e.g.,][]{McLure2013}. Thus, our understanding of merging is currently incomplete and controversial at best.

One of the most direct measurements one can perform in order to infer how galaxies form and evolve through mergers is to measure the fraction of galaxies undergoing such an event. This provides a path to derive the integrated effect of mergers for specific populations of galaxies. This has previously been achieved at many redshift regimes using two main methodologies. Where high resolution, high signal-to-noise (S/N) imaging exists, selecting mergers through some combination of morphological indicators is popular (e.g., concentration, asymmetry and clumpiness (CAS): \citealt{Conselice2003, Jogee2009, Lopez-Sanjuan2009, Conselice2014b}; or Gini and $M_{20}$: \citealt{Lotz2004,Lotz2008}). These selections are confirmed to almost always probe \textit{ongoing} merging events \citep{Conselice2003, Conselice2008}. Such analysis has even been used to select galaxies at specific stages after coalescence has occurred \citep{Pawlik2016}. However, the requirement for high resolution and high S/N necessarily means that expensive space-based observations are the only route to performing morphological analysis at $z > 1$. The small volumes and thus number densities of galaxies supplied by such campaigns represent a significant source of uncertainty in the robust study of merger histories. The second approach is to select galaxies with small projected separations --- close-pairs --- on the sky \citep[e.g.,][]{Carlberg1994,Patton1997,Patton2000,Kartaltepe2007}. Although selection of close-pairs does not directly trace merging events, it has been shown that galaxies within some small separation are more likely than not to merge in the relatively near future \citep{Mihos1995, Patton1997, Patton2002, Kitzbichler2008}.

While much progress has been made in the literature, various complications exist when attempting to compare measures of merger fractions from different studies. Indeed many studies also find an increasing merger fraction with redshift \citep{LeFevre2000,Bluck2009}, while others find a relatively flat slope or a plateau at high redshift \citep{Williams2011,Newman2012}. At low redshift ($z < 0.2$) studies generally agree on a merger fraction of the order of less than a few percent \citep[e.g.,][]{DePropris2007}. On the other hand, agreement is generally not reached at high redshift ($z > 1$), where merger fractions up to one third \citep[e.g.,][]{LeFevre2000, Bluck2009} have been measured. It has been comprehensively shown that measurements made using stellar mass or luminosity selected samples result in stark differences between the normalisation and measured slopes of the merger fraction \citep{Man2014}. These differences go some but not all the way to reconciling the results from different studies. What is clear is that a consistent picture of galaxy mergers has not been painted over the majority of the history of the Universe.

Deep near-IR imagery combined with complimentary multi-wavelength observations is required to accurately probe the stellar populations at high redshift $z > 1$. Such data allow for photometric redshifts reaching precisions of $\sim0.01(1+z)$ \citep[e.g.,][]{Ilbert2009, Hartley2013,Mortlock2013, Muzzin2013b}, and stellar population parameters including stellar mass to be estimated out to the furthest redshifts \citep[e.g.,][]{Duncan2014}. Modern wide-area, deep surveys represent the only way to observe the merger histories of massive galaxies with any statistical significance across cosmic time. To this end, this paper, in combination with Duncan et al. (\textit{in prep}), herein D17, who study objects at $z > 2$ within the CANDELS field, presents a new method to measure stellar mass selected merger fractions across a large redshift range, exploiting the statistical power of large multi-wavelength datasets. For the first time, we can measure the major and minor merger fractions at $0.005 < z < 6$ consistently using a combination of ground- and space-based observations, providing the first consistent picture of galaxy mergers to within the first Gyr of cosmic time. In this paper, we present merger fractions and derive merger rates of massive galaxies ($\log(\mathcal{M}_* / \mathrm{M}_\odot) > 10$) at $z < 3.5$ using a combination of three square-degree-sized, deep near-IR surveys (totalling 3 square degrees), the publicly available Galaxy And Mass Assembly (GAMA) second data release (DR2) (totalling 144 square degrees), and multiple CANDELS regions (totalling 0.26 square degrees).

This paper is organised as follows: In Section \S\ref{sec:datamethods} we describe the various data used in this work; in Section \S\ref{sec:pairs} we detail the method with which we measure close-pairs of stellar mass selected galaxies; in Section \S\ref{sec:merger-fractions} we explore the measured major merger \textit{fractions}; in Section \S\ref{sec:merger-rates} we derive and compare merger \textit{rates} and discuss our results throughout; in Section \S\ref{sec:discussion} we discuss our results and the tests applied to them; and in Section \S\ref{sec:summary} we summarise the results of this work. Throughout we quote magnitudes in the AB system \citep{Oke1983} unless otherwise stated, stellar masses are calculated using a \citet{Chabrier2003} initial mass function (IMF) and we utilise a $\Lambda$CDM cosmology with $\Omega_{\mathrm{m},0} = 0.3$, $\mathrm{H}_0 = 70$ km s$^{-1}$ Mpc$^{-1}$ and $\Omega_\Lambda = 1 - \Omega_\mathrm{m}$.
\section{Data \& Data Products}
\label{sec:datamethods}

We utilise the deepest and widest surveys of the low and high redshift Universe available today. A combination of Galaxy And Mass Assembly (GAMA), the UKIDSS Ultra Deep Survey (UDS), VIDEO and UltraVISTA provides 144 square degrees at $z < 0.2$ and 3.25 square degrees at $0.2 < z < 3.5$. The depth and wavelength of the surveys used in this work allows us to study the distant Universe with fewer biases against red and dusty galaxies, which could otherwise be completely missed in ultraviolet (UV) and optically selected surveys. While details on how photometric redshift and stellar masses are estimated are given in Section \S\ref{sec:photozs} and \S\ref{sec:masses}, below we discuss the survey fields used in this work.

\subsection{UKIDSS Ultra Deep Survey (UDS)}
\label{sec:uds}

This work employs the eighth data release (DR8) of the UKIDSS UDS (Almaini et al. \textit{in prep}). The UDS is the deepest of the UKIRT (United Kingdom Infra-Red Telescope) Infra-Red Deep Sky Survey \citep[UKIDSS; ][]{Lawrence2007} projects, covering $0.77$ square degrees. Deep photometry is obtained in $J$, $H$ and $K$ to limiting AB magnitudes of 24.9, 24.2 and 24.6 in 2'' apertures. It is currently the deepest near-IR survey ever undertaken over such an area. Complementary multi-wavelength observations exist in the form of $u$-band data obtained from CFHT Megacam; $B$, $V$, $R$, $i$ and $z$-band data from the Subaru-XMM Deep Survey \citep{Furusawa2008}; $Y$-band data from the ESO VISTA Survey Telescope; and IR photometry from the Spitzer Legacy Program (SpUDS, PI: Dunlop). These observations over the wavelength range $0.3 \mathrm{\mu m} < \lambda < 4.6 \mathrm{\mu m}$ are vital for the computation of accurate photometric redshifts, stellar masses and rest-frame magnitudes out to the highest redshifts we probe in this work. We utilise a galaxy catalogue selected in the $K$-band containing approximately 90,000 galaxies out to $z \sim 3.5$, reaching a $99\%$ completeness depth of $K = 24.3$ with an effective area of 0.63 square degrees. We use a combination of spectroscopic redshifts from archival sources as well as the UDSz \citep{Curtis-Lake2012,Bradshaw2013}, which provide 2292 high quality spectroscopic redshifts at $0 < z < 4.5$ (90\% at $z < 2$) in the UDS region.

\subsection{UltraVISTA}
\label{sec:uvista}

We use the publicly available $K_s$-band selected UltraVISTA catalogue produced by \citet{Muzzin2013a}. The UltraVISTA survey observes the COSMOS field \citep{Scoville2007b} with the ESO Visible and Infrared Survey Telescope for Astronomy (VISTA) survey telescope, covering an effective area of $1.62$ square degrees. The catalogue provides PSF-matched $2.1''$ aperture photometry across 30 bands covering the wavelength range $0.15 \mathrm{\mu m} < \lambda < 24 \mathrm{\mu m}$ down to a limiting $90\%$ completeness magnitude of $K_s = 23.4$. Only sources above this detection limit with reliable photometry are used in this work. We do not use the MIPS photometry in this paper as it is uncertain how well models reproduce this regime of a galaxy spectrum. Furthermore, we produce our own photometric redshifts and stellar masses, as described in \S\ref{sec:photozs} and \S\ref{sec:masses}. The catalogue includes GALEX \citep{Martin2005}, CFHT/Subaru \citep{Capak2007}, S-COSMOS \citep{Sanders2007} and UltraVISTA \citep{McCracken2012} photometry as well as the zCOSMOS Bright \citep{Lilly2007a} spectroscopic dataset, providing 5467 high quality spectroscopic redshifts at $z < 2.5$. The vast majority (99\%) of these spectroscopic redshifts are at $z < 1$ and 50\% are at $z < 0.5$.

\subsection{VIDEO}
\label{sec:video}

The VISTA Deep Extragalactic Observations (VIDEO) survey \citep{Jarvis2012} is a $\sim$12 square degree survey in the near-infrared $Z$, $Y$, $J$, $H$ and $K_s$ bands, specifically designed to enable the evolution of galaxies and large structures to be traced as a function of both epoch and environment from the present day out to $z = 4$, and active galactic nuclei (AGNs) and the most massive galaxies up to and into the epoch of reionization. In this work we use observations matched to those of the Canada-France-Hawaii Telescope Legacy Survey Deep-1 field (CFHTLS-D1) providing multi-wavelength ($0.3\mathrm{\mu{m}} < \lambda < 2.1\mathrm{\mu{m}}$) coverage over a total of 1 square degree down to a 90\% completeness magnitude of $K_s = 22.5$. We perform comprehensive simulations to calculate the completeness level as a function of total $K$-band magnitude which are described in Appendix \ref{sec:appendix-video-completeness}.

For the purpose of this work, we utilise a $K_s$-selected catalogue (released in June 2015) containing 54,373 sources after star/galaxy separation using a $uJK$ colour selection, magnitude cuts, star masking, and selecting only sources with a detection signal-to-noise $> 2$. Bright stars and areas visibly contaminated with starlight are manually masked out using the VIDEO $K_s$-band image. Objects within these masked regions are flagged and discarded from the sample. A spectroscopic sample of galaxies is constructed from the latest VIMOS VLT Deep Survey \citep[VVDS;][]{LeFevre2005} and the VIMOS Public Extragalactic Redshift Survey \citep[VIPERS;][]{Garilli2014} data releases. We match the most secure redshifts (quality flags 3 and 4) within one arcsecond of our $K_s$-band sources, providing 4,382 spectroscopic redshifts over the range $0 < z < 4.5$. The vast majority (90\%) of this sample is below $z < 1.5$ however.

\subsection{GAMA}
\label{sec:gama}
In order to obtain a measurement of the merger fraction at redshifts where we are restricted by volume in other fields, we utilise the second data release (DR2) of the Galaxy And Mass Assembly (GAMA) campaign \citep{Driver2009,Liske2015}. This release provides multi-wavelength photometry in 9 filters over three fields totalling 144 square degrees. Complimenting this data, 98\% of the detections are provided with secure spectroscopic redshifts. GAMA therefore represents a large and unique dataset with which to probe galaxy evolution at low redshift.

In this paper we utilise combined data from all three GAMA fields (G09, G12 and G15), herein collectively referred to as the GAMA region, included in the DR2 release. When calculating stellar masses in this region we apply the recommended photometric zero-point offsets\footnote{\url{http://www.gama-survey.org/dr2/schema/table.php?id=168}} and stellar mass scaling factors \citep{Taylor2011} provided with the release documentation. What differentiates this dataset from the others used in this paper is the unprecedented spectroscopic coverage. Combining the three aforementioned GAMA regions yields 55,199 objects with good quality spectroscopic redshift (quality flag $\mathrm{nQ} > 2$) and $z_\mathrm{spec} > 0.005$, which minimises contamination from stars (visual inspection of a $u-J$ vs $J-K$ plot reveals this cut removes the stellar locus), representing 97 per cent of the total number of objects down to a limiting Petrosian $r$-band magnitude of $m_r = 19$. This allows us to perform our analysis in two ways: photometrically and spectroscopically, which we discuss in Section \S\ref{sec:comp-spec-photo-fm}.

\subsection{Simulated Data}
\label{sec:simdata}

Models of galaxy formation and evolution have advanced dramatically over the last few decades. Semi-analytic models (SAMs) aim to reproduce and predict the statistical properties of galaxy populations, historically at low redshift. We use the latest development in the Munich `family' of models \citep[e.g.,][]{Croton2006,DeLucia2006,Guo2011}, as described in \citet{Henriques2014}, herein \citetalias{Henriques2014}, to provide predictions of the pair fraction. This model is applied to the output of The Millennium Simulation \citep{Springel2005}, scaled to a Planck cosmology \citep{Planck2014params}. We downloaded all 24 mock lightcones from the German Astrophysical Virtual Observatory \citep[GAVO;][]{Lemson2006} which we reduce in size from a circular aperture of two degrees diameter to a square field-of-view with an area of one square degree. Doing so allows us to quantify the expected variance between surveys similar in size to those used in this study. Furthermore, we explore and compare the results of the merger fractions obtained using the \citetalias{Henriques2014} model in Section \S\ref{sec:merger-fractions}. Furthermore, we also compare results of the merger rate to that within the Illustris simulation \citep{Vogelsberger2014,Vogelsberger2014a,Rodriguez-Gomez2015} in Section \S\ref{sec:merger-rates}.

\subsection{Photometric redshift probability distributions}
\label{sec:photozs}

Photometric redshift probability distributions are calculated for all sources using the \eazy\ photometric redshift code \citep{Brammer2008a}. \eazy\ determines the $z_\text{phot}$ for a galaxy by fitting a spectral energy distribution (SED) produced by a linear combination of templates to a set of photometric measurements. It has been shown that the default set of six templates, derived from the PEGASE models \citep{Fioc1999}, in combination with an additional red template from the \citet{Maraston2005} models, and a 1 Gyr-old single-burst \citet{Bruzual2003} template are required to provide robust SED fits to the zoo of observed galaxies in modern surveys \citep[e.g.,][]{Onodera2012, Muzzin2013a}.

As such, we use this set of templates to calculate photometric redshifts and photometric redshift probability distributions (PDFs). The PDF is constructed for each galaxy from its $\chi^2(z)$ distribution following $P(z) \propto \exp(-\chi^2(z)/2)$, after convolution with a photometric prior. We now discuss the use of a photometric prior in these calculations and the ability of the resulting PDFs to accurately reproduce photometric redshift confidence intervals.

\subsubsection{Photometric redshift prior}

In calculating galaxy PDFs and best-fit photometric redshifts, many studies make use of a luminosity or colour dependent redshift prior. The use of such priors have been shown to improve best-fit solutions when compared to spectroscopic redshift measurements \citep[e.g.,][]{Benitez2000,Brammer2008a}. However the use of such priors may introduce bias into the measurement of close pairs. As an example, let us consider two galaxies at the same redshift with identical properties except for stellar mass (luminosity). A luminosity based prior will influence the probability distribution of each galaxy and, in the example, the higher mass system will have its PDF biased towards lower redshifts, and vice-versa for the second galaxy. Furthermore, priors are necessarily based on simulations. At higher redshifts ($z > 2$) these may deviate from the true distribution of galaxies, however at lower redshift they are much more constrained and in agreement with observations.

\begin{figure}
    \includegraphics[width=0.475\textwidth]{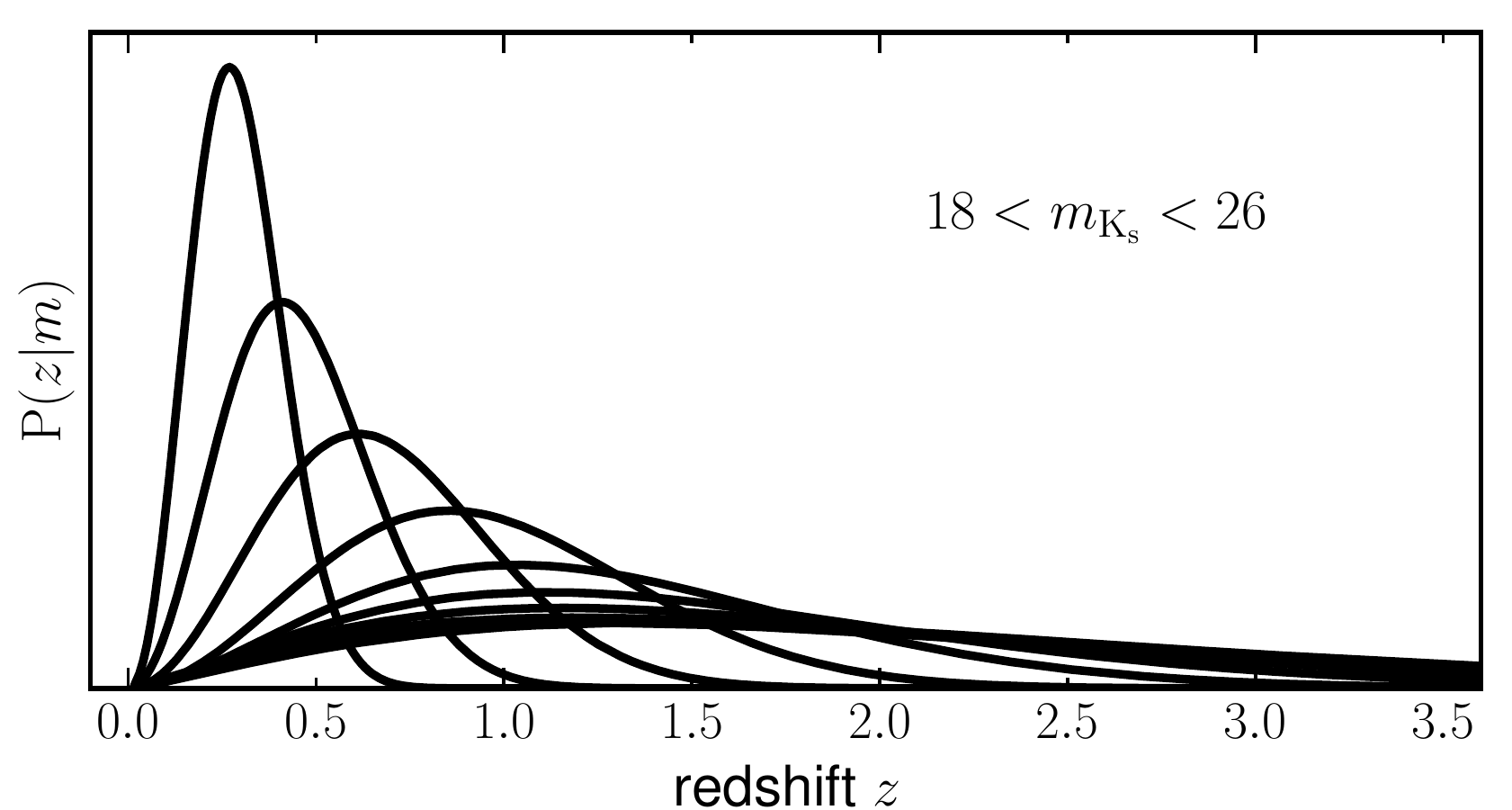}
    \caption{Relative prior probabilities, $\mathrm{P}(z|m_K)$, as a function of apparent $K_s$-band magnitude extracted from semi-analytic light cones \citep{Henriques2014}. Plotted probability densities in steps of $\Delta m_K = 1$ over the magnitude range $18 < m_K < 26$, normalised such that $\int \mathrm{P}(z|m_K)\ dz = 1$, with $\mathrm{P}(z|m_K)$ given by Equation \ref{eqn:prior}.}
    \label{fig:prior-probs}
\end{figure}

We therefore construct a new luminosity prior $\text{P}(z|m)$, which denotes the probability of a galaxy with apparent $K$-band magnitude $m$ being found at redshift $z$, by extracting galaxy number counts from the \citetalias{Henriques2014} semi-analytic model using 24 independent light cones. This model has been shown to accurately reproduce the observed number densities of galaxies out to $z \sim 3$, and thus is perfect to construct a prior from. This is achieved in the same manner as \citet{Brammer2008a} and \citet{Benitez2000}, parametrising each magnitude bin $i$ as
\begin{equation} \label{eqn:prior}
	P(z|m_{K,i}) \propto z^{\gamma_i} \times \exp( -(z/z_i)^{\gamma_i} ),
\end{equation}
where $\gamma_i$ and $z_i$ are fit to the redshift distribution in each magnitude bin. This is done to ensure that the prior is smooth over the redshift range of interest. We calculate these distributions over the redshift range $0 < z < 7$ and apparent magnitude range $17 < m_K < 27$. Calculated fitting parameters are displayed in Figure \ref{fig:prior-probs} which shows the calculated prior probabilities as a function of apparent magnitude. We find that pair fractions obtained using photometric redshifts calculated with and without a prior are indistinguishable within the calculated uncertainties, however the prior is used in this work because it improves the best-fit $z_\text{phot}$ estimates and reduces the number of catastrophic outliers (see Section \S\ref{sec:photz:bestfit}). The default EAZY $r$-band prior is used when calculating photometric data products for the GAMA survey region as this data is $r$-band selected.

\subsubsection{Photometric redshift confidence intervals}
\label{sec:photoz-sigma}
Redshift probability distributions output by photometric redshift codes are often unable to accurately represent photometric redshift confidence intervals \citep[e.g.,][]{Hildebrandt2008,Dahlen2013}. The causes include, but are not limited to, inaccurate photometry errors or the choice of template set. Although average agreement between best-fit $z_\text{phot}$ and $z_\text{spec}$ can be excellent, 1$\sigma$ and 2$\sigma$ confidence intervals can be significantly over- or under-estimated.

Analysing the photometric redshift probability distributions output by \eazy, discussed in Section \S\ref{sec:photozs}, we observe that the confidence intervals are indeed incorrect. Using high quality spectroscopically obtained redshifts for a subset of galaxies in each field we find that 72\%, 71\%, 81\% and 50\% of $z_\text{spec}$ are found within the $1\sigma$ photometric PDF interval for the UDS, VIDEO, COSMOS and GAMA regions, respectively. In order to address this we sharpen PDFs that overestimate the confidence intervals. This is done as in \citet{Dahlen2013}, however we briefly outline the method here.

\begin{figure}
	\includegraphics[width=0.475\textwidth]{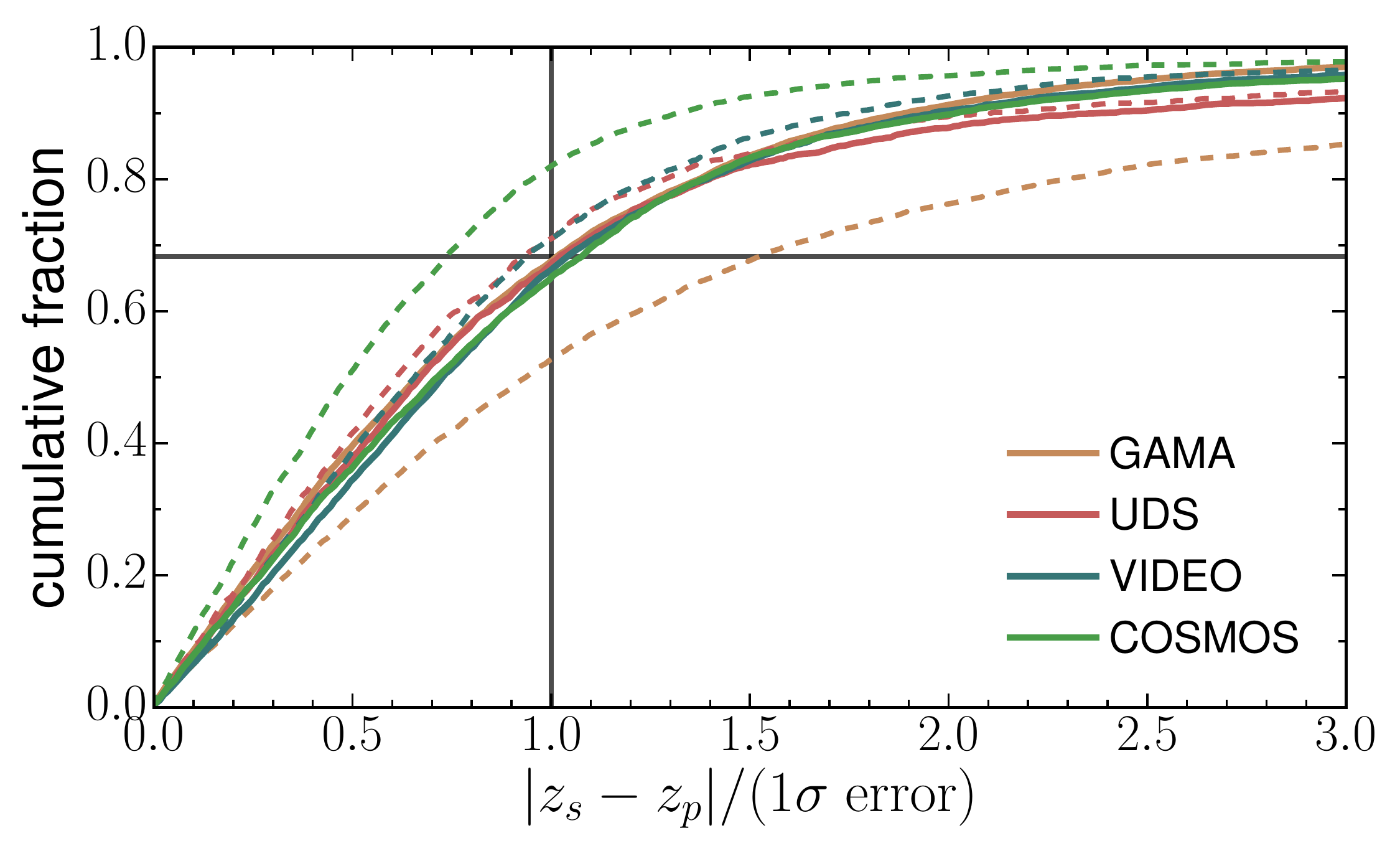}
    \caption{Cumulative distribution of the $|z_p − z_s|/(1\sigma ~\textrm{error})$ for the GAMA (gold), UDS (red), VIDEO (blue) and COSMOS (green) survey regions. Dashed lines indicate the distributions found before the corrections described in Section \S\ref{sec:photoz-sigma} while solid lines represent the corrected distributions. The cross-hair represents the expected 68.3\% of sources at $|z_s - z_p|/(1\sigma ~\textrm{error}) = 1$.}
    \label{fig:before-after-pdfs}
\end{figure}

To sharpen, the PDFs are replaced with $P(z_i) = P(z_i)_{0}^{1/\alpha}$ until the value of alpha gives the correct fraction of 68.3\%. To smooth, the PDFs are convolved with a kernel of $[0.25, 0.5, 0.25]$ until the correct fraction of 68.3\% is recovered. The same process is then applied to the entire sample. In doing so, we obtain values of $\alpha = 0.832$, $0.818$, $0.482$ for the UDS, VIDEO and COSMOS fields, respectively. The GAMA field required $N=350$ smoothing iterations to match the same requirements. The cumulative distribution of $|z_s - z_p| / (1\sigma ~\textrm{error})$ is shown in Figure \ref{fig:before-after-pdfs} both before and after these corrections for sources with spectroscopic observations in all fields. This figure shows that the corrections applied provide the expected $\sim 68\%$ of sources with a spectroscopic redshift within $1\sigma$ of the calculated photometric redshift.

\subsubsection{Best-fit solutions}
\label{sec:photz:bestfit}
\begin{figure}
    \includegraphics[width=0.475\textwidth]{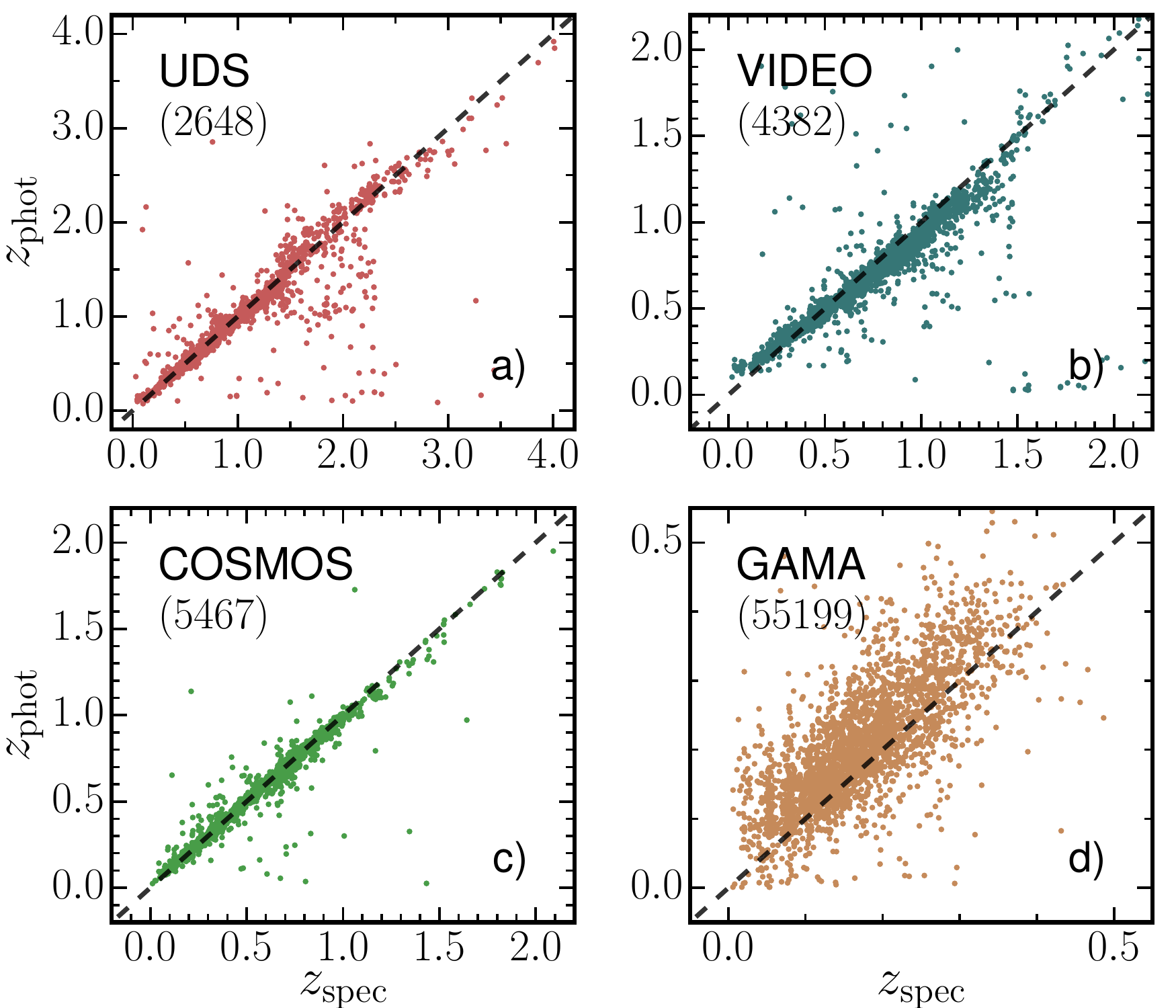}
	\caption{Comparison between best-fit photometrically derived redshifts, $z_\mathrm{phot}$, and spectroscopically measured redshifts, $z_\mathrm{spec}$, in the a) UDS, b) VIDEO, c) COSMOS, and d) GAMA regions. Numbers within parenthesis denote the number of science-quality spectroscopic redshifts within each field. Due to the extremely large number of sources within the GAMA region, a randomly selected sample of 5\% is displayed for this field only. The normalised median absolute deviation, average offset and outlier fraction of our photometric redshifts are listed in Table \ref{tab:photo-z-results} for each region.}
    \label{fig:photoz}
\end{figure}

While we are interested in the PDFs associated with each galaxy, it is useful to compare best-fit photometric redshift solutions with spectroscopically obtained values. Various measures exist to quantify the agreement between photometric and spectroscopic redshifts and here we report the normalised median absolute deviation (NMAD), mean $|\Delta{z}| / (1+z_\text{spec})$, where $\Delta{z} = (z_\text{spec} - z_\text{phot})$, and outlier fraction, defined in two ways. These measures of photometric redshift quality are provided in Table \ref{tab:photo-z-results}, and a visual comparison between spectroscopic and photometric redshifts within all regions is shown in Figure \ref{fig:photoz}. We note that all fields except for GAMA possess averages biases of $z_\text{spec} - z_\text{phot} \approx 0$. As is apparent in Figure \ref{fig:photoz}, We note that there is a relatively large apparent bias in our photometric redshifts within the GAMA region whereby our photometric redshifts tend to be larger than the spectroscopic redshift by $\Delta{z} = 0.02$ on average. This is the largest bias we observe in the datasets we use. We note that if the brightest 10\% (25\%) of objects in the GAMA region are analysed, this bias is reduced by a factor of $\sim3$ ($\sim2$), suggesting that fainter ($r > 18$) objects are more affected by this bias. Such an effect would not be seen in the other regions as their spectroscopic samples are typically biased towards the brightest objects in the field. However, as we do not observe any suggestion of stellar mass dependence (see Section \S\ref{sec:merger-fractions}) in the pair fractions, this issue is not expected to affect the results presented herein.

We find that the use of a photometric prior typically reduces the difference between photometric and spectroscopic redshifts, whilst also reducing the fraction of catastrophic failures. Furthermore, we find that the COSMOS region provides the most accurate photometric redshifts when compared to a subset of spectroscopic redshifts. However, spectroscopic redshift samples that are co-spatial with deep, wide near-IR surveys like UltraVISTA/COSMOS are often heavily biased towards the nearest and brightest systems. With a 97\% completeness fraction the spectroscopic sample in the GAMA region is undoubtedly unbiased and is arguably a better indicator of photometric redshift efficacy. Here the prior reduces the NMAD and the mean offset by 18\% and 15\%, respectively. 

\begin{figure}
    \includegraphics[width=0.475\textwidth]{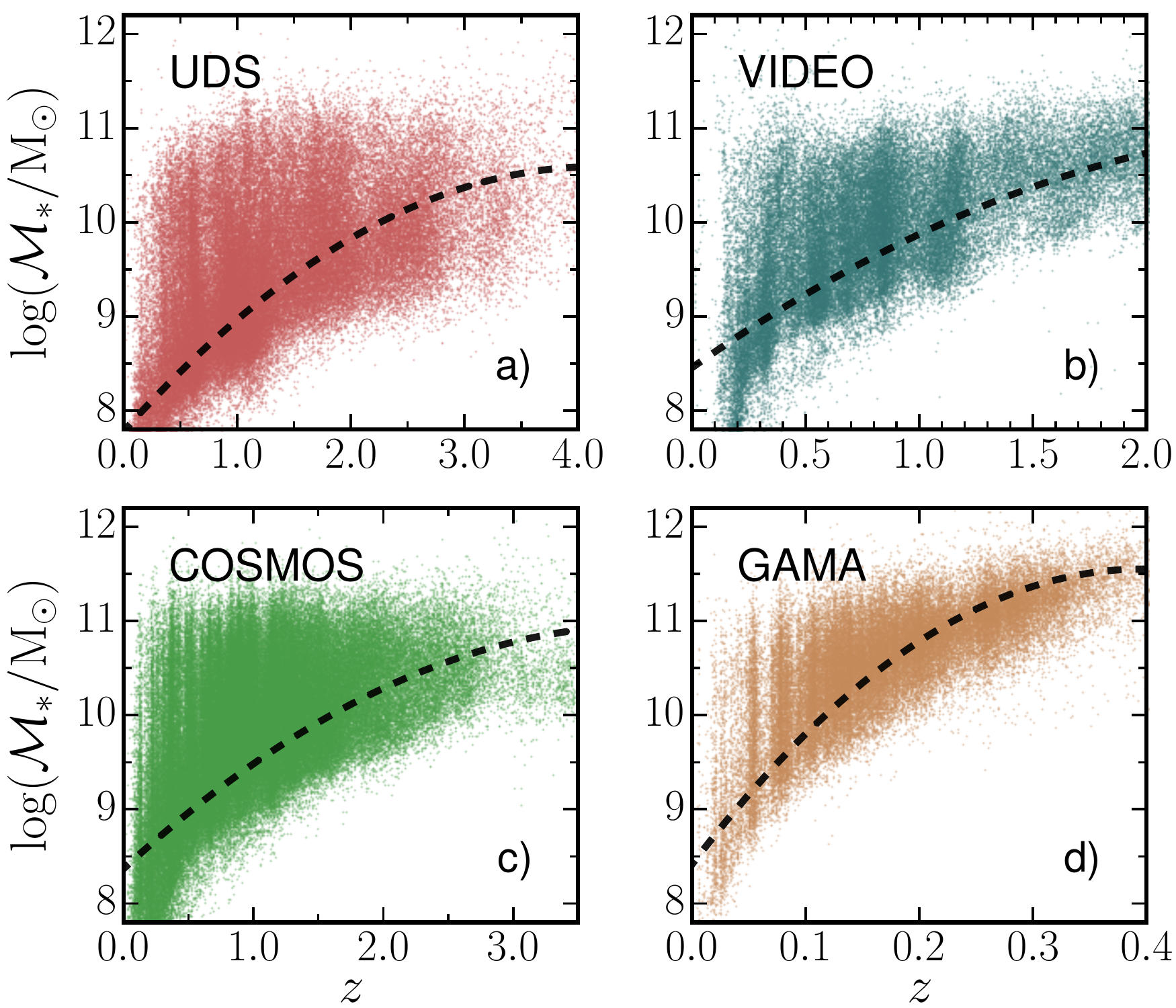}
	\caption{Redshift versus stellar mass distributions in the in the a) UDS, b) VIDEO, c) COSMOS, and d) GAMA regions. Redshifts presented in the GAMA region are spectroscopic ($z_\mathrm{spec}$) while those displayed in other regions are photometric ($z_\mathrm{phot}$). 90\% stellar mass completeness limits, $\mathcal{M}_*^{90}(z)$, within each region, determined using magnitude limits of $r = 19.0$ and $K = 24.3, 22.5, 23.4$, respectively, are given by the dashed black lines.}
    \label{fig:masses-comp}
\end{figure}

\begin{table}
\begin{minipage}{0.475\textwidth}
\centering
\caption{Best-fit photometric redshift (with and without prior) comparison with high quality spectroscopic sample outlined in Section \S\ref{sec:datamethods}. For each field we list the number of secure spectroscopic redshifts available ($N_s$), the normalised median absolute deviation ($\sigma_{_\text{NMAD}}$), mean $|\Delta{z}| / (1+z_s)$, average bias $\Delta{z} = z_\text{spec} - z_\text{phot}$, and fraction of catastrophic outliers ($\eta_{_1}$ and $\eta_{_2}$) defined in two ways.}
\label{tab:photo-z-results}
\begin{tabular}{@{}lcccccc} 
\hline
Field	&	$N_\text{s}$ & $\sigma_{_\text{NMAD}}$	&	$\frac{|\Delta{z}|}{(1+z_\text{s})}$	&	$\Delta{z}$ & $\eta_{_1}$\footnote{Catastrophic outliers determined as $|\Delta{z}|/(1+z_\text{spec}) > 0.15$.} & $\eta_{_2}$\footnote{Catastrophic outliers determined as $|\Delta{z}|/(1+z_\text{spec}) > 3 \times \sigma_{_\text{NMAD}}$.} \\
\hline
\multicolumn{7}{c}{WITH MAGNITUDE PRIOR} \\
\hline
UDS & 2648 & 0.053 & 0.045 & 0.01 & 5.3\% & 5.0\% \\
VIDEO & 4382 & 0.044 & 0.038 & 0.01 & 2.9\% & 3.3\% \\
COSMOS & 5467 & 0.013 & 0.010 & 0.00 & 0.5\% & 2.5\% \\
GAMA & 55199 & 0.049 & 0.044 & -0.02 & 2.4\% & 2.5\% \\
\hline
\multicolumn{7}{c}{WITHOUT MAGNITUDE PRIOR} \\
\hline
UDS & 2648 & 0.051 & 0.045 & 0.01 & 5.3\% & 5.3\% \\
VIDEO & 4382 & 0.048 & 0.042 & 0.02& 3.4\% & 3.5\% \\
COSMOS & 5467 & 0.013 & 0.011 & 0.00 & 0.5\% & 3.2\% \\
GAMA & 55199 & 0.060 & 0.052 & -0.03 & 3.4\% & 1.7\% \\
\hline
\end{tabular}
\end{minipage}
\end{table}

Applying the corrections described in Section \S\ref{sec:photoz-sigma}, results in PDFs which accurately represent the probability of every galaxy at every redshift over the range $0 < z < 6$. The integral of the PDF over some redshift range measures the probability of the galaxy being found within said redshift range.

\begin{figure}
	\includegraphics[width=0.48\textwidth]{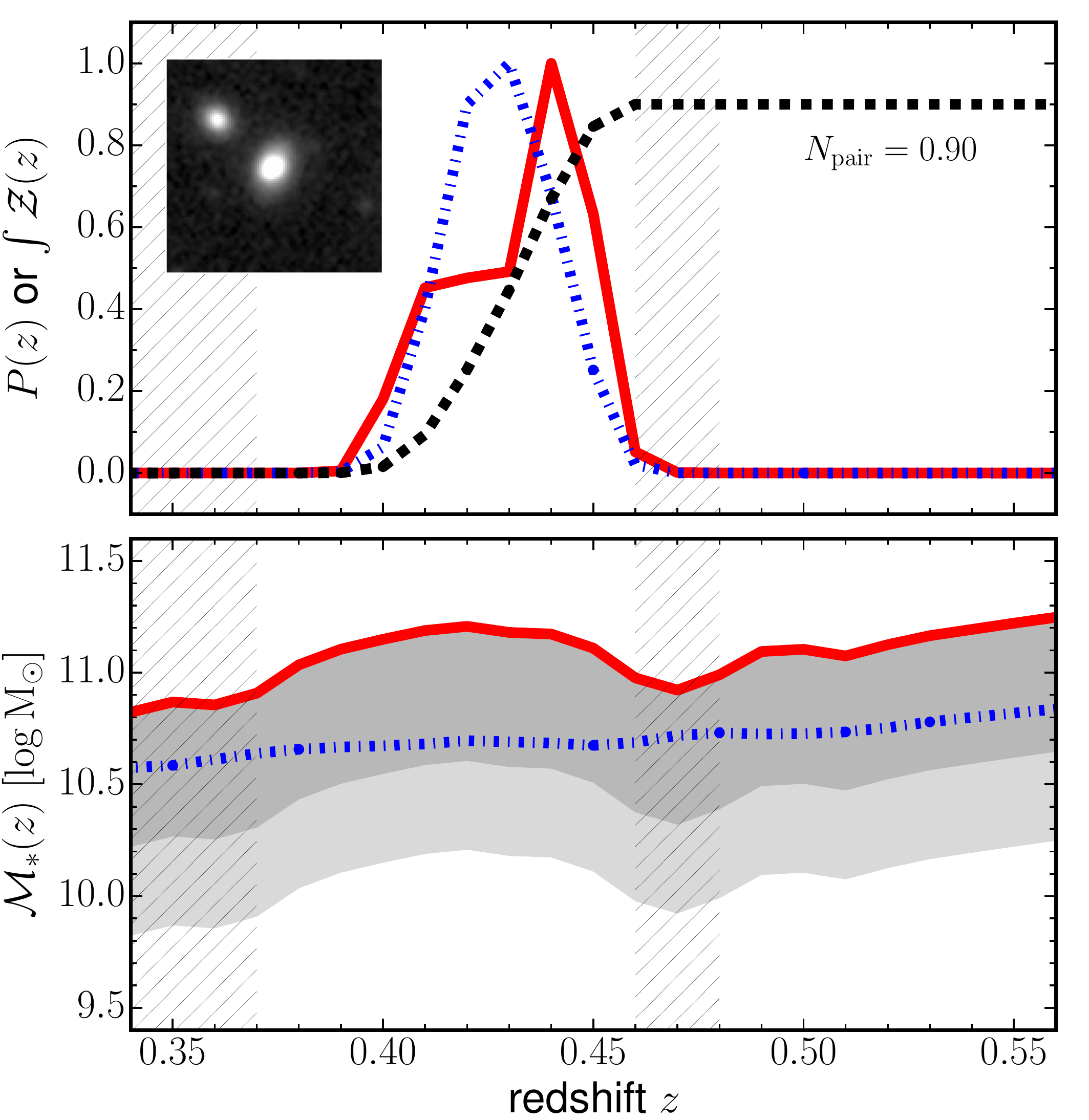}
    \caption{\textit{Top:} Computed redshift probability distributions, $P(z)$, for an identified close-pair system with a primary galaxy (solid red line) at best-fit redshift $z_\text{peak} = 0.44$ and secondary galaxy (dashed dotted blue line) at best-fit redshift $z_\text{peak} = 0.43$. A greyscale $K_s$-band image of the pair, of side length 20'', is shown inset. The integrated cumulative probability function (Equation \ref{eqn:z}) of the system is given by the dashed black line. \textit{Bottom:} The stellar mass as a function of redshift, via SED-fitting, for the primary and secondary galaxies. At their best-fit $z_\text{peak}$, the primary and secondary galaxies possess stellar masses of $\log (\mathcal{M}_* / \mathrm{M}_\odot) = 11.2$ and $10.7$, respectively. The major merger mass ratio (1:4) is given by the dark shaded region while the minor merger mass ratio (1:10) is given by the light shaded region. The hatched regions represent redshift ranges where the close-pair system is not considered as the primary galaxy does not meet the criteria of $\log (\mathcal{M}_* / \mathrm{M}_\odot) > 11$.}
    \label{fig:pair-example}
\end{figure}

\subsection{Stellar masses}
\label{sec:masses}

Stellar masses are calculated using {\tt{smpy}}, a custom spectral energy distribution (SED) fitting code, first introduced in \citet{Duncan2014} and available online\footnote{https://www.github.com/dunkenj/smpy/}. We use \citet[][BC03]{Bruzual2003} stellar population synthesis models with a \citet{Chabrier2003} IMF. Model ages are allowed to vary between 0.01--13.7 Gyr. Star-formation histories are described by a simple $\tau$-model and are allowed to be exponentially increasing or decreasing with values of $|\tau|$ allowed between 0.01--13.7 Gyr, plus an option for a constant star-formation history. The effects of dust are parametrised as in \citet{Calzetti2000}, with an extinction ($A_V$) allowed to vary between 0 -- 4 magnitudes. Stellar metallicity is allowed in the range $0.005 < Z/Z_\odot < 2.5$. We do not include nebular emission. In short, at every redshift the stellar mass is calculated as the mean stellar mass summed over all template fits, weighted by the goodness of fit. All available photometry are fit to a library of 34,803 synthetic SEDs simultaneously to achieve this. Stellar mass as a function of redshift within each region is shown in Figure \ref{fig:masses-comp}.
\section{Counting galaxy pairs}
\label{sec:pairs}

Modern multi-wavelength, deep photometric surveys offer a wealth of data from which the distances to, and physical properties of large galaxy samples can be calculated. Arguably the most fundamental properties of a galaxy that can be calculated from this data are the redshift and stellar mass. For the purposes of this work, the measurement we ultimately make is the fraction of galaxies in the process of merging, $f_\text{merge}$. To this end, we analyse galaxy close pairs, and perform a measurement of the pair fraction, $f_\text{pair}$, through the use of photometric redshift probability distributions (PDFs) and stellar mass-redshift functions, $\mathcal{M}_*(z)$. Use of the PDF allows uncertainty in galaxy redshifts to be taken into account when selecting galaxy pairs. The full code we have developed for this work, named {\tt{Pyrus}} (Pyrus being the genus of tree on which pears grow), is available freely online\footnote{\url{http://www.github.com/ppxcjm/Pyrus}} with accompanying documentation. We describe the conversion of the pair fraction to the merger fraction in Section \S\ref{sec:merger-rates}. Our method builds upon the photometric pair method described by \citet{Lopez-Sanjuan2014} to allow for pair fraction measurements of stellar mass selected samples of galaxies constructed from flux-limited catalogues. We refer the interested reader to this paper, however we describe the method in full below.

\begin{figure*}
  \includegraphics[width=0.995\textwidth]{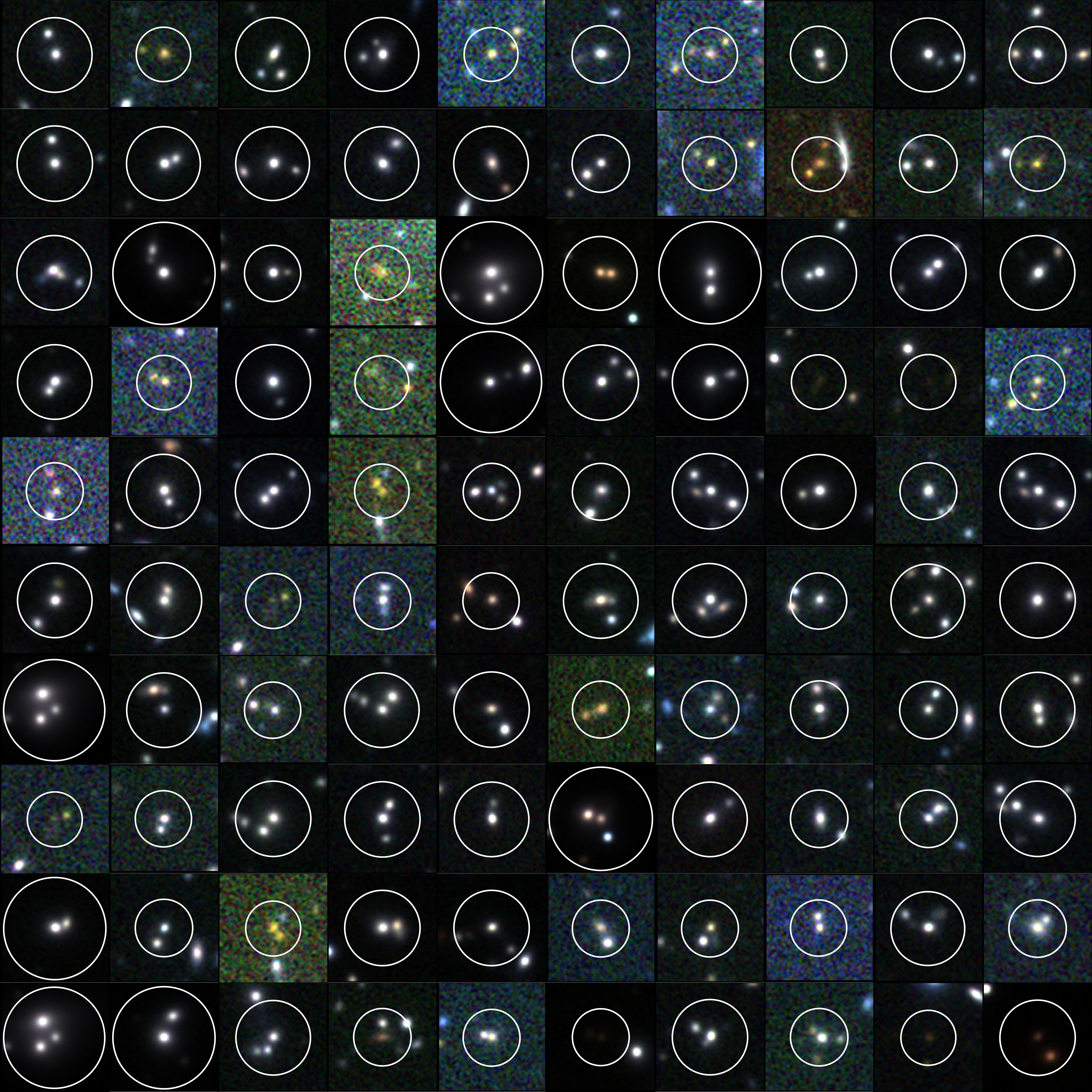}
  \caption{Three-colour image using the UltraVISTA DR1 $J$, $H$, and $K_s$-band images of close-pairs at $0.3 < z < 3.0$ that contribute $\mathrm{N}_\mathrm{pair} > 0.7$ after weightings are applied. Each postage stamp is centred on the primary (most massive) galaxy and the outer white circles represent a physical search radius of 30 kpc around each centred primary galaxy. Colour scaling is done automatically to highlight the often faint galaxies of interest. A range of morphologies, colours and galaxy sizes are apparent.}
  \label{fig:rgb-pairs}
\end{figure*}

Figure \ref{fig:pair-example} illustrates the resulting photometric redshift PDF (top panel) and estimated stellar mass (bottom panel) as a function of redshift for an identified close-pair in the COSMOS region whose primary galaxy is found to be at $z_\mathrm{phot} = 0.44$ with a stellar mass of $\log(\mathcal{M}_*/\mathrm{M}_\odot) = 11.2$. Further examples of probable close-pairs ($N_\text{pair} > 0.7$) identified in the COSMOS region are shown in Figure \ref{fig:rgb-pairs}.

\subsection{Close-pair selection}
\label{sec:pair-selection}

Using the science catalogues within each survey region, an initial list of \textit{projected} galaxy close pairs is constructed. Based on the desired physical separation limits, the minimum and maximum considered angular separations are calculated using the extremes of the redshift range being probed. In this paper we look at the merger histories of galaxies with $\mathcal{M}_* > 10^{10}\mathrm{M}_\odot$ through close-pairs at physical separations between 5--30 kpc and a stellar mass ratio of $\mu > 1/4$, i.e. major mergers.

Next, each pair has their PDFs convolved and normalised such that the integral of the resulting PDF can maximally contribute a single close pair to the final analysis. This combined redshift probability function, $\mathcal{Z}(z)$, is defined as 
\begin{equation}
\label{eqn:z}
  \mathcal{Z}(z) = \frac{2 \times P_{1}(z) \times P_{2}(z)}{P_{1}(z) + P_{2}(z)} = \frac{P_{1}(z) \times P_{2}(z)}{N(z)}.
\end{equation}
\noindent Here $P_{1}(z)$ and $P_{2}(z)$ represent the PDFs of the primary and secondary galaxy within each \textit{projected} close pair. It follows from this prescription that $\mathcal{Z}(z)$ represents the number of close pairs contributed by each \textit{projected} pair at redshift $z$ and can necessarily range only between 0 and 1. Close pairs with $\int_{0}^{\infty}\mathcal{Z}(z) dz = 0$ are discarded from the subsequent analysis.

Additional selection criteria are enforced using binary redshift masks. These are 0 when criteria are not met, and 1 otherwise. Firstly, the use of physical separations to define close pairs translates into angular separation conditions that are a function of redshift. Thus, an angular separation mask, $\mathscr{M}^{\theta}(z)$, is calculated for each pair. This is defined as
\begin{equation}
\label{eqn:pair-mask-theta}
	\mathscr{M}^\theta(z) = 
    \left\{\begin{matrix*}[l]
      1, & \text{it}\ \theta_\text{min}(z) \leq \theta \leq \theta_\text{max}(z) \\ 
      0, & \text{otherwise} \\
    \end{matrix*}\right.
\end{equation}
where $\theta$ is the projected separation on the sky between two galaxies, $\theta_\text{min}(z) = r_\text{min} / d_A(z_\text{max})$, and $\theta_\text{min}(z) = r_\text{max} / d_A(z_\text{min})$, where $d_A(z)$ is the angular diameter distance. For the purposes of this paper we choose $r_\text{min} = 5$ kpc and $r_\text{max} = 20$ or $30$ kpc in order to maximise opportunities for comparison with previous literature studies. A similar mask is defined to enforce the stellar mass conditions required to label two galaxies as a close-pair. This pair selection mask is defined as
\begin{equation}
\label{eqn:pair-mask-selection}
	\mathscr{M}^{\text{pair}}(z) = 
    \left\{\begin{matrix*}[l]
    1, & \text{if}\ \mathcal{M}_{*,1}^{\text{lim}}(z) \leq \mathcal{M}_{*,1}(z) \leq \mathcal{M}_*^\text{max} \\
       & \text{and}\ \mathcal{M}_{*,2}^{\text{lim}}(z) \leq \mathcal{M}_{*,2}(z)\\
    0, & \text{otherwise} \\
    \end{matrix*}\right.
\end{equation}
where $\mathcal{M}_{*,1}(z)$ and $\mathcal{M}_{*,2}(z)$ are the stellar masses of the primary and secondary galaxies, respectively. The stellar mass limits in the above equation are defined as
\begin{equation}
\label{eqn:mlim1}
	\mathcal{M}_{*,1}^{\text{lim}}(z) = \max\{\mathcal{M}_*^\text{min}(z), \mathcal{M}_*^\text{comp}(z)\}
\end{equation}
\noindent and 
\begin{equation}
\label{eqn:mlim2}
	\mathcal{M}_{*,2}^{\text{lim}}(z) = \max\{\mu\mathcal{M}_*^1(z), \mathcal{M}_*^\text{comp}(z)\}
\end{equation}
respectively, where $\mathcal{M}_*^\text{comp}(z)$ is the stellar mass completeness limit at redshift $z$ for the survey region the galaxies belong to, $\mathcal{M}_*^\text{min}(z)$ is the lower stellar mass limit for the primary sample, and $\mathcal{M}_*^\text{max}(z)$ is the upper stellar mass limit for the primary sample. Application of this mask ensures that: (i) the primary galaxy is within the stellar mass range being probed; (ii) that the correct stellar mass ratio between the primary and secondary galaxy is enforced at every redshift, and (iii) both galaxies are above the stellar mass completeness limits of their respective survey region.

With these properties in hand for each \textit{projected} pair, the pair probability function, $\mathrm{PPF}(z)$, is then defined as
\begin{equation}\label{eqn:ppf}
\text{PPF}(z) = \mathcal{Z}(z) \times \mathscr{M}^{\theta}(z) \times \mathscr{M}^\text{pair}(z).
\end{equation}
The integral of the PPF provides the \textit{unweighted} number of close-pairs (as defined by the chosen selection criteria) that two galaxies contribute to the measured pair fraction.

\subsection{Close-pair weightings}
\label{sec:weights}

The $\mathrm{PPF}$ in Equation \ref{eqn:ppf} is affected by three selection effects: (i) incompleteness in the projected spatial search area around primary galaxies; (ii) the difference in quality of the photometric redshifts between survey regions; and (iii) the stellar mass search area found beyond the completeness limit. The corrections we make for these issues are explained in the following sections.

\subsubsection{Stellar mass (in)completeness}
\label{sec:mass-completeness}

The various limiting fluxes of the surveys used in this work correspond to redshift dependent stellar mass completeness limits. As we have a statistically large number of galaxies at every redshift within each of the surveys used, we follow \citet{Pozzetti2010} in calculating the empirical $90\%$ stellar mass completeness limit, $\mstar^{\text{90}}(z)$ for each survey. This is found by scaling the stellar masses of the faintest $20\%$ of sources to that which they would have at the flux limit (survey completeness magnitude) of the survey. The $90\%$ stellar mass completeness limit is taken as the 90th per centile of the resulting scaled mass distribution. Stellar mass completeness limits for all fields are shown in Figure \ref{fig:masses-comp} for comparison. We find that the UDS, VIDEO and COSMOS fields are complete at stellar masses above $10^{10} \msun$ ($10^{11} \msun$) below redshift 2.3, 1.0, 1.5 (3.5, 2.0, 3.0), respectively, while the GAMA region is found to be complete at redshift 0.2 (0.2).

Selecting galaxies by their stellar mass requires us to take into account scenarios where a search for close-pair companions falls below the known completeness stellar mass. A primary galaxy with a stellar mass, $\mathcal{M}_{*,1}(z)$, close to the redshift dependant stellar mass completeness limit may have a reduced mass range within which to search for secondary galaxies, for example if $\mu \mathcal{M}_{*,1}(z) < \mathcal{M}_*^\text{lim}(z)$. The weighting we prescribe can be written as the inverse of the fraction of the stellar mass search area above the stellar mass completeness limit. This weighting is applied to all secondary galaxies around a primary galaxy and is defined as
\begin{equation}
	w^\text{comp}_2(z) = \Bigg[ \frac{\int_{\mathcal{M}_*^\textrm{lim}(z)}^{\mathcal{M}_{1}}\ \phi(\mstar,z)\ \textrm{d}\mstar}{\int_{\mu \mathcal{M}_{1}}^{\mathcal{M}_{1}}\ \phi(\mstar,z)\ \textrm{d}\mstar} \Bigg]^{-1},
\end{equation}
where $\phi(\mstar,z)$ represents the GSMF at the appropriate redshift. Making this correction we recover pair statistics corresponding to a volume limited study. These secondary weights are a stellar mass version of the luminosity weights presented in \citet{Patton2000}. Additional weights are assigned to the primary galaxies, as in \citet{Patton2000}, to minimise the error from galaxies that are close to the flux limit which will have fewer numbers of observed pairs. The primary completeness weight, $w_1^\text{comp}(z)$, is given by
\begin{equation}
	w^\text{comp}_1(z) = \frac{\int_{\mathcal{M}_*^{\text{lim},1}(z)}^{\mathcal{M}_*^\text{max}}\ \phi(\mathcal{M}_*,z)\ \mathrm{d}\mathcal{M}_*}{\int_{\mathcal{M}_*^\text{min}}^{\mathcal{M}_*^\text{max}}\ \phi(\mathcal{M}_*,z)\ \mathrm{d}\mathcal{M}_*}
\end{equation}
where $\mathcal{M}_*^\text{min}$ and $\mathcal{M}_*^\text{max}$ are the lower and upper stellar mass limits of the primary sample and $\mathcal{M}_{*,1}^{\text{lim}}(z)$ is defined in Equation \ref{eqn:mlim1}.

\subsubsection{Masked areas}
\label{sec:area-weights}

Primary galaxies which lie close to the boundaries of the survey may have their spatial search area reduced, finding fewer pair galaxies as a result. This is also the case for galaxies near survey areas masked out due to contamination, from bright stars for example. As the search area depends on the fixed physical search radius, this correction is also a function of redshift and must be calculated for every redshift of interest. The area around each primary galaxy that may be excluded by these effects is calculated by performing photometry on the mask image. We use the {\tt{photutils}}\footnote{\url{http://photutils.readthedocs.org}} (v0.2) Python package for this task. Each secondary galaxy is then weighted by the inverse of the fraction of the search area available around its primary host and is defined as
\begin{equation}
\label{eqn:area-weight}
	w^\text{area}(z) = \frac{1}{f_\text{area}(z)},
\end{equation}
where $f_\text{area}(z)$ is the sum of the mask image within the annulus at a given redshift, divided by the sum over the same area in an image of equal size with all values equal to unity. This method automatically accounts for irregular survey shapes and small calculation errors from finite pixel sizes.

\subsubsection{Photometric redshift quality}
\label{sec:photoz-quality-weight}

As in \citetalias{Lopez-Sanjuan2014} we apply a selection in the odds parameter $\mathcal{O}$ \citep{Benitez2000,Molino2013} which represents the photometric redshift quality. The odds parameter encodes the probability of a galaxy being found within some redshift interval centred on its best-fit value. The odds sampling rate (OSR) for galaxies with apparent magnitude $m$ is defined as 
\begin{equation}
\label{eqn:osr}
\text{OSR}(m) = \frac{\sum N_{\mathcal{O}\geq0.3}}{\sum N_{\mathcal{O}\geq0}},
\end{equation}
the ratio between the number of galaxies with $\mathcal{O} \geq 0.3$ and the total number of galaxies with magnitude $m$. The choice of cut in the Odds parameter is explored in Appendix \ref{appendix:odds}. We calculate this quantity in bins of width $\Delta m = 0.5$ and linearly interpolate these values to define the OSR at every possible magnitude. Figure \ref{fig:osr} shows the odds sampling rate for the surveys used in this work, showing clearly the differing flux limits. We find that no more than 5 per cent of galaxies in each magnitude bin fall below this cut, even at the respective magnitude limits in each region. Both the primary and secondary galaxy in a close-pair are then weighted by 
\begin{equation}
\label{eqn:osr-weight}
	w^\text{OSR} = \frac{1}{\text{OSR}(m)},
\end{equation}
where $m$ is the apparent magnitude of the galaxy in the detection filter (e.g. the $K_s$-band for the UDS, VIDEO and COSMOS fields, and the $r$-band for the GAMA field).

\subsubsection{Final weightings}
\label{sec:total-weights}

Taking all of the aforementioned weights into account, the total weight applied to each secondary galaxy around a given primary galaxy is given by
\begin{equation}
\label{eqn:sec-weight}
	w_2(z) = w^\text{area}_1(z) \times w_1^\text{comp}(z) \times w_2^\text{comp}(z) \times w_1^\text{OSR} \times w_2^\text{OSR}.
\end{equation}
Similarly, the weight applied to every primary galaxy is given by
\begin{equation}
\label{eqn:pri-weight}
	w_1(z) = w_1^\text{comp}(z) \times w_1^\text{OSR}.
\end{equation}
The application of these weightings allows for the measurement of volume-limited pair fractions. In this work, however, we are careful to only make use of measured pair fractions where we are complete in stellar mass, and so $w_1^\text{comp} = w_2^\text{comp} = 1$.

\subsection{The pair fraction}
\label{sec:pair-fraction}

Here we describe how the pair fraction is calculated. The number of associated close-pairs with each galaxy, $i$, in the primary sample over the redshift range $z_\text{min} < z < z_\text{max}$ is given by
\begin{equation}
	N_\text{pair}^i = \sum_j\ \int_{z_\text{min}}^{z_\text{max}}\ w_2^j(z) \times \text{PPF}_j(z)\ \mathrm{d}z,
\end{equation}
where $j$ indexes the secondary galaxies associated with each primary galaxy, $\text{PPF}_j(z)$ the corresponding pair probability function (see Section \S\ref{sec:pair-selection}), and $w_2^j(z)$ the pair weight given by Equation \ref{eqn:sec-weight}. The number of primary sample galaxies each galaxy contributes within the same redshift range is similarly given by
\begin{equation}
	N_1^i = \sum_i\ \int_{z_\text{min}}^{z_\text{max}}\ w_1^i(z) \times P_i(z) \times \mathcal{S}_1^i(z)\ \mathrm{d}z
\end{equation} 
where $\mathcal{S}_1^i(z)$ is the selection function for the primary sample of galaxies, $P_i(z)$ is the normalised redshift probability density function, and $w_1^i$ the primary galaxy's weighting as given in Equation \ref{eqn:pri-weight}. The selection function simply enforces the criteria for a galaxy to be included in the primary sample and is defined as
\begin{equation}
\label{eqn:selection-fn}
	\mathcal{S}_1(z) = 
    \left\{\begin{matrix*}[l]
		1, & \text{if}\ \mathcal{M}_{*,1}^{\text{lim}}(z) \leq \mathcal{M}_{*,1}(z) \leq \mathcal{M}_*^\text{max} \\
        0, & \text{otherwise} \\
	\end{matrix*}\right.
\end{equation}
where the functions have the same definitions as above. Let us consider the example of attempting to measure the pair fraction for a sample of galaxies with stellar mass $\log(\mathcal{M}_*/\mathrm{M}_\odot) > 11$, as in this work. In this particular case we would arrive at $\mathcal{M}_{*,1}^{\text{lim}} = 10^{11} \mathrm{M}_\odot$ and $\mathcal{M}_*^\text{max} = 10^{12} \mathrm{M}_\odot$. This upper limit is chosen to exclude unphysical stellar mass estimates, and remove rare galaxies with bad SED-fitting results from consideration.

It follows that the pair fraction, $f_\text{pair}$, is simply defined as the number of observed close-pairs divided by the total number of galaxies in the primary sample. Over the same redshift range as above this is given by
\begin{equation}
\label{eqn:fpair}
	f_\text{pair} = \frac{N_\text{pair}}{N_\text{tot}} = \frac{\sum_i N_\text{pair}^i}{\sum_i N_1^i}.
\end{equation}
The conversion of this \textit{pair} fraction to a \textit{merger} fraction, and subsequently a merger \textit{rate}, is discussed in Section \S\ref{sec:merger-rates}.
\section{Observed pair fractions}
\label{sec:merger-fractions}

\begin{figure*}
\includegraphics[width=0.975\textwidth]{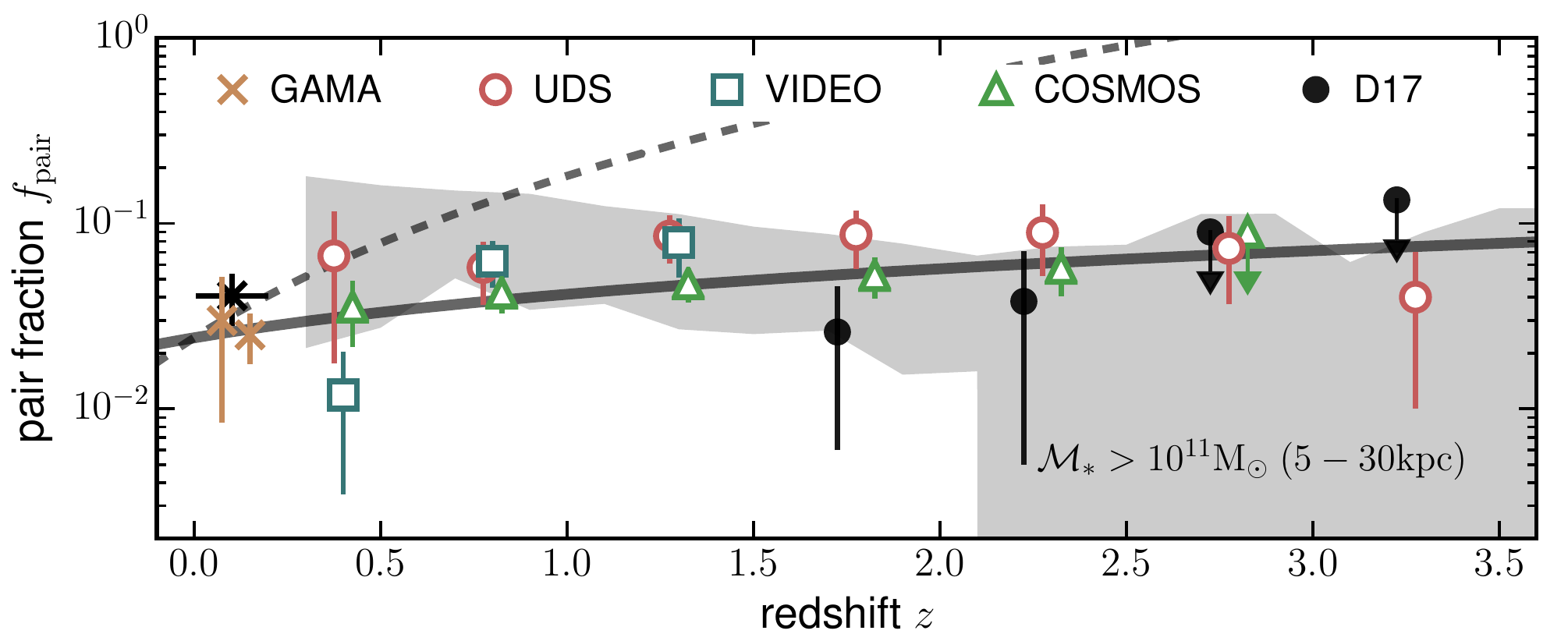}
\caption{The measured major merger ($\mu > 1/4$) pair fraction, $f_\text{pair}$, for galaxies with $\log(\mathcal{M}_*/\mathrm{M}_\odot) > 11$ at physical separations of 5--30 kpc as a function of redshift in the GAMA (gold and black crosses), UDS (red circles), COSMOS (green triangles) and VIDEO (blue squares) fields. The black crosses with horizontal error bars are points measured using the GAMA spectroscopic sample, including Poisson errors and cosmic variance estimates. Results from a complimentary study within the CANDELS fields (Duncan et al., \textit{in prep}) are presented as solid black circles. Upper limits on the pair fraction are given by points with solid filled arrows. The best-fit to all the data, as provided in Table \ref{tab:fm:fits}, is shown as a solid grey line. The grey shaded area represents the $1\sigma$ variation in the pair fraction as measured using 24 light cones based on the \citetalias{Henriques2014} semi-analytic model.}
\label{fig:fm-11-30kpc}
\end{figure*}

\begin{table}
\caption{Major merger ($\mu > 1/4$) pair fractions, $f_\text{pair}$, and associated errors calculated using {\tt{Pyrus}} for constant stellar mass selected samples. Fractions are listed by each survey region, separated by stellar mass and physical search radius parameters. Errors include contributions from cosmic variance, bootstrap error analysis and Poisson errors. Pair fractions in GAMA at $0 < z < 0.2$ are measured using spectroscopic redshifts.}
\label{tab:fm}
\centering
\begin{tabular}{ccccc}
\hline
$z$ & GAMA & UDS & VIDEO & COSMOS \\
\hline
\multicolumn{5}{c}{$\mathcal{M}_* > 10^{10}\mathrm{M}_\odot\ (5-20\mathrm{kpc})$}\\
\hline
0.0 -- 0.2 & 0.011$\pm$0.002 & - & - & - \\
0.0 -- 0.1 & 0.020$\pm$0.005 & - & - & - \\ 
0.1 -- 0.2 & 0.014$\pm$0.002 & - & - & - \\  
0.2 -- 0.5 & - & 0.018$\pm$0.006 & 0.019$\pm$0.005 & 0.015$\pm$0.004 \\ 
0.5 -- 1.0 & - & 0.042$\pm$0.007 & $<$0.036      & 0.023$\pm$0.003 \\ 
1.0 -- 1.5 & - & 0.057$\pm$0.008 & -               & $<$0.036      \\
1.5 -- 2.0 & - & $<$0.099      & -               & -               \\
\hline
\multicolumn{5}{c}{$\mathcal{M}_* > 10^{10}\mathrm{M}_\odot\ (5-30\mathrm{kpc})$}\\
\hline
0.0 -- 0.2 & 0.019$\pm$0.004 & - & - & - \\
0.0 -- 0.1 & 0.035$\pm$0.008 & - & - & - \\ 
0.1 -- 0.2 & 0.025$\pm$0.003 & - & - & - \\  
0.2 -- 0.5 & - & 0.042$\pm$0.013 & 0.042$\pm$0.010 & 0.029$\pm$0.007 \\
0.5 -- 1.0 & - & 0.075$\pm$0.012 & $<$0.089      & 0.055$\pm$0.007 \\
1.0 -- 1.5 & - & 0.101$\pm$0.014 & -               & $<$0.087      \\
1.5 -- 2.0 & - & $<$0.176      & -               & -               \\
\hline
\multicolumn{5}{c}{$\mathcal{M}_* > 10^{11}\mathrm{M}_\odot\ (5-20\mathrm{kpc})$}\\
\hline
0.0 -- 0.2 & 0.022$\pm$0.007 & - & - & - \\
0.0 -- 0.1 & 0.023$\pm$0.018 & - & - & - \\ 
0.1 -- 0.2 & 0.017$\pm$0.006 & - & - & - \\  
0.2 -- 0.5 & - & 0.010$\pm$0.016  & 0.002$\pm$0.003 & 0.022$\pm$0.010 \\
0.5 -- 1.0 & - & 0.034$\pm$0.011  & 0.033$\pm$0.011 & 0.018$\pm$0.005 \\
1.0 -- 1.5 & - & 0.053$\pm$0.019  & 0.028$\pm$0.010 & 0.021$\pm$0.006 \\
1.5 -- 2.0 & - & 0.101$\pm$0.060  & -               & 0.026$\pm$0.008 \\
2.0 -- 2.5 & - & 0.065$\pm$0.030  & -               & 0.034$\pm$0.011 \\
2.5 -- 3.0 & - & 0.057$\pm$0.030  & -               & $<$0.057      \\
3.0 -- 3.5 & - & 0.033$\pm$0.026  & -               & -               \\
\hline
\multicolumn{5}{c}{$\mathcal{M}_* > 10^{11}\mathrm{M}_\odot\ (5-30\mathrm{kpc})$}\\
\hline
0.0 -- 0.2 & 0.041$\pm$0.013 & - & - & - \\
0.0 -- 0.1 & 0.030$\pm$0.022 & - & - & - \\ 
0.1 -- 0.2 & 0.025$\pm$0.008 & - & - & - \\  
0.2 -- 0.5 & - & 0.067$\pm$0.049 & 0.012$\pm$0.008 & 0.035$\pm$0.014 \\
0.5 -- 1.0 & - & 0.058$\pm$0.022 & 0.063$\pm$0.018 & 0.042$\pm$0.009 \\
1.0 -- 1.5 & - & 0.086$\pm$0.025 & 0.079$\pm$0.028 & 0.047$\pm$0.010 \\
1.5 -- 2.0 & - & 0.087$\pm$0.030 & -               & 0.053$\pm$0.013 \\
2.0 -- 2.5 & - & 0.089$\pm$0.037 & - 			   & 0.057$\pm$0.017 \\
2.5 -- 3.0 & - & 0.073$\pm$0.037 & -               & $<$0.090       \\
3.0 -- 3.5 & - & 0.040$\pm$0.030 & -               & -               \\
\hline
\end{tabular}
\end{table}

In this section we detail the measured pair fractions obtained for various primary samples. These are chosen in order to enable comparison of their derived merger rates with previous works in Section \S\ref{sec:merger-rates}. As previously mentioned, we perform the close-pair analysis in the GAMA region in two ways: photometrically and spectroscopically. For the latter, we enforce the condition that \textit{projected} close-pairs must be within $\Delta v < 500$ km/s ($\Delta z = 0.0017$) of each other. A combination of mass (in)completeness and the potential to miss a large population of massive galaxies at faint magnitudes \citep{Caputi2015} limit our study to $z < 3.5$ in the deepest near-IR survey region. In Section \S\ref{sec:major-mergers-fm} we describe pair fractions obtained for constant stellar mass selected samples, and in Section \S\ref{sec:fpair:n} we report pair fractions for samples of galaxies selected at a constant cumulative comoving number density.

Firstly we justify our choice of parameters. We define the minimum physical separation of a close-pair as 5 kpc in order to minimise the influence of objects whose photometry has become blended and to ensure the host galaxy is not counted as its own companion. This physical separation translates into angular separations between 0.7--1.5 arcseconds at the redshift ranges probed in this study. The pixel scales in the UDS (0.27$"$/pix), VIDEO (0.19$"$/pix) and COSMOS (0.15$"$/pix) images, from which the catalogues were produced, represent minimum separations of 3, 3 and 5 pixels, respectively.

\subsection{Constant stellar mass selected samples}
\label{sec:major-mergers-fm}

The volume afforded by square degree-sized surveys allows the most massive galaxies ($\mathcal{M}_* > 10^{11} \mathrm{M}_\odot$) to be probed across cosmic time. We obtain major merger fractions for two stellar mass selections at two physical separations purely for comparison with previous literature works. These fractions are tabulated for reference in Table \ref{tab:fm} and we subsequently derive major merger rates in Section \S\ref{sec:merger-rates}.

\subsubsection{Massive galaxies ($\mathcal{M}_* > 10^{11} \mathrm{M}_\odot$)}
\label{sec:fm:11}

We measure the pair fraction for a sample of galaxies defined by the limit $\mathcal{M}_* > 10^{11} \mathrm{M}_\odot$. We calculate this pair fraction at maximum physical separations of 20 kpc and 30 kpc to enable comparison with previous works. Obtained fractions and estimated errors at both separations are given in Table \ref{tab:fm}, however we only discuss those at 30 kpc due to the larger sample sizes obtained using this larger separation. Results of $f_\text{pair}$ at this separation in the GAMA, UDS, VIDEO and COSMOS regions are shown in Figure \ref{fig:fm-11-30kpc} as gold and black crosses, red circles, blue squares and green triangles, respectively. Results from a complimentary study within the CANDELS fields (Duncan et al., \textit{in prep}) are shown as filled black circles. Where the primary sample is complete (in stellar mass) but the companion search area is $>50\%$ complete, one sigma upper limits on $f_\text{pair}$ are denoted by symbols with a filled arrow of the same colour. Errors include contributions from cosmic variance estimates \citep{Moster2011}, Poisson statistics and a bootstrap error analysis. These contributions are summed in quadrature.

Towards higher redshift the UDS, VIDEO and COSMOS fields provide an insight into the evolution of the pair fraction to within the first 2 Gyr of cosmic time. Pair fractions measured in the lowest redshift bin ($0.2 < z < 0.5$) exhibit a large scatter between fields and possess large uncertainties. This is attributed to the relatively small volumes in this redshift bin which translates into a small sample of massive galaxies. However, all three fields report values of $f_\text{pair}$ that agree to within the errors. At $z > 0.5$ we observe consensus that $f_\text{pair}$ evolves very little at $z < 3.5$. The measurements within the VIDEO region are found to be consistent with those obtained in the UDS region, however stellar mass completeness limits our comparison to $z < 1.5$ in this region.

As in previous works we fit our pair fraction results via a least-squares fitting routine with a simple power law of the form $f_\text{pair} = f_0(1+z)^m$ \citep[e.g.,][]{Patton2002,Conselice2003,Bridge2007}, which describes the pair fraction at $z=0$ and the slope of the pair fraction with redshift. In general we find a weakly increasing pair fraction with redshift. A similar evolution is found by \citet[][see their Fig. 5]{Lopez-Sanjuan2009} at $0.2 < z < 1$, \citet[][see their Fig. 11]{Lopez-Sanjuan2014} at $0 < z < 1$, and \citet[][see their Fig. 14]{Conselice2003} at $1.4 < z < 3.4$, albeit with slightly varying selections and methodologies.

Performing the fitting procedure to the data from all \textit{observationally} determined pair fractions shown in Figure \ref{fig:fm-11-30kpc} and find \[f_\text{pair}(z) = 0.024\pm0.004 \times (1+z)^{0.78\pm0.20}\] for close-pairs selected at 5--30 kpc. This is plotted as a solid black line. Fitting parameters for close-pairs selected at 5--20 kpc, at lower stellar masses, and using different combinations of data are presented in Table \ref{tab:fm:fits}. Our data are complimented by pair fraction measurements within the CANDELS fields at $z > 1.5$ presented in Duncan et al. (\textit{in prep}). The relative scarcity of high mass galaxies combined with the small volumes probed by the CANDELS fields result in upper limits (solid black circles with a downward pointing solid black arrow) of the pair fraction at $z > 2.5$ although they are consistent with measurements in the UDS and COSMOS regions of this work. If we just consider the GAMA, UDS, VIDEO and COSMOS data we find a very similar evolution in the pair fraction of \[f_\text{pair}(z) = 0.024\pm0.004 \times (1+z)^{0.85^{+0.19}_{-0.20}},\] which is found to be in excellent agreement with the fit obtained when considering the CANDELS data at high redshift.

\begin{table}
\centering
\caption{Major merger ($\mu > 1/4$) fraction fitting parameters for combinations of survey regions, for a parametrisation of the form $f_\text{pair}(z) = f_0(1+z)^{m}$. Fitting is performed on $f_\text{pair}$ measurements up to the redshifts reported in Table \ref{tab:fm}. Errors are determined using a bootstrap analysis and the resulting parameter distributions of 10,000 realisations. The number of merging events, $N_\text{merg}$, a galaxy undergoes at $0 < z < 3.5$, given by the integral in Equation \ref{eqn:n_merg}, is provided in the far right column.}
\label{tab:fm:fits}


\begin{tabular}{lcccc}
\hline
Survey Region & $f_0$ & $m$ & $N_\text{merg}$ \\
\hline
\multicolumn{4}{c}{$\mathcal{M}_* > 10^{10}\mathrm{M}_\odot\ (5-20\mathrm{kpc})$}\\
\hline
UDS    & $0.012^{+0.005}_{-0.004}$ & $1.99^{+0.56}_{-0.51}$ & $1.1^{+1.1}_{-0.6}$ \\
COSMOS & $0.009^{+0.006}_{-0.004}$ & $1.78^{+1.20}_{-1.02}$ & $0.7^{+1.7}_{-0.5}$ \\
All    & $0.006^{+0.003}_{-0.002}$ & $2.68^{+0.59}_{-0.59}$ & $1.1^{+1.3}_{-0.6}$ \\
All + GAMA & $0.010^{+0.002}_{-0.002}$ & $1.82^{+0.37}_{-0.34}$ & $0.8^{+0.6}_{-0.3}$ \\
\hline
\multicolumn{4}{c}{$\mathcal{M}_* > 10^{10}\mathrm{M}_\odot\ (5-30\mathrm{kpc})$}\\
\hline
UDS    & $0.029^{+0.012}_{-0.009}$ & $1.57^{+0.55}_{-0.50}$ & $1.0^{+1.0}_{-0.5}$ \\
COSMOS & $0.014^{+0.008}_{-0.006}$ & $2.50^{+1.08}_{-0.84}$ & $1.1^{+2.4}_{-0.8}$ \\
All    & $0.018^{+0.005}_{-0.004}$ & $2.14^{+0.40}_{-0.41}$ & $1.1^{+0.8}_{-0.5}$ \\
All + GAMA & $0.020^{+0.003}_{-0.003}$ & $1.97^{+0.26}_{-0.25}$ & $1.0^{+0.6}_{-0.3}$ \\
All + GAMA + D17 & $0.028^{+0.002}_{-0.002}$ & $0.80^{+0.09}_{-0.09}$ & $0.5^{+0.3}_{-0.1}$ \\
\hline
\multicolumn{4}{c}{$\mathcal{M}_* > 10^{11}\mathrm{M}_\odot\ (5-20\mathrm{kpc})$}\\
\hline
UDS    & $0.022^{+0.011}_{-0.009}$ & $0.68^{+0.42}_{-0.40}$ & $0.7^{+0.7}_{-0.4}$ \\
COSMOS & $0.013^{+0.009}_{-0.006}$ & $0.70^{+0.67}_{-0.65}$ & $0.5^{+0.5}_{-0.3}$ \\
All    & $0.014^{+0.006}_{-0.004}$ & $0.75^{+0.41}_{-0.43}$ & $0.5^{+0.4}_{-0.2}$ \\
All + GAMA & $0.015^{+0.004}_{-0.004}$ & $0.66^{+0.34}_{-0.34}$ & $0.5^{+0.3}_{-0.2}$ \\
\hline
\multicolumn{4}{c}{$\mathcal{M}_* > 10^{11}\mathrm{M}_\odot\ (5-30\mathrm{kpc})$}\\
\hline
UDS    & $0.073^{+0.034}_{-0.025}$ & $-0.05^{+0.40}_{-0.40}$ & $0.8^{+0.7}_{-0.4}$ \\
VIDEO  & $0.007^{+0.005}_{-0.003}$ & $3.22^{+0.97}_{-0.87}$ & $1.1^{+2.9}_{-0.8}$ \\
COSMOS & $0.031^{+0.013}_{-0.010}$ & $0.52^{+0.44}_{-0.41}$ & $0.5^{+0.4}_{-0.2}$ \\
All    & $0.025^{+0.005}_{-0.005}$ & $0.82^{+0.22}_{-0.22}$ & $0.5^{+0.3}_{-0.2}$ \\
All + GAMA & $0.024^{+0.004}_{-0.004}$ & $0.85^{+0.19}_{-0.20}$ & $0.5^{+0.3}_{-0.2}$ \\
All + GAMA + D17 & $0.024^{+0.004}_{-0.004}$ & $0.78^{+0.20}_{-0.20}$ & $0.5^{+0.3}_{-0.1}$\\
\hline
\multicolumn{4}{c}{$n(>\mathcal{M}_*) = 1 \times 10^{-4}\ $Mpc$^{-3}\ (5-30\mathrm{kpc})$}\\
\hline
All & $0.023^{+0.015}_{-0.010}$ & $0.95^{+0.65}_{-0.61}$ & $0.5^{+0.6}_{-0.3}$ \\
All + GAMA & $0.019^{+0.007}_{-0.006}$ & $1.16^{+0.42}_{-0.37}$ & $0.5^{+0.4}_{-0.2}$ \\
\hline
\multicolumn{4}{c}{$n(>\mathcal{M}_*) = 5 \times 10^{-4}\ $Mpc$^{-3}\ (5-30\mathrm{kpc})$}\\
\hline
All & $0.027^{+0.012}_{-0.009}$ & $1.01^{+0.59}_{-0.55}$ & $0.6^{+0.6}_{-0.3}$ \\
All + GAMA & $0.023^{+0.005}_{-0.004}$ & $1.22^{+0.31}_{-0.31}$ & $0.6^{+0.4}_{-0.2}$ \\
\hline
\end{tabular}
\end{table}

As mentioned previously, it is difficult to compare merger fractions measured between different studies. Thus we do not attempt to compare our fitted value of $f_\text{pair}(z=0) = f_0$ with previous work, however we are able to draw comparisons between the calculated slope of the evolution in $f_\text{pair}$. Our value of $m = 0.78\pm0.20$ is in agreement with that found by \citet{Conselice2003} for a primary sample of $\mathrm{M}_\mathrm{B} > -20$ over a similar redshift range. On the other hand, the major merger fraction slope of $m = 2.9 \pm 0.4$ found for galaxies with $\log (\mathcal{M}_*/\mathrm{M}_\odot) > 11$ in \citet{Bluck2009} is seemingly at odds with the measurement presented in this work. However, their fit is anchored by the $z=0$ point of \citet{DePropris2007} which was measured using  different selection criteria to the $z > 0.5$ data. Re-fitting to just the high redshift data presented in Fig. 1 of \citet{Bluck2009} results in a significantly shallower slope of $m = 0.48\pm0.41$, in agreement with our result.

\subsubsection{Intermediate mass galaxies ($\mathcal{M}_* > 10^{10} \mathrm{M}_\odot$)}
\label{sec:fm:10}

\begin{figure*}
\includegraphics[width=0.975\textwidth]{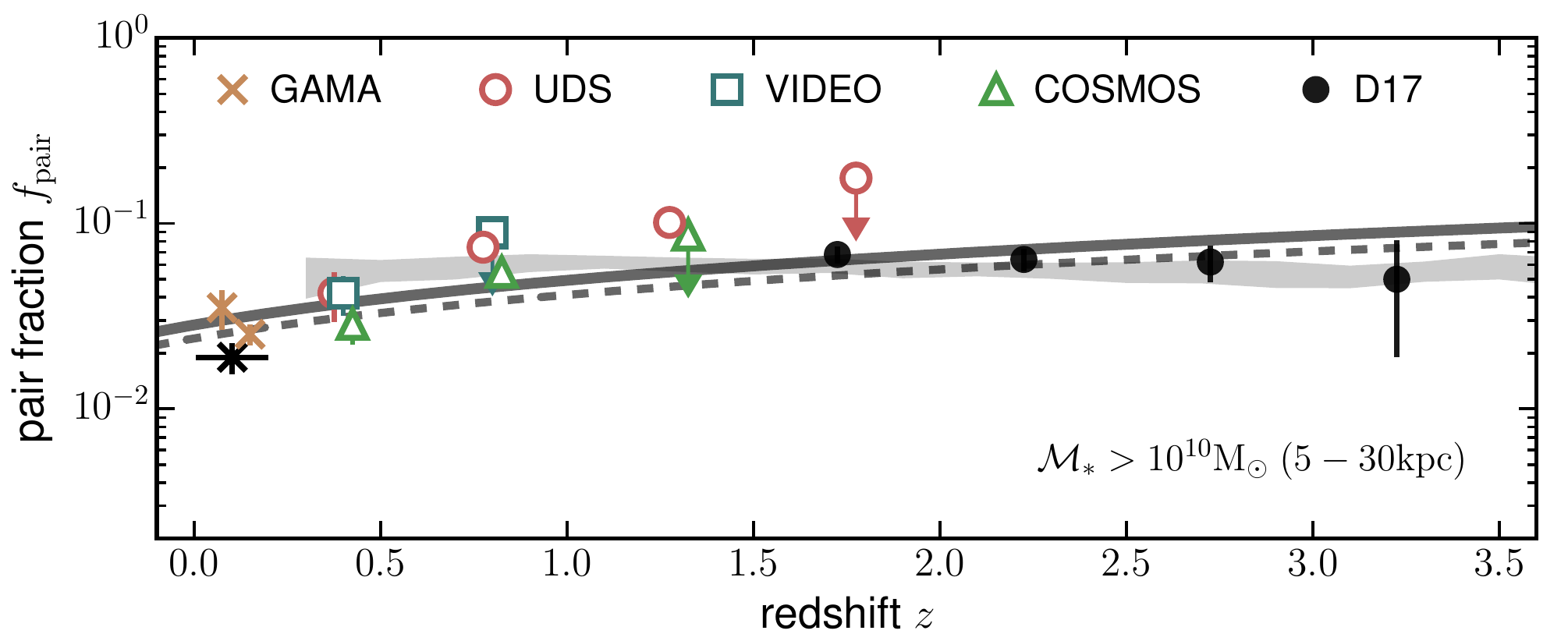}
\caption{The measured major merger ($\mu > 1/4$) pair fraction $f_\text{pair}$ for galaxies with $\log(\mathcal{M}_* / \mathrm{M}_\odot) > 10$ at physical separations of 5--30 kpc as a function of redshift in the GAMA (gold and black crosses), UDS (red circles), COSMOS (green triangles) and VIDEO (blue squares) fields. The black crosses with horizontal error bars are points measured using the GAMA spectroscopic sample, including Poisson errors and cosmic variance estimates. Results from a complimentary study within the CANDELS fields (Duncan et al., \textit{in prep}) are presented as solid black circles. Upper limits on the merger fraction are given by points with solid filled arrows. The best-fit $f_\text{pair}(z)$ for galaxies with $\log(\mathcal{M}_*/\mathrm{M}_\odot) > 11$ (Figure \ref{fig:fm-11-30kpc}) is shown as a dashed grey line for comparison. The grey shaded area represents the $1\sigma$ variation in the merger fraction as measured using 24 light cones based on the \citetalias{Henriques2014} semi-analytic model.}
\label{fig:fm-10-30kpc}
\end{figure*}

We perform the same measurement for a sample of lower stellar mass systems with $\log (\mathcal{M}_*/\mathrm{M}_\odot) > 10$. Stellar mass completeness considerations limit our measurements to $z < 1.5$. As displayed in Figure \ref{fig:fm-10-30kpc}, we find an increase in the pair fraction from $f_\text{pair} \sim 0.03$ at $z \sim 0.1$ to $f_\text{pair} \sim 0.1$ at $z \sim 1.25$. If the results of Duncan et al. (\textit{in prep}) are considered at $1.5 < z < 3.5$ in addition to those at $z < 1.5$, we find that the pair fraction remains roughly constant ($f_\text{pair} = 0.06$) to high redshift. Fitting these data, as in Section \S\ref{sec:fm:11}, we obtain \[f_\text{pair} = (0.028\pm0.002) \times (1+z)^{0.80\pm0.09}.\] When the calculated uncertainties are considered, the evolution of the pair fraction for intermediate mass galaxies is entirely consistent with that measured for the most massive galaxies in Section \S\ref{sec:fm:11}. The fit for this higher mass selection is illustrated in Figure \ref{fig:fm-10-30kpc} as the dashed black line.

The measured pair fractions in this work compare favourably to those in previous studies. Using sources in the GOODS-S and GOODS-N fields, \citet{Bundy2009} find $f_\text{pair} = 5\pm2\%$, $7\pm3\%$, $9\pm2\%$ at $z =$ 0.4--0.7, 0.7--0.9, 0.9--1.4 for galaxies with $>10^{11}\ \mathrm{M}_\odot$. \citet{Man2012} find pair fractions that fall from $15\pm8$\% at $1.7 < z < 3$ to $8\pm5$\% at $0 < z < 1$ using the COSMOS survey. Although these fractions are larger than the best-fit pair fraction found in this work, they agree within error and agree especially well with measurements within the UDS and VIDEO fields found in this work. Similarly, our results for massive galaxies are in good agreement with those of \citet{Lopez-Sanjuan2012} who measure pair fractions of 3--6\% at $0.3 < z < 0.9$. Probing galaxies with $> 2 \times 10^{11}\ \mathrm{M}_\odot$, \citet{Ruiz2014} find satellite fractions of $\sim10\%$ for at similar merger ratios within the SDSS. \citet{Bluck2009} use a morphologically selected sample of galaxies and find $f_\text{pair} = 0.29\pm0.06$ at $1.7 < z < 3$. This measurement is larger than the results found in this work by a factor of $\sim4$. This can be attributed to the selection in morphology rather than luminosity or stellar mass \citet{Man2014}. At masses $>10^{10}\ \mathrm{M}_\odot$, \citeauthor{Bundy2009} find smaller pair fractions of $f_\text{pair} = 3\pm2\%$, $5\pm3\%$, $6\pm2\%$ in the same redshift bins. This work's results over the same redshift and mass regime (3-10\%) are therefore in good agreement. While comparisons of measured pair fractions between studies can be useful, the reader is cautioned about making direct comparison without reviewing the methodologies and parameter choices employed between studies.

Additionally, the \citetalias{Henriques2014} semi-analytic model light cones predict pair fractions (solid grey shaded region in Figure \ref{fig:fm-10-30kpc}) in excellent agreement with all observations at $z > 0.3$. As with more massive samples, the cosmic variance between the light cones also appears to be reproduced. This agreement also extends to pair fractions measured at the smaller separation of 5--20 kpc.

\subsection{Constant number density selected samples}
\label{sec:fpair:n}

Selecting samples of galaxies at a constant cumulative comoving number density has been used to connect samples of galaxies across time \citep[e.g.,][]{Papovich2011, Conselice2013, Ownsworth2014, Torrey2015, Ownsworth2016}, and has been shown to be more successful at tracing galaxy populations than a selection above a constant stellar mass with redshift \citep{Behroozi2013,Leja2013, Mundy2015, Jaacks2016}. 

To provide the best estimate of the evolution of the merger histories of the progenitors of today's most massive galaxies, we measure the pair fraction for a sample of galaxies selected at a constant cumulative comoving number density of $n = 5 \times 10^{-4}$ Mpc$^{-3}$ which provides a sample of galaxies with $\mathcal{M}_* > 10^{11} \mathrm{M}_\odot$ at $z\approx0$, and galaxies with $> 10^{9.5}\mathrm{M}_\odot$ at $z\sim 3.25$. We calculate the corresponding stellar mass limit at every redshift using the galaxy stellar mass function, described further in Section \S\ref{sec:gsmf}. Making this selection, we are directly probing the progenitors of these galaxies at higher redshift \citep{Mundy2015}. This choice of number density is a trade off between satisfactory sample sizes at low redshift and avoiding mass completeness issues at high redshift.

The pair fraction evolution from this number density selection, measured at a separation of 5--30 kpc, is found to have a similar $z=0$ normalisation compared to the pair fractions measured for constant stellar mass selected samples. However, the measured slope is a factor of $\sim 2$ larger compared to galaxies at $>10^{11}\ \mathrm{M}_\odot$, and a factor of $\sim 1.5$ compared to galaxies at $>10^{10}\ \mathrm{M}_\odot$. The fitting procedure parametrises the pair fraction for this selection as \[f_\text{pair}(z) = (0.023^{+0.005}_{-0.004}) \times (1+z)^{1.22\pm0.31}.\] This fit is obtained using the pair fraction measurements in the GAMA, UDS, VIDEO and COSMOS fields at $z < 1.5$. We have measured the pair fraction on a finer redshift grid in the VIDEO field to constrain the slope of the pair fraction over this small redshift range. Measured pair fractions for this selection are listed in Table \ref{tab:fm-n}.

\begin{figure*}
\includegraphics[width=0.975\textwidth]{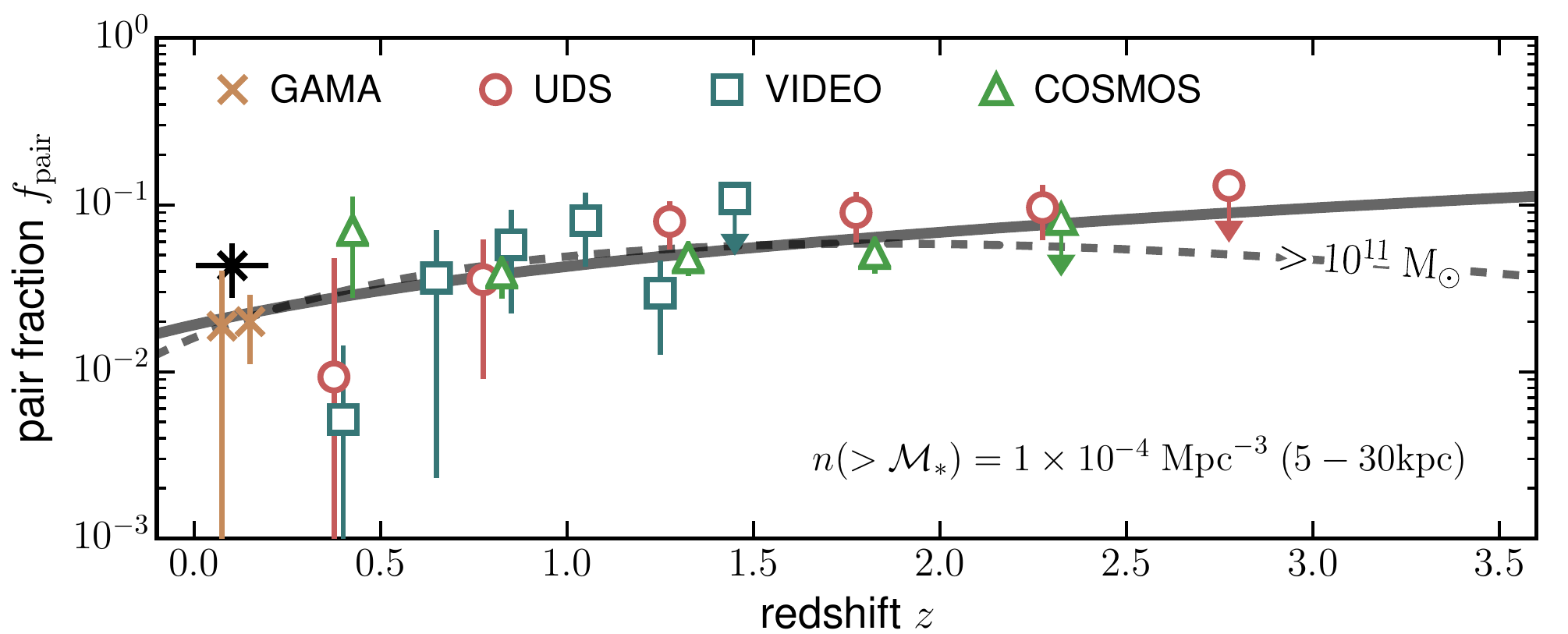}
\caption{The measured major merger ($\mu > 1/4$) pair fraction, $f_\text{pair}$, for galaxies selected at a constant cumulative number density of $n(>\mathcal{M}_*) = 1\times10^{-4}$ Mpc$^{-3}$ at physical separations of 5--30 kpc as a function of redshift in the GAMA (gold and black crosses), UDS (red circles), COSMOS (green triangles) and VIDEO (blue squares) fields. The black crosses with horizontal error bars are points measured using the GAMA spectroscopic sample, including Poisson errors and cosmic variance estimates. Upper limits on the merger fraction are given by points with solid filled arrows. The best-fit $f_\text{pair}(z)$ for galaxies with $\log(\mathcal{M}_*/\mathrm{M}_\odot) > 11$ (Figure \ref{fig:fm-11-30kpc}) is shown as a dashed grey line for comparison. Pair fractions in the VIDEO region have been measured in a finer redshift grid to provide better constraints on the slope of the pair fraction.}
\label{fig:fm-1em4-30kpc}
\end{figure*}

Probing a smaller number density selection of $n = 1 \times 10^{-4}$ Mpc$^{-3}$ provides a sample of galaxies at $0 < z < 0.2$ with stellar mass $\log( \mathcal{M}_* / \mathrm{M}_\odot) > 11.2$ and allows us to probe the progenitors of such galaxies out to a higher redshift of $z = 2.5$. Measured pair fractions are shown in Figure \ref{fig:fm-1em4-30kpc} with the best-fit parametrisation given by the solid grey curve, and the best-fit pair fraction for galaxies at $>10^{11}\ \mathrm{M}_\odot$ shown as a dashed grey line. We find a similar value for $f(z=0$) as the larger number density, but a slightly shallower evolution with redshift becoming only slightly steeper (but agreeing within the errors) than $f_\text{pair}$ measured for constant stellar mass selections at $>10^{10}\ \mathrm{M}_\odot$. Fitting the data, we find that \[f_\text{pair}(z) = (0.019^{+0.007}_{-0.006}) \times (1+z)^{1.16^{+0.42}_{-0.37}}.\] While the best-fit parameters predict a steeper evolution with increasing redshift, once the uncertainties are considered the evolution is consistent with that found for the larger number density of $n = 5 \times 10^{-4}$ Mpc$^{-3}$. Therefore we do not detect any significant change in the pair fraction evolution between these two selections. Additionally, there is no significant difference between the evolution of the pair fraction in these selections with those of a constant mass selection when the same redshift range and datasets are considered. Further exploration at higher redshift is needed to constrain this evolution and make a comparison at higher redshift.

\begin{table*}
\begin{minipage}{\textwidth}
\caption[Major merger ($\mu > 1/4$) pair fractions for constant cumulative number density selections]{Calculated major merger ($\mu > 1/4$) pair fractions, $f_\text{pair}$, and associated errors for a constant cumulative comoving number density, $n$, selected sample of galaxies. The stellar mass limit, $\mathcal{M}_*^\text{lim}$, at the corresponding number density and redshift is calculated by integrating the appropriate galaxy stellar mass function. Errors include contributions from cosmic variance, bootstrap error analysis and Poisson errors.}
\label{tab:fm-n}
\centering


\begin{tabular}{cccccc}
\hline
$z$ & $\mathcal{M}_*^\text{lim}$ & GAMA & UDS & VIDEO & COSMOS \\
 & ($\log\mathrm{M}_\odot$) & & & & \\
\hline
\multicolumn{6}{c}{$n = 5 \times 10^{-4}\ \mathrm{Mpc}^{-3}\ (5-30\mathrm{kpc})$}\\
\hline
0.0 -- 0.1 & 10.8$\pm0.1$ & $0.034\pm0.017$ & - & - & - \\
0.1 -- 0.2 & 10.8$\pm0.1$ & $0.025\pm0.006$ & - & - & - \\
0.2 -- 0.5 & 10.9$\pm0.1$ & - & $0.057\pm0.046$ & $0.011\pm0.008$ & $0.034\pm0.013$ \\
0.5 -- 0.7 & 10.9$\pm0.1$ & - &  -               & $0.053\pm0.022$ & -               \\
0.7 -- 0.9 & 10.9$\pm0.1$ & - &  -               & $0.065\pm0.022$ & -               \\
0.5 -- 1.0 & 10.9$\pm0.1$ & - & $0.065\pm0.019$ & -               & $0.041\pm0.009$ \\
0.9 -- 1.1 & 10.9$\pm0.1$ & - &  -               & $0.073\pm0.019$ & -               \\
1.1 -- 1.3 & 10.8$\pm0.1$ & - &  -               & $<0.083$ & -               \\
1.0 -- 1.5 & 10.8$\pm0.1$ & - & $0.093\pm0.022$ & -               & $0.048\pm0.008$ \\
1.5 -- 2.0 & 10.6$\pm0.1$ & - & $0.111\pm0.030$ & -               & $<0.078$ \\
2.0 -- 2.5 & 10.7$^{+0.1}_{-0.2}$ & - & $<0.319$ & - & \\
\hline
\multicolumn{6}{c}{$n = 1 \times 10^{-4}\ \mathrm{Mpc}^{-3}\ (5-30\mathrm{kpc})$}\\
\hline
0.0 -- 0.1 & 11.2$\pm0.1$ & $0.019\pm0.022$ & - & - & - \\
0.1 -- 0.2 & 11.2$\pm0.1$ & $0.020\pm0.009$ & - & - & - \\
0.2 -- 0.5 & 11.3$\pm0.1$ & - & $0.009\pm0.039$ & $0.005\pm0.009$ & $0.070\pm0.043$ \\
0.5 -- 0.7 & 11.2$\pm0.1$ & - &  -              & $0.037\pm0.034$ & -               \\
0.7 -- 0.9 & 11.2$\pm0.1$ & - &  -              & $0.058\pm0.035$ & -               \\
0.5 -- 1.0 & 11.2$\pm0.1$ & - & $0.036\pm0.027$ & -               & $0.039\pm0.012$ \\
0.9 -- 1.1 & 11.2$\pm0.1$ & - &  -              & $0.081\pm0.038$ & -               \\
1.1 -- 1.3 & 11.1$\pm0.1$ & - &  -              & $0.030\pm0.017$ & -               \\
1.0 -- 1.5 & 11.1$\pm0.1$ & - & $0.080\pm0.026$ & -               & $0.048\pm0.011$ \\
1.3 -- 1.5 & 10.9$\pm0.1$ & - &  -              & $<0.109$ & -               \\
1.5 -- 2.0 & 10.9$\pm0.1$ & - & $0.090\pm0.030$ & -               & $0.051\pm0.013$ \\
2.0 -- 2.5 & 10.7$^{+0.1}_{-0.2}$ & - & $0.096\pm0.035$ & -               & $<0.082$ \\
2.5 -- 3.0 & 10.6$\pm0.2$ & - & $<0.139$ & -               & -               \\
\hline
\end{tabular}
\end{minipage}
\end{table*}

\subsection{Comparison between spectroscopically and photometrically determined merger fractions}
\label{sec:comp-spec-photo-fm}

The extraordinarily high spectroscopic completeness ($>97$\%) of the GAMA region (see \citealt{Baldry2010,Baldry2014}, \citealt{Robotham2010} and \citealt{Hopkins2013} for details on the spectroscopic targeting campaign and subsequent analysis) allowed us to perform several tests. We compared measured merger fractions in the GAMA region at $z < 0.2$ in two ways: spectroscopically and photometrically. To perform the measurement spectroscopically some tolerance in redshift must be chosen, translating to a cut in relative velocities between the galaxies in a close-pair system. Previous studies have chosen a relative velocity offset of $|\Delta v| < 500$ km/s \citep[e.g.,][]{Patton2000,Lin2004,Lin2008,DeRavel2009,Lopez-Sanjuan2012} in order to select close-pairs with a high probability of coalescence. We therefore enforce this condition when measuring the spectroscopic pair fraction in the GAMA region (see black crosses with horizontal error bars in Figures \ref{fig:fm-11-30kpc} and \ref{fig:fm-10-30kpc}).

Pair fractions measured with photometric and spectroscopic redshifts for massive galaxies ($\mathcal{M}_* > 10^{11} \mathrm{M}_\odot$; Section \S\ref{sec:fm:11}) are found to be in excellent agreement. Using the calculated GAMA photometric redshifts we find a pair fraction of $f_\text{pair} = 0.030\pm0.022$ at $0.005 < z < 0.1$, and $f_\text{pair} = 0.025\pm0.008$ at $0.1 < z < 0.2$. Performing the analysis at $0.005 < z < 0.2$ using the available spectroscopic redshifts instead, we obtain $f_\text{pair} = 0.041\pm0.013$, in good agreement with the photometric analysis. Intermediate mass galaxies ($\mathcal{M}_* > 10^{10} \mathrm{M}_\odot$; Section \S\ref{sec:fm:10}) possess photometric pair fractions of $f_\text{pair} = 0.035\pm0.008$ and $f_\text{pair} = 0.025\pm0.003$ within the same redshift bins. Again performing the analysis spectroscopically we find $f_\text{pair} = 0.019\pm0.004$. This close agreement suggests that the criteria we enforce on the photometric redshift probability distributions of the galaxies is equivalent to enforcing a cut of $\Delta v < 500$ km/s in relative velocity. Similar agreement is also seen at the smaller separation of 5--20 kpc, and for selections of galaxies at the number densities probed in this work. The observed consistency between the results of performing the analysis photometrically and spectroscopically suggests that the two methods perform equivalent measurements.

\section{Major merger rates}
\label{sec:merger-rates}

While the fraction of galaxies undergoing a merger event within a particular sample is a useful quantity, the ultimate goal is to measure the rate at which a typical galaxy (or population of galaxies) undergo merging events. To achieve this, the merger fraction must be converted to a merger rate. The following section describes the process we follow to do this.

\subsection{Calculating the merger rate from the pair fraction}
\label{sec:rates:m}

Whereas merger fractions obtained via different methods may not necessarily be directly comparable, derived merger rates are if the typical timescale over which each method can observe a galaxy merger is known. The conversion to merger rates is strongly dependent on the method of choice (e.g. close-pairs) and is sensitive to various parameter choices (e.g. physical separation). We follow \citet{Lotz2011} in deriving merger rates from the merger fractions presented in Section \S\ref{sec:merger-fractions} and we refer to interested reader to this paper for a concise and thorough introduction to the topic.

Two measures of the merger rate are often used in the literature. These are the volume-averaged galaxy merger rate, $\Gamma_\mathrm{merg}$, and the fractional galaxy merger rate, $\mathcal{R}_\mathrm{merg}$. The difference between these two quantities is important: $\Gamma(z)$ traces the number of merging events per unit comoving volume above a mass limit, while $\mathcal{R}(z)$ encodes the number of mergers per massive galaxy \citep{Lotz2011}. The volume-averaged merger rate is defined as 
\begin{equation}
	\label{eqn:gamma}
	\Gamma_\mathrm{merg}(z) = \frac{\phi_\mathrm{merg}(z)}{\left< T_\mathrm{obs} \right>} = \frac{f_\text{merg}(z) n_1(z)}{\left< T_\mathrm{obs} \right>}, \hspace{5mm} [\text{Mpc}^{-3}\ \text{Gyr}^{-1}]
\end{equation}
and the fractional merger rate defined as 
\begin{equation}
	\label{eqn:R}
	\mathcal{R}_\mathrm{merg}(z) = \frac{f_\mathrm{merg}(z)}{\left< T_\mathrm{obs} \right>},  \hspace{5mm} [\text{Gyr}^{-1}]
\end{equation}
where ${\left< T_\mathrm{obs} \right>}$ is the average timescale during which a merger can be observed given the method used to identify it, $n_1(z)$ is the number density of the primary sample, ${\phi_\text{merg}}$ is the number density of mergers, and ${f_\mathrm{merg}}$ is the merger fraction. As we directly measure only the \textit{pair} fraction, a correction must be made such that $f_\mathrm{merg} = C_\textrm{merg} f_\mathrm{pair}$, where $C_\mathrm{merg}$ is the fraction of pairs that will eventually result in a merger event. This is typically taken to be $C_\text{merg} = 0.6$ \citep{Lotz2011} and so we continue this convention, however we note the large uncertainty on this value and its origin going forward. This number expresses our uncertainty in the fraction of galaxies in pairs which will eventually merge.

The number density of the primary sample, $n_1(z)$, is calculated by integrating the GSMF at the appropriate redshift between the stellar mass limits of $\mathcal{M}_*^\text{min}(z) < \mathcal{M}_*(z) < \mathcal{M}_*^\text{max}(z)$, where the maximum stellar mass considered is $10^{12} \mathrm{M}_\odot$.

We assume values of ${\left< T_\mathrm{obs} \right>} = 0.60$ Gyr for close-pairs selected at 5--30 kpc, and ${\left< T_\mathrm{obs} \right>} = 0.32$ Gyr for close-pairs selected at 5--20 kpc \citep{Lotz2011}. Like the value of $C_\textrm{merg}$,  the value of ${\left< T_\mathrm{obs} \right>}$ is uncertain.  It is relatively simple, however, to correct the results presented in this work to other combinations of $C_\text{merg}$ and $\left<T_\text{obs}\right>$, as these values are simply constants in any integrations performed. For this purpose, we define the ratio of these two quantities as \[\eta_\text{merg} = \frac{C_\text{merg}}{\left<T_\text{obs}\right>}. \hspace{5mm} [\text{Gyr}^{-1}] \] The $r_p < 30$ kpc merger \textit{rates} used in this work therefore correspond to $\eta_\text{merg} = 1$, while the $r_p < 20$ kpc merger rates assume $\eta_\text{merg} = 1.875$. If one then wished to correct the merger rates, the estimated number of major mergers undergone by a galaxy, or even the stellar mass accrued through major mergers for a different value of $\eta$, simply multiple the values quoted in this paper by a factor of $\eta_\text{new} / \eta_\text{old}$. We note that the timescales assumed in this work differ to those described by \citet{Kitzbichler2008}, a commonly used reference in the galaxy merger literature. Using their equation 9 to calculate the merger timescale, $T_\text{merge}$, for the stellar mass regimes probed in this work results in $T_\text{merge} =$ 1.1--1.8 Gyr. Note, however, that this is a merger timescale and not an \textit{observability} timescale. Additionally, the former quantity inherently includes the probability of a merger between close-pairs and therefore $T_\text{merge}$ is not directly comparable to this work's definition of $\eta$. 

Using these values for the observability timescale, which are measured using a suite of simulations, we find remarkable agreement between derived merger rates of both 20 and 30 kpc separations. For the sake of brevity, and the advantage of larger number statistics we only report merger rates derived from 5--30 kpc pair fractions in the text and figures. We fit the derived merger rate points with either a simple power law of the same form as fitted to the pair fraction, and with a combined power law and exponential. The choice of fitting form is determined using the $\chi^2$ goodness-of-fit parameter. Fitted volume-averaged and fractional merger rates at both separations are listed in Table \ref{tab:rates:gamma} and Table \ref{tab:rates:rmerg}.

\begin{table}
\caption{Fitting parameters for the volume-averaged merger rate, $\Gamma_\text{merg}(z)$, as given in Equation \ref{eqn:gamma}, for various combinations of surveys used within this work. Fits with two parameters are of the form $\Gamma_\text{merg}(z) = \Gamma_0 (1+z)^{m_\Gamma}$, while those with three parameters are of the form $\Gamma_\text{merg}(z) = \Gamma_0 (1+z)^{m_\Gamma} \exp(-{c_\Gamma}z)$. Appropriate fitting forms are decided by comparing the goodness of fit using the $\chi^2$. Parameters and their associated uncertainties are calculating using a bootstrap technique, accounting for uncertainties on the pair fraction and GSMF.}
\label{tab:rates:gamma}
\centering
\begin{tabular}{lccc}
\hline
Survey & $\Gamma_0$ & $m_\Gamma$ & $c_\Gamma$ \\
       & (Mpc$^{-3}$ Gyr$^{-1}$) &  & \\
\hline
\multicolumn{4}{c}{$\mathcal{M}_* > 10^{10}\mathrm{M}_\odot\ (5-20\mathrm{kpc})$} \\
\hline
All & $6.45^{+5.57}_{-3.33} \times 10^{-5}$ & $1.35^{+1.14}_{-1.12}$ & - \\
All + GAMA & $1.64_{-0.41}^{+0.58} \times 10^{-4}$ & $0.48^{+1.00}_{-1.15}$ & - \\

\hline
\multicolumn{4}{c}{$\mathcal{M}_* > 10^{10}\mathrm{M}_\odot\ (5-30\mathrm{kpc})$} \\
\hline
All & $0.55^{+1.84}_{-0.43} \times 10^{-4}$ & $6.66^{+12.80}_{-13.40}$ & $3.37^{+7.75}_{-7.41}$ \\
All + GAMA & $1.11^{+0.47}_{-0.35} \times 10^{-4}$ & $0.56^{+0.77}_{-0.87}$  & - \\
All + GAMA + D17 & $1.00^{+0.64}_{-0.52} \times 10^{-4}$ & $4.22^{+5.00}_{-3.73}$ &  - \\

\hline
\multicolumn{4}{c}{$\mathcal{M}_* > 10^{11}\mathrm{M}_\odot\ (5-20\mathrm{kpc})$} \\
\hline
All & $1.10^{+2.12}_{-0.89} \times 10^{-5}$ & $3.15^{+9.18}_{-5.48}$ & $2.40^{+2.65}_{-4.70}$ \\
All + GAMA & $1.05^{+0.71}_{-0.58} \times 10^{-5}$ & $3.53^{+5.95}_{-3.83}$ & $2.61^{+2.03}_{-3.26}$ \\

\hline
\multicolumn{4}{c}{$\mathcal{M}_* > 10^{11}\mathrm{M}_\odot\ (5-30\mathrm{kpc})$} \\
\hline
All & $0.85^{+2.06}_{-0.73} \times 10^{-5}$ & $6.58^{+12.01}_{-6.53}$ & $4.10^{+3.09}_{-6.28}$ \\
All + GAMA & $4.22^{+4.39}_{-2.90} \times 10^{-6}$ & $7.34^{+7.25}_{-4.55}$ & $4.20^{+2.20}_{-3.75}$ \\
All + GAMA + D17 & $6.61^{+5.27}_{-4.56} \times 10^{-6}$ & $9.21^{+8.87}_{-4.75}$ & $5.62^{+2.61}_{-5.08}$ \\

\hline
\multicolumn{4}{c}{$n(>\mathcal{M}_*) = 1 \times 10^{-4}\ \mathrm{Mpc}^{-3} (5-30\mathrm{kpc})$} \\
\hline
All & $1.29^{+0.94}_{-0.64} \times 10^{-6}$ & $1.62^{+0.81}_{-0.75}$ & - \\
All + GAMA & $1.46_{-0.64}^{+0.74} \times 10^{-6}$ & $1.45^{+0.70}_{-0.59}$ & - \\

\hline
\multicolumn{4}{c}{$n(>\mathcal{M}_*) = 5 \times 10^{-4}\ \mathrm{Mpc}^{-3} (5-30\mathrm{kpc})$} \\
\hline
All & $1.44^{+0.12}_{-0.72} \times 10^{-5}$ & $1.05^{+1.00}_{-0.95}$ & - \\
All + GAMA & $1.26^{+0.51}_{-0.43} \times 10^{-5}$ & $1.23^{+0.63}_{-0.60}$ & - \\

\hline
\end{tabular}
\end{table}
\begin{table}
\caption{Fitting parameters for the fractional merger rate, $\mathcal{R}_\text{merg}(z)$, as given in Equation \ref{eqn:R}, for various combinations of surveys used within this work. Fits with two parameters are of the form $\mathcal{R}_\text{merg}(z) = \mathcal{R}_0 (1+z)^{m_\mathcal{R}}$, while those with three parameters are of the form $\mathcal{R}_\text{merg}(z) = \mathcal{R}_0 (1+z)^{m_\mathcal{R}} \exp(-{c_\mathcal{R}}z)$. Appropriate fitting forms are decided by comparing the goodness of fit using the $\chi^2$. Parameters and their associated uncertainties are calculating using a bootstrap technique.}
\label{tab:rates:rmerg}
\centering


\begin{tabular}{lccc}
\hline
Survey & $\mathcal{R}_0$ & $m_\mathcal{R}$ & $c_\mathcal{R}$ \\
       & (Gyr$^{-1}$) &  & \\
\hline
\multicolumn{4}{c}{$\mathcal{M}_* > 10^{10}\mathrm{M}_\odot\ (5-20\mathrm{kpc})$} \\
\hline
All & $9.86^{+2.73}_{-1.90} \times 10^{-3}$ & $2.87^{+0.35}_{-0.35}$ & - \\
All + GAMA & $1.82^{+0.22}_{-0.21} \times 10^{-2}$ & $1.87^{+0.23}_{-0.22}$ & - \\

\hline
\multicolumn{4}{c}{$\mathcal{M}_* > 10^{10}\mathrm{M}_\odot\ (5-30\mathrm{kpc})$} \\
\hline
All & $1.73^{+0.22}_{-0.19} \times 10^{-2}$ & $2.17^{+0.19}_{-0.20}$ & - \\
All + GAMA & $1.94^{+0.13}_{-0.12} \times 10^{-2}$ & $1.9^{+0.12}_{-0.12}$ & - \\
All + GAMA + D17 & $1.73^{+0.15}_{-0.14} \times 10^{-2}$ & $4.13^{+0.49}_{-0.49}$ & $1.41^{+0.26}_{-0.26}$ \\

\hline
\multicolumn{4}{c}{$\mathcal{M}_* > 10^{11}\mathrm{M}_\odot\ (5-20\mathrm{kpc})$} \\
\hline
All & $2.55^{+0.87}_{-0.69} \times 10^{-2}$ & $0.79^{+0.35}_{-0.35}$ & - \\
All + GAMA & $2.83^{+0.62}_{-0.57} \times 10^{-2}$ & $0.68^{+0.27}_{-0.25}$ & - \\

\hline
\multicolumn{4}{c}{$\mathcal{M}_* > 10^{11}\mathrm{M}_\odot\ (5-30\mathrm{kpc})$} \\
\hline
All & $1.30^{+0.58}_{-0.43} \times 10^{-2}$ & $3.83^{+1.49}_{-1.38}$ & $1.32^{+0.57}_{-0.65}$ \\
All + GAMA & $1.76^{+0.41}_{-0.40} \times 10^{-2}$ & $2.87^{+1.06}_{-0.92}$ & $0.95^{+0.42}_{-0.49}$ \\
All + GAMA + D17 & $1.79^{+0.44}_{-0.38} \times 10^{-2}$ & $2.79^{+1.05}_{-0.95}$ & $0.93^{+0.44}_{-0.49}$ \\

\hline
\multicolumn{4}{c}{$n(>\mathcal{M}_*) = 1 \times 10^{-4}\ \mathrm{Mpc}^{-3} (5-30\mathrm{kpc})$} \\
\hline
All & $1.82^{+0.91}_{-0.65} \times 10^{-2}$ & $1.22^{+0.47}_{-0.45}$ & - \\
All + GAMA & $1.76^{+0.57}_{-0.50} \times 10^{-2}$ & $1.26^{+0.37}_{-0.32}$ & - \\

\hline
\multicolumn{4}{c}{$n(>\mathcal{M}_*) = 5 \times 10^{-4}\ \mathrm{Mpc}^{-3} (5-30\mathrm{kpc})$} \\
\hline
All & $1.84^{+0.58}_{-0.47} \times 10^{-2}$ & $1.51^{+0.38}_{-0.36}$ & - \\
All + GAMA & $2.02^{+0.36}_{-0.33} \times 10^{-2}$ & $1.39^{+0.24}_{-0.21}$ & - \\
 
\hline
\end{tabular}
\end{table}

\begin{figure*}
\includegraphics[width=0.9\textwidth]{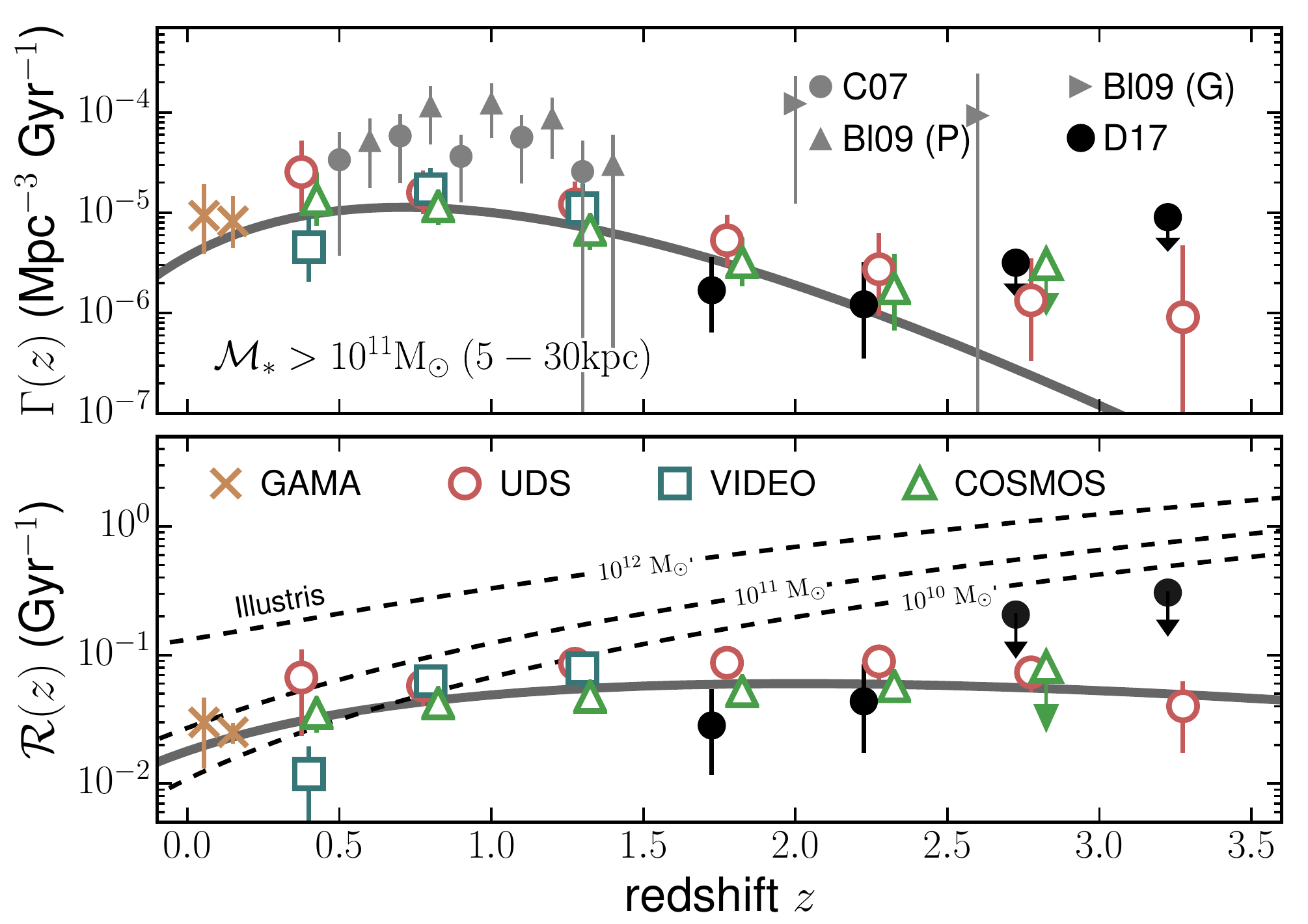}
\caption{Derived volume-averaged (top) and fractional (bottom) major merger rates for galaxies at $\log(\mathcal{M}_* / \mathrm{M}_\odot) >  11$ at $5 < r\ [\mathrm{kpc}] < 30$ in the GAMA (gold crosses), UDS (red circles), VIDEO (blue squares) and COSMOS (green triangles) regions. Error bars include contributions from a bootstrap error analysis, cosmic variance estimates and Poisson statistics, combined in quadrature. Data points from \citet{Conselice2007} and \citet[][P: close-pair data; G: morphological data]{Bluck2009} are shown for comparison. Illustris major merger rates for galaxies with stellar masses of $10^{10}$, $10^{11}$ and $10^{12}$ $\mathrm{M}_\odot$ are shown as dashed black lines.}
\label{fig:mr-11-30kpc}
\end{figure*}

Galaxies at $\log(\mathcal{M}_*/\mathrm{M}_\odot) > 11$ exhibit a constant volume-averaged merger rate (top panel in Figure \ref{fig:mr-11-30kpc}) of $\Gamma \sim 10^{-5}$ Mpc$^{-3}$ Gyr$^{-1}$ at $z < 1.5$, which declines steadily by a factor of $\sim10$ towards higher redshift such that, at $z = 3.25$, we find $\Gamma \sim 10^{-6}$ Mpc$^{-3}$ Gyr$^{-1}$. This is attributed to the decrease in the number density of such massive galaxies. \citet{Conselice2007} estimate the merger rate of a morphologically selected sample using the same stellar mass criteria at $0.4 < z < 1.4$ as $2.0^{+3.0}_{-1.6} \times 10^{-5}$ Mpc$^{-3}$ Gyr$^{-1}$ which is consistent with our findings. \citet{Bluck2009} measure merger rates for a similar sample at high redshift and find a merger rate of $\Gamma < 1.2 \times 10^{-4}$ Mpc$^{-3}$ Gyr$^{-1}$ at $z = 0.5$, and at $z = 2.6$ find $\Gamma < 5 \times 10^{-4}$ Mpc$^{-3}$ Gyr$^{-1}$. These upper limits are consistent with the results presented here. As seen in Figure \ref{fig:mr-11-30kpc} we find our derived merger rates at $z<1.5$ to be a factor of $\sim 2$ smaller than those described in the aforementioned literature sources, although we note that we are typically consistent within $2\sigma$. This discrepancy is attributed to a number of factors. \citet{Bluck2009} find significantly higher pair fractions than this work; approximately $\sim5\%$ at $0.5 < z < 1.5$, and $\sim30\%$ at $2 < z < 3$. These are a factor of $\sim 2$ and $\sim 4$ larger, respectively, which, coupled with merger timescales of 0.4$\pm$0.2 Gyr (close-pair sample) and 1.0$\pm$0.2 Gyr (CAS sample) that \citeauthor{Bluck2009} adopt, makes their derived merger rates a factor of $\sim2$ larger at low redshift, and a significant factor larger at high redshift (see their Section 3.2). 

\begin{figure*}
\includegraphics[width=0.9\textwidth]{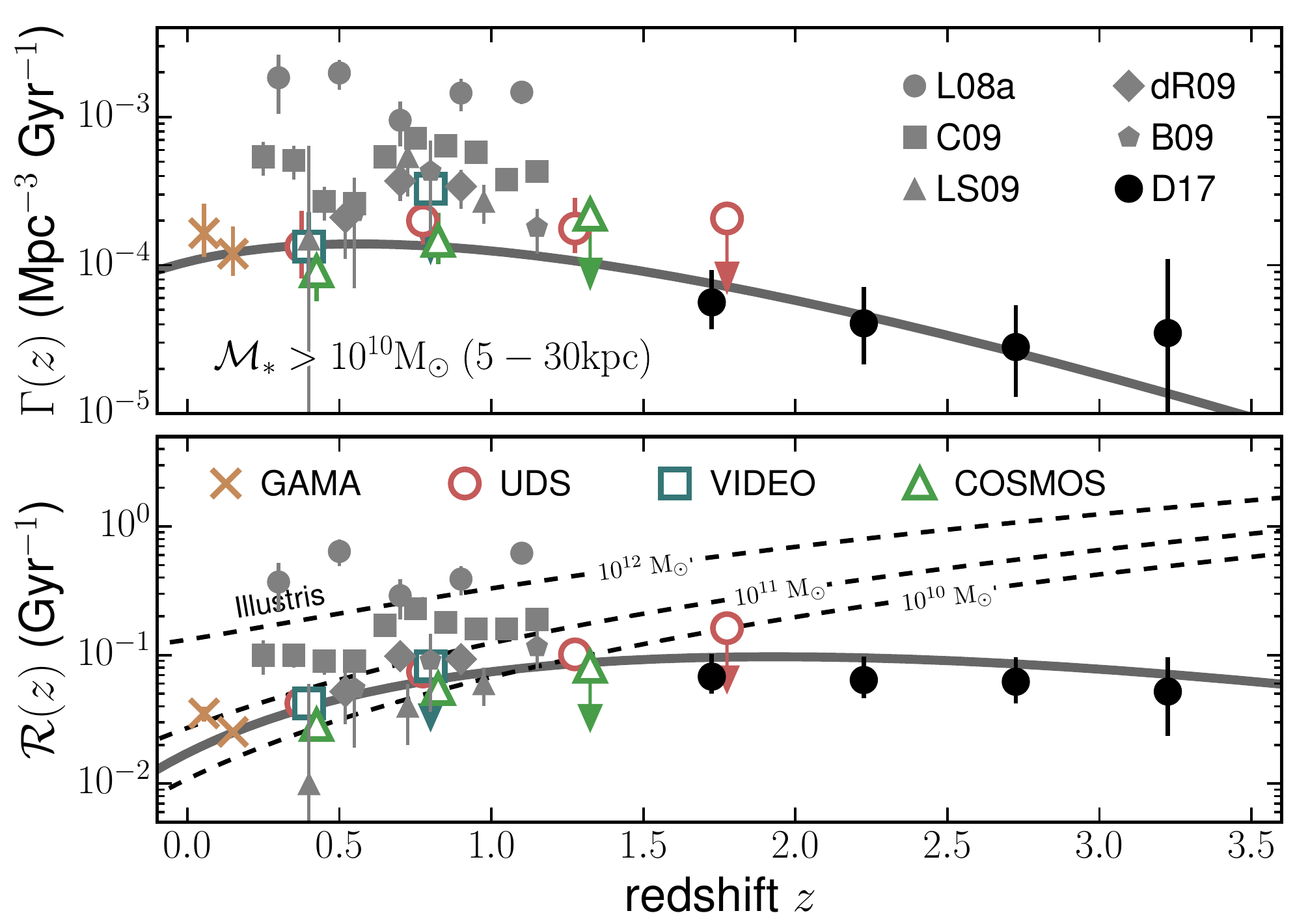}

\caption{Derived volume-averaged (top) and fractional (bottom) major merger rates for galaxies at $\log(\mathcal{M}_* / \mathrm{M}_\odot) >  10$ in the GAMA (gold crosses), UDS (red circles), VIDEO (blue squares) and COSMOS (green triangles) regions. Error bars include contributions from a bootstrap error analysis, cosmic variance estimates and Poisson statistics, combined in quadrature. Data points, compiled in \citet{Lotz2011}, from \citet{Lotz2008,Conselice2009a,Lopez-Sanjuan2009,DeRavel2009} and \citet{Bundy2009} are shown as grey symbols for comparison. Illustris major merger rates for galaxies with stellar masses of $10^{10}$, $10^{11}$ and $10^{12}$ $\mathrm{M}_\odot$ are shown as dashed black lines.}
\label{fig:mr-10-30kpc}
\end{figure*}

Galaxies with $\log(\mathcal{M}_*/\mathrm{M}_\odot) > 10$ exhibit a qualitatively similar evolution of the volume-averaged merger rate, shown in the top panel of Figure \ref{fig:mr-10-30kpc}. However the rate is typically an order of magnitude greater than that derived for the higher stellar mass sample. At $z < 1.5$ we find an approximately constant $\Gamma \sim 10^{-4}$ Mpc$^{-3}$ Gyr$^{-1}$. Considering the derived merger rates using the pair fractions obtained by Duncan et al. (\textit{in prep}) in the CANDELS fields extends the measurement at this stellar mass range to $z = 3.25$. We find a steep decline of $\Gamma$ at $z > 1.5$ such that at $z = 3.25$, $\Gamma \sim 3 \times 10^{-5}$ Mpc$^{-3}$ Gyr$^{-1}$, albeit with an uncertainty of a factor of $\sim 5$. We compare our derived merger rates with a selection of literature rates \citep{Lotz2008, Bluck2009, Conselice2009a, DeRavel2009,Lopez-Sanjuan2009} compiled in \citet{Lotz2011}. These are shown in Figure \ref{fig:mr-11-30kpc} as solid grey markers. Our results are consistent with rates derived in \citet{Bluck2009}, \citet{Lopez-Sanjuan2009} and \citet{DeRavel2009}, however our best-fit rates are consistently a factor of $\sim 2$ smaller than the average literature merger rate. The derived fractional merger rate shows a clear evolution to larger values with increasing redshift and is consistent with the results of \citet{Lopez-Sanjuan2009} and \citet{DeRavel2009}, where overlap allows comparison. We also find that the discrepancy between COSMOS and the other survey regions is reduced when probing this stellar mass range, suggesting the cause of the discrepancy seen in Figure \ref{fig:fm-11-30kpc} is limited to higher mass galaxies. Cosmic variance likely contributes to the observed discrepancy, as it affects observations of the most massive objects more \citep{Somerville2004,Driver2010,Moster2011}. However, it most likely cannot explain the systematic offset of the COSMOS field. We discuss this issue and the steps taken to identify the cause further in Section \S\ref{sec:fm-tests}.

\subsection{Number of merger events at $z < 3.5$}
\label{sec:merger-rates:nmerg}
The number of merger events a typical galaxy within each primary sample goes through between two redshifts can be approximated by integrating over the average time between merger events with respect to time. This typical timescale is given by $\left< T_\mathrm{obs} \right> / f_\text{merg}(z) = \mathcal{R}_\text{merg}(z)^{-1}$, where $\left< T_\mathrm{obs} \right>$ is the average time during which a merger can be observed, as in Equations \ref{eqn:gamma} and \ref{eqn:R}. The number of mergers, $N_\text{merg}$, between two redshift bins is then simply given by 
\begin{equation}
	\label{eqn:n_merg}
	N_\text{merg} = \int_{t_1}^{t_2} \mathcal{R}_\text{merg}(z)\ dt = \int_{z_1}^{z_2} \frac{\mathcal{R}_\text{merg}(z)}{(1+z)H(z)}\ \mathrm{d}z,
\end{equation}
where the substitution $\mathrm{d}t = \mathrm{d}z / (1+z)H(z)$ has been made. Here $H(z)$ is the Hubble constant at redshift $z$, alternatively defined as $H(z) = H_0(\Omega_M(1+z)^3 + \Omega_\Lambda)^{1/2}$. 

Performing this integration between $0 < z < 3.5$ and assuming a conservative $30\%$ uncertainty on the value of $\left< T_\mathrm{obs} \right>$, we find that a galaxy with $\log(\mathcal{M}_*/\mathrm{M}_\odot) > 11$ undergoes $0.5^{+0.3}_{-0.1}$ major mergers between these times. Lower stellar mass galaxies, with $\log(\mathcal{M}_*/\mathrm{M}_\odot) > 10$ undergo $0.5^{+0.3}_{-0.1}$ major mergers, approximately the same as higher mass galaxies. This means that, on average, one out of every two galaxies with $>10^{10} \mathrm{M}_\odot$ has undergone a single major merger over the last 12 Gyr.

For the most massive galaxies, our calculated value of $N_\text{merg}$ is a factor of $\sim2$ smaller than that reported in \citet{Ownsworth2014}, which calculated $N_\text{merg} = 1.2\pm0.5$ using a fit to merger fractions from several literature sources \citep{Bluck2009, Bundy2009, Xu2012, Ruiz2014} who employ a range of values of $C_\text{merg} \approx 0.5 - 1.0$. Furthermore, their fitting parameters are driven by the large merger fractions at high-redshift ($z > 1.5$) from \citet{Bluck2009} and the $z = 0$ point of \citet{Xu2012}, and are obtained from works with various definitions and sample selections. Our calculated value for the number of mergers experienced by the most massive galaxies is also at odds with that calculated in \citet{Man2014}, who for galaxies at $>10^{10.8}\ \mathrm{M}_\odot$ find $N_\text{merg} = 0.9\pm0.2$ ($1.1\pm0.4$) at $0.1 < z < 2.5$ using UltraVISTA (3DHST + CANDELS) at 10--30 kpc $h^{-1}$. This larger value is attributed to the larger observed pair fractions of 5--10\% at $0.5 < z < 2.5$, a factor of $\sim 2$ larger than found in this work's best fit merger fraction parametrisation.

\subsection{Stellar mass added by mergers}
\label{sec:merger-rates:mass-added}

Ultimately, we wish to uncover the role of galaxy mergers in the grander picture of galaxy formation. The stellar mass accrued through major mergers is an important quantity that allows comparisons to be made between other pathways of stellar mass growth such as star-formation, however knowing the rate at which a merger event occurs for a given sample of galaxies is not enough to calculate this quantity between two redshifts. The average stellar mass of a companion galaxy must be known as well. With this information, the additional stellar mass from mergers, $\mathcal{M}_*^{+}$, for a typical primary sample galaxy between two redshifts can be estimated as 
\begin{equation}
\label{eqn:added_mass}
	\mathcal{M}_*^{+} = \int_{t_1}^{t_2} \mathcal{R}_\text{merg}(z)\ \mathcal{M}_{*,2}(z)\ \mathrm{d}t,
\end{equation}
where $\mathcal{R}_\text{merg}$ is the fractional merger rate, defined in Equation \ref{eqn:R} in terms of the pair fraction, and $\mathcal{M}_{*,2}(z)$ is the average stellar mass of a close-pair companion at redshift $z$. We use the GSMF to calculate these quantities as it minimises the effects of cosmic variance that may be present if it is measured by simply taking an average of stellar masses using the datasets used in this work. Using the GSMF thus allows for a statistical, cosmologically-averaged analysis to be performed. For completeness, we note that calculating the average stellar mass both ways results in values that agree within $\leq 0.1$ dex at high masses, and $\leq 0.3$ dex at lower masses.

Within any redshift bin the GSMF, $\phi(z, \mathcal{M}_*)$, can be used to calculate the average stellar mass of a galaxy within the primary sample, and is defined as
\begin{equation}
\label{eqn:pri_mass}
	\mathcal{M}_{*,1}(z) = \frac{\int^{\mathcal{M}_{*,1}^{\text{max}}}_{\mathcal{M}_{*,1}^{\text{min}}} \phi(z, \mathcal{M}_*)\ \mathcal{M}_*\ \mathrm{d}\mathcal{M}_*}{\int^{\mathcal{M}_{*,1}^{\text{max}}}_{\mathcal{M}_{*,1}^{\text{min}}} \phi(z, \mathcal{M}_*)\ \mathrm{d}\mathcal{M}_*},
\end{equation}
where $\mathcal{M}_{*,1}^{\text{max}}$ and $\mathcal{M}_{*,1}^{\text{min}}$ are the maximum and minimum stellar masses of the primary galaxy sample, respectively. A similar integration is performed to calculate the average stellar mass of a companion galaxy, $\mathcal{M}_{*,2}(z)$, whereby the integration in Equation \ref{eqn:pri_mass} is instead performed between the stellar mass limits of $\mathcal{M}_{*,1}$ and $\mu \mathcal{M}_{*,1}$. Armed with this information, we calculate the stellar mass added through major mergers alone. Uncertainties are estimated using a bootstrap approach, accounting for errors on the galaxy stellar mass function parameters and the uncertainty in the fit of $f_\text{pair}$. We note that the values of $\mathcal{M}_*^+$, and the major merger stellar mass accretion rate density presented in Section \S\ref{sec:rho_major_merger}, do not account for new stars created in star-formation episodes triggered by mergers.

Using the fitted pair fractions (see Table \ref{tab:fm:fits}) using data points from GAMA, UDS, VIDEO, COSMOS and D17, we find  that galaxies with $\mathcal{M}_* > 10^{11} \mathrm{M}_\odot$ are estimated to accrete an average stellar mass of $\log(\mathcal{M}_*^{+}/\mathrm{M}_\odot) = 10.6\pm0.2$ at $0 < z < 3.5$. This is in excellent agreement with \citet{Man2014} who find galaxies at $>10^{10.8}\ \mathrm{M}_\odot$ achieve a stellar mass growth of $4\times10^{10}\ \mathrm{M}_\odot$ at $0.1 < z < 2.5$ from major mergers. Similarly, over the same redshift range, galaxies with $\mathcal{M}_* > 10^{10} \mathrm{M}_\odot$ are expected to accrete $\log(\mathcal{M}_*^{+}/\mathrm{M}_\odot) = 10.1^{+0.2}_{-0.1}$ from major mergers. As the typical galaxy in each of these samples at $z \sim 3.25$ is $\log(\mathcal{M}_* / \mathrm{M}_\odot) = 11.2\pm0.1$ and $\log(\mathcal{M}_* / \mathrm{M}_\odot) = 10.5\pm0.1$, respectively, this represents an average increase in stellar mass of $23^{+14}_{-10}\%$, and $36^{+23}_{-18}\%$, respectively, due solely to major mergers. The average stellar mass of primary and secondary samples and the stellar mass gained through major mergers are tabulated in Table \ref{tab:arates}.

\subsection{Merger rates at a constant cumulative number density}
\label{sec:rates:n}

\begin{figure*}
\includegraphics[width=0.975\textwidth]{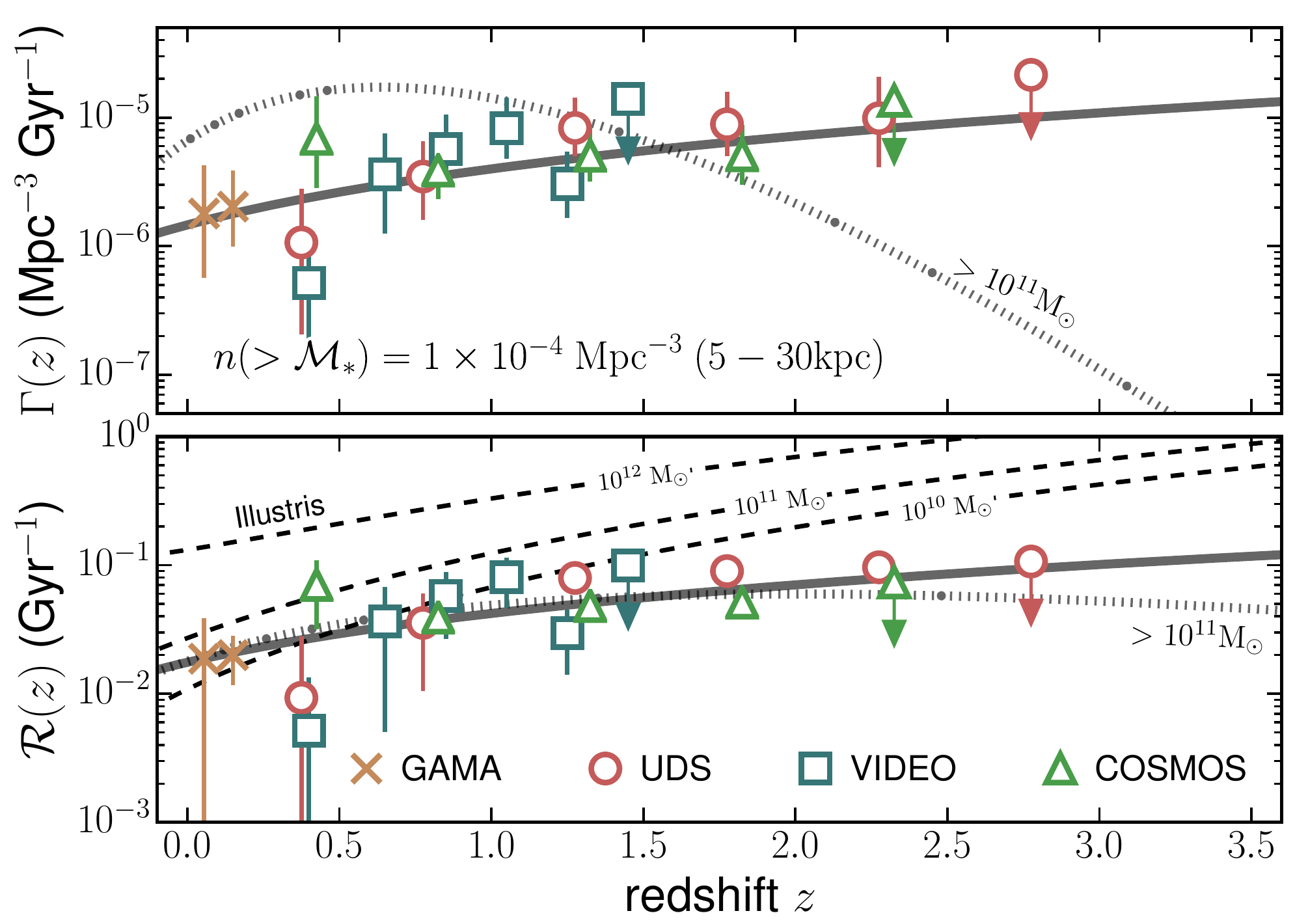} 

\caption{Derived volume-averaged (top) and fractional (bottom) major merger rates for galaxies at a constant cumulative comoving number density of $n(>\mathcal{M}_*) = 1 \times 10^{-4}$ Mpc$^{-3}$ at 5--30 kpc in the GAMA (gold crosses), UDS (red circles), VIDEO (blue squares) and COSMOS (green triangles) regions. Error bars include contributions from a bootstrap error analysis, cosmic variance estimates and Poisson statistics, combined in quadrature. Illustris major merger rates at $10^{10}$, $10^{11}$ and $10^{12}$ $\mathrm{M}_\odot$ are shown as dashed black lines. These are obtained by integrating the galaxy-galaxy merger rate parametrisation given in Table 1 of \citet{Rodriguez-Gomez2015} with respect to the stellar mass merger ratio between $0.25 < \mu < 1.0$. Best-fit relations, as described in the text, are shown as solid grey lines. The dotted grey line represents the derived merger gates for galaxies with $>10^{11} \mathrm{M}_\odot$.}
\label{fig:fm:rates:1em4}
\end{figure*}

We calculate merger rates for the two number density selections first introduced in Section \S\ref{sec:fpair:n}. Both number density selections exhibit a slight increase in the volume-averaged merger rate over the studied redshift range. This is in contrast to the declining rate observed in constant stellar mass selections for galaxies at $>10^{10}\ \mathrm{M}_\odot$. The fractional merger rates, however, are extremely similar to those calculated for the constant stellar mass selected samples (shown in Figure \ref{fig:fm:rates:1em4} as a dashed black line). For a selection at $n(>\mathcal{M}_*) = 5 \times 10^{-4}$ Mpc$^{-3}$ we find the evolution of the volume-averaged merger rate can be parametrised as \[\Gamma(z) = (1.3^{+0.5}_{-0.4}\times10^{-5}) \times (1+z)^{1.2^{+0.6}_{-0.6}},\] and the fractional merger rate for the same selection is given by \[\mathcal{R}(z) = (2.0^{+0.4}_{-0.3} \times 10^{-2}) \times (1+z)^{1.4^{+0.2}_{-0.2}}.\] Similarly, for the smaller choice of number density, $n(>\mathcal{M}_*) = 1 \times 10^{-4}$ Mpc$^{-3}$, we find \[\Gamma(z) = (1.5^{+0.7}_{-0.6} \times 10^{-6}) \times (1+z)^{1.5^{+0.7}_{-0.6}},\] and \[\mathcal{R}(z) = (1.8^{+0.6}_{-0.5} \times 10^{-2}) \times (1+z)^{1.3^{+0.4}_{-0.3}}.\] Individual merger rate data points and the best fitting parametrisation for the latter number density choice are shown in Figure \ref{fig:fm:rates:1em4}, as this extends further in redshift than the former number density. Our fit is thus better constrained for this number density choice and, as has been mentioned, is not significantly different to the larger number density. Merger rate fits for both number density selections are shown in Table \ref{tab:rates:gamma} and Table \ref{tab:rates:rmerg}.

In contrast to the derived merger rates of constant stellar mass selections (see Section \S\ref{sec:rates:m}), we observe no evidence for a turnover in either merger rate, which are consistent with remaining approximately constant at $z < 2.5$. These data suggest then that the merger rate of galaxies as they evolve over time have remained approximately constant, however further exploration is needed at high redshift to determine this (see Duncan et al. \textit{in prep}).

From these rates, we estimate selections at $n = 5 \times 10^{-4}$ Mpc$^{-3}$ and $n = 1 \times 10^{-4}$ Mpc$^{-3}$ undergo $0.6^{+0.4}_{-0.2}$ and $0.5^{+0.4}_{-0.2}$ major mergers since $z = 3.5$, respectively. This equates to total accreted stellar masses of $\log(\mathcal{M}_*/\mathrm{M}_\odot) = 10.4\pm0.2$ and $10.6\pm0.3$, respectively. Using the average stellar mass of these samples at $z \sim 0$, we find that major mergers account for $22^{+16}_{-12}\%$ and $24^{+20}_{-14}\%$, respectively, of the in-situ stellar mass at this redshift. For the smaller number density choice, this is in excellent agreement with \citet{Ownsworth2014} who find major mergers responsible for $17\pm15\%$ of the accumulated stellar mass in a typical $z = 0.3$ massive galaxy. Taking the average stellar masses of these samples at $z \sim 3.25$, we find that major mergers represent an increase in stellar mass of a factor of $1.8^{+0.5}_{-0.4}$ and $1.5\pm0.5$, respectively.

\subsection{Major merger stellar mass accretion rate density}
\label{sec:rho_major_merger}

\begin{figure} 
	\includegraphics[width=0.475\textwidth]{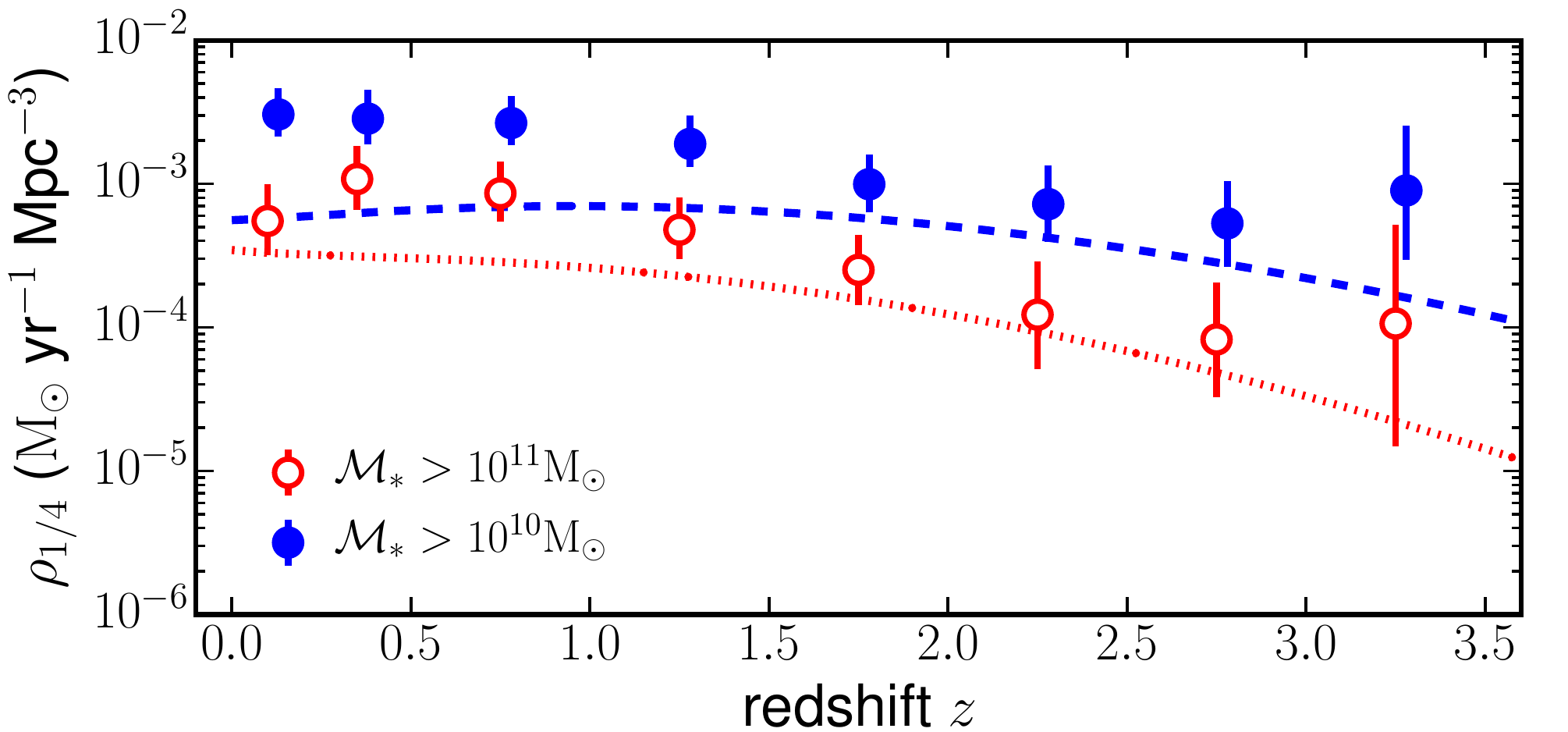}
    \caption{The major merger accretion rate density, $\rho_{1/4}$, for galaxies selected at $\mathcal{M}_* > 10^{10}\ \mathrm{M}_\odot$ (blue filled circles) and $\mathcal{M}_* > 10^{11}\ \mathrm{M}_\odot$ (red open circles). Error bars represent the $1\sigma$ uncertainty on this quantity, including errors from GSMF parameters, errors from fits on the pair fraction, and a 33\% uncertainty on the observability timescale. Dashed blue and dotted red lines indicate the derived $\rho_{1/4}$ values from Illustris using Equation \ref{eqn:illustris_rho}.}
    \label{fig:arate:m}
\end{figure}

\begin{figure} 
    \includegraphics[width=0.475\textwidth]{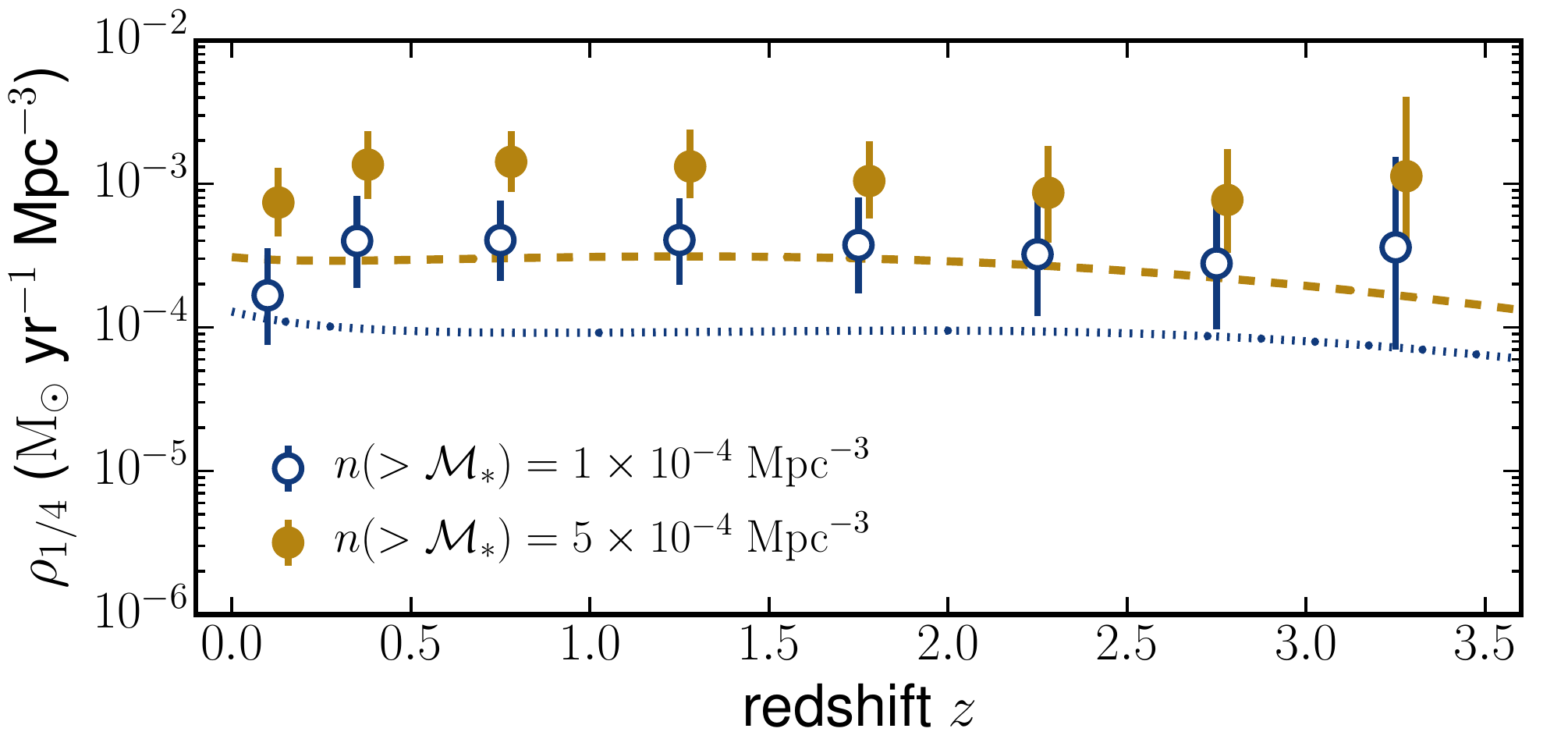}
    \caption{The major merger accretion rate density, $\rho_{1/4}$, for galaxies selected at $n(>\mathcal{M}_*) = 1\times10^{-4}$ Mpc$^{-3}$ (blue open circles) and $n(>\mathcal{M}_*) = 5\times10^{-4}$ Mpc$^{-3}$ (gold filled circles). Error bars represent the $1\sigma$ uncertainty on this quantity, including errors from GSMF parameters, errors from fits on the pair fraction, and a 33\% uncertainty on the observability timescale. Dashed gold and dotted blue lines indicate the derived $\rho_{1/4}$ values from Illustris using Equation \ref{eqn:illustris_rho}.}
    \label{fig:arate:n}
\end{figure}

It is then trivial to calculate the major merger stellar mass accretion rate density, $\rho_{1/4}$, for each of the selected samples presented in this paper. This quantity represents the stellar mass gained through major mergers per unit time and per unit volume for a particular stellar mass (or constant number density) selected population of galaxies, and can be considered the major merger analogue of the well-studied star-formation rate density. Figure \ref{fig:arate:m} displays this quantity for galaxies selected at $\mathcal{M}_* > 10^{10}\ \mathrm{M}_\odot$ (blue dashed line and shaded area) and $\mathcal{M}_* > 10^{11}\ \mathrm{M}_\odot$ (red solid line and shaded area), while Figure \ref{fig:arate:n} displays results for galaxies selected at $n(>\mathcal{M}_*) = 1\times10^{-4}$ Mpc$^{-3}$ (blue solid line and shaded area) and $n(>\mathcal{M}_*) = 5\times10^{-4}$ Mpc$^{-3}$ (gold dashed line and shaded area). We find that derived major merger accretion rate densities change significantly within the redshift range probed. From $z=3.25$ to $z=0.1$, we see an increase in $\rho_{1/4}$ at $z < 3.5$ of a factor of $\sim5$ ($\sim4$) for the high (low) stellar mass selected sample, respectively.

\begin{table}
\begin{minipage}{0.475\textwidth}
\caption{Estimated average stellar mass of a primary, $\left<\mathcal{M}_{*,1}\right>$, and secondary, $\left<\mathcal{M}_{*,2}\right>$, galaxies, the average stellar mass gained through major mergers, $\mathcal{M}_*^+$, and the number density of the primary sample, $\left<n_1\right>$, at the redshifts probed in this work. Values derived use pair fraction fits to GAMA, UDS, VIDEO, COSMOS and D17 data points at 5--30 kpc. Uncertainties include contributions from GSMF parameter and pair fraction fit errors (see Table \ref{tab:fm:fits}). Redshift bins with a superscript \ddag denote redshifts where the quoted values have been derived from extrapolations of the fit to $f_\text{pair}(z)$ where we have no measurements.}
\label{tab:arates}
\centering


\begin{tabular}{ccccc}
\hline
$z$ & $\left<\mathcal{M}_{*,1}\right>$ & $\left<\mathcal{M}_{*,2}\right>$ & $\mathcal{M}_*^+$ & $\left< n_1 \right>$ \\
    & $(\log\ \mathrm{M}_\odot)$ & $(\log\ \mathrm{M}_\odot)$ & $(\log\ \mathrm{M}_\odot)$ & $(10^{-4}\ \mathrm{Mpc}^{-3})$ \\
\hline
\multicolumn{5}{c}{$\mathcal{M}_* > 10^{10} \mathrm{M}_\odot$} \\
\hline
0.0 -- 0.2 & 10.6$^{+0.1}_{-0.1}$ & 10.3$^{+0.1}_{-0.1}$ & $9.2^{+0.1}_{-0.2}$ & 46.7$^{+6.1}_{-5.5}$ \\
0.2 -- 0.5 & 10.7$^{+0.1}_{-0.1}$ & 10.4$^{+0.1}_{-0.1}$ & $9.4^{+0.1}_{-0.2}$ & 31.9$^{+7.5}_{-7.2}$ \\
0.5 -- 1.0 & 10.7$^{+0.1}_{-0.1}$ & 10.4$^{+0.1}_{-0.1}$ & $9.4^{+0.1}_{-0.2}$ & 26.8$^{+3.9}_{-3.7}$ \\
1.0 -- 1.5 & 10.6$^{+0.1}_{-0.1}$ & 10.3$^{+0.1}_{-0.1}$ & $9.2^{+0.1}_{-0.2}$ & 17.3$^{+3.0}_{-2.8}$ \\
1.5 -- 2.0 & 10.6$^{+0.1}_{-0.1}$ & 10.3$^{+0.1}_{-0.1}$ & $9.1^{+0.1}_{-0.2}$ & 8.3$^{+2.3}_{-2.0}$ \\
2.0 -- 2.5 & 10.5$^{+0.1}_{-0.1}$ & 10.2$^{+0.1}_{-0.1}$ & $8.9^{+0.2}_{-0.2}$ & 6.2$^{+3.0}_{-2.5}$ \\
2.5 -- 3.0 & 10.5$^{+0.1}_{-0.1}$ & 10.2$^{+0.1}_{-0.1}$ & $8.8^{+0.2}_{-0.2}$ & 4.4$^{+2.6}_{-2.1}$ \\
3.0 -- 3.5 & 10.5$^{+0.1}_{-0.1}$ & 10.1$^{+0.1}_{-0.1}$ & $8.6^{+0.2}_{-0.2}$ & 7.0$^{+10.0}_{-4.3}$ \\
\hline
\multicolumn{5}{c}{$\mathcal{M}_* > 10^{11} \mathrm{M}_\odot$} \\
\hline
0.0 -- 0.2 & 11.2$^{+0.1}_{-0.1}$ & 10.8$^{+0.1}_{-0.1}$ & $9.6^{+0.2}_{-0.2}$ & 3.2$^{+1.3}_{-1.0}$ \\
0.2 -- 0.5 & 11.3$^{+0.1}_{-0.1}$ & 11.0$^{+0.1}_{-0.1}$ & $9.9^{+0.2}_{-0.2}$ & 3.8$^{+1.3}_{-1.0}$ \\
0.5 -- 1.0 & 11.3$^{+0.1}_{-0.1}$ & 10.9$^{+0.1}_{-0.1}$ & $9.9^{+0.2}_{-0.2}$ & 2.7$^{+0.6}_{-0.5}$ \\
1.0 -- 1.5 & 11.2$^{+0.1}_{-0.1}$ & 10.9$^{+0.1}_{-0.1}$ & $9.7^{+0.2}_{-0.2}$ & 1.4$^{+0.3}_{-0.3}$ \\
1.5 -- 2.0 & 11.2$^{+0.1}_{-0.1}$ & 10.9$^{+0.1}_{-0.1}$ & $9.6^{+0.2}_{-0.2}$ & 0.6$^{+0.2}_{-0.2}$ \\
2.0 -- 2.5 & 11.2$^{+0.1}_{-0.1}$ & 10.8$^{+0.1}_{-0.1}$ & $9.4^{+0.2}_{-0.2}$ & 0.3$^{+0.2}_{-0.2}$ \\
2.5 -- 3.0 & 11.2$^{+0.1}_{-0.1}$ & 10.8$^{+0.1}_{-0.1}$ & $9.3^{+0.2}_{-0.2}$ & 0.2$^{+0.2}_{-0.1}$ \\
3.0 -- 3.5 & 11.2$^{+0.1}_{-0.1}$ & 10.8$^{+0.1}_{-0.1}$ & $9.2^{+0.2}_{-0.3}$ & 0.2$^{+0.6}_{-0.2}$ \\
\hline
\multicolumn{5}{c}{$n(>\mathcal{M}_*) = 1 \times 10^{-4}\ \mathrm{Mpc}^{-3}$} \\
\hline
0.0 -- 0.2 & 11.3$^{+0.1}_{-0.1}$ & 10.9$^{+0.1}_{-0.1}$ & $9.6^{+0.2}_{-0.2}$ & 1.0$^{+0.6}_{-0.4}$ \\
0.2 -- 0.5 & 11.5$^{+0.1}_{-0.1}$ & 11.2$^{+0.1}_{-0.1}$ & $10.0^{+0.2}_{-0.2}$ & 1.0$^{+0.5}_{-0.3}$ \\
0.5 -- 1.0 & 11.4$^{+0.1}_{-0.1}$ & 11.1$^{+0.1}_{-0.1}$ & $10.1^{+0.2}_{-0.2}$ & 1.0$^{+0.3}_{-0.2}$ \\
1.0 -- 1.5 & 11.3$^{+0.1}_{-0.1}$ & 10.9$^{+0.1}_{-0.1}$ & $9.8^{+0.3}_{-0.3}$ & 1.0$^{+0.3}_{-0.2}$ \\
1.5 -- 2.0 & 11.1$^{+0.1}_{-0.1}$ & 10.8$^{+0.1}_{-0.1}$ & $9.6^{+0.3}_{-0.3}$ & 1.0$^{+0.3}_{-0.3}$ \\
2.0 -- 2.5 & 11.0$^{+0.1}_{-0.1}$ & 10.6$^{+0.1}_{-0.1}$ & $9.3^{+0.3}_{-0.3}$ & 1.0$^{+0.6}_{-0.5}$ \\
2.5 -- 3.0$^\ddag$ & 10.9$^{+0.1}_{-0.1}$ & 10.5$^{+0.1}_{-0.1}$ & $9.2^{+0.3}_{-0.3}$ & 1.0$^{+0.7}_{-0.5}$ \\
3.0 -- 3.5$^\ddag$ & 10.9$^{+0.1}_{-0.1}$ & 10.6$^{+0.1}_{-0.1}$ & $9.1^{+0.4}_{-0.4}$ & 1.0$^{+1.8}_{-0.7}$ \\
\hline
\multicolumn{5}{c}{$n(>\mathcal{M}_*) = 5 \times 10^{-4}\ \mathrm{Mpc}^{-3}$} \\
\hline
0.0 -- 0.2 & 11.1$^{+0.1}_{-0.1}$ & 10.8$^{+0.1}_{-0.1}$ & $9.6^{+0.2}_{-0.2}$ & 5.0$^{+1.7}_{-1.3}$ \\
0.2 -- 0.5 & 11.2$^{+0.1}_{-0.1}$ & 10.9$^{+0.1}_{-0.1}$ & $9.8^{+0.2}_{-0.2}$ & 5.0$^{+1.6}_{-1.3}$ \\
0.5 -- 1.0 & 11.1$^{+0.1}_{-0.1}$ & 10.8$^{+0.1}_{-0.1}$ & $9.9^{+0,2}_{-0.2}$ & 5.0$^{+0.9}_{-0.8}$ \\
1.0 -- 1.5 & 11.0$^{+0.1}_{-0.1}$ & 10.7$^{+0.1}_{-0.1}$ & $9.6^{+0.2}_{-0.2}$ & 5.0$^{+1.0}_{-0.9}$ \\
1.5 -- 2.0$^\ddag$ & 10.8$^{+0.1}_{-0.1}$ & 10.4$^{+0.1}_{-0.1}$ & $9.3^{+0.2}_{-0.2}$ & 5.0$^{+1.4}_{-1.3}$ \\
2.0 -- 2.5$^\ddag$ & 10.6$^{+0.1}_{-0.1}$ & 10.3$^{+0.1}_{-0.1}$ & $9.1^{+0.2}_{-0.3}$ & 5.0$^{+2.5}_{-2.1}$ \\
2.5 -- 3.0$^\ddag$ & 10.5$^{+0.1}_{-0.1}$ & 10.1$^{+0.1}_{-0.1}$ & $8.9^{+0.3}_{-0.3}$ & 5.0$^{+2.9}_{-2.3}$ \\
3.0 -- 3.5$^\ddag$ & 10.5$^{+0.1}_{-0.1}$ & 10.2$^{+0.1}_{-0.1}$ & $8.9^{+0.3}_{-0.3}$ & 5.0$^{+7.1}_{-3.2}$ \\
\hline
\end{tabular}
\end{minipage}
\end{table}

Figure \ref{fig:arate:n} displays the major merger accretion rate density, $\rho_{1/4}$, for galaxies selected at $n(>\mathcal{M}_*) = 1\times10^{-4}$ Mpc$^{-3}$ (blue solid line and shaded area) and $n(>\mathcal{M}_*) = 5\times10^{-4}$ Mpc$^{-3}$ (gold dashed line and shaded area). In contrast to constant stellar mass selections (see Section \S\ref{sec:merger-rates:mass-added}), there is no observed evolution of $\rho_{1/4}$ for galaxies selected at a constant cumulative number density. Populations selected at the smaller (larger) number density possess major merger accretion rate densities of $\approx 2\times10^{-4}\ \mathrm{M}_\odot$ yr$^{-1}$ Mpc$^{-3}$ ($\approx 1\times10^{-3}\ \mathrm{M}_\odot$ yr$^{-1}$ Mpc$^{-3}$).

\subsection{Comparing the role of major mergers and star-formation}
\label{sec:sfrd}

In order to compare the stellar mass accreted through major mergers with that produced via the process of star-formation, we must calculate the star-formation rate density, $\rho_\Psi$, of the same stellar mass selected samples of galaxies we use to calculate the merger fractions and subsequent merger rates. Here we describe the steps taken to achieve this.

We take estimated stellar masses and total (UV + IR) star-formation rates from \citet[][see their Section 5.5]{Muzzin2013a} and observe the distribution of star-formation rates, $\Psi$, within discrete bins of stellar mass and redshift. \citeauthor{Muzzin2013a} determines $L_{2800}$ using {\tt{EAZY}} and converts this UV luminosity into a SFR using the standard conversion factors \citep{Kennicutt1998, Bell2005}. Similarly, the IR luminosity, $L_\text{IR}$, is estimated using the templates of \citet{Chary2001} and converted to a SFR using similar conversions as with $L_{2800}$. Stellar mass bins over the range $9.5 < \log\mathcal{M}_* < 11.5$ with a width of 0.25 dex are used, while redshift bins are chosen to be the same as the redshift bins seen in Figure \ref{fig:fm-11-30kpc}. We fit the resulting distributions of $\log\Psi$ with a combination of two Gaussian functions, representing a `red' and a `blue' population, respectively. We note that it is important only that the total SFR distribution is well reproduced at every stellar mass and redshift bin and that a combination of two Gaussian distributions achieves this. These fitted distributions are normalised such that a galaxy with a particular $(\mathcal{M}_*, z)$ must have a star-formation rate in the range $-3 < \log\Psi\ [ \mathrm{M}_\odot\ \mathrm{yr}^{-1} ] < 5$.

In order to estimate $\rho_\Psi$ we substitute stellar mass for star-formation rate in the GSMF, effectively transforming the GSMF into a star-formation rate function (SFRF), $\phi_\Psi$. The SFRF evaluates the number density of objects with a particular star-formation rate at a particular redshift. At fixed redshift this substitute is performed by sampling a star-formation rate from the appropriate distribution at every stellar mass. The star-formation rate density can then be estimated by integrating the sampled $\phi_\Psi$ with respect to stellar mass. This can be written as 
\begin{equation}
	\rho_\Psi = \int_{\mathcal{M}_{*,1}}^{\mathcal{M}_{*,2}}\ \phi_\Psi(\mathcal{M}_*, z)\ \times\ \Psi(\mathcal{M}_*, z)\ \mathrm{d}\mathcal{M}_*.
\label{eqn:rho_psi}
\end{equation}
Performing a large number of these $\rho_\Psi$ `realisations', we obtain a distribution of values for the star-formation rate density of a mass selected sample of galaxies from which the most likely value and associated uncertainties are extracted.

\begin{figure}
\centering
\begin{subfigure}[]{0.975\textwidth} 
	\includegraphics[width=0.49\textwidth]{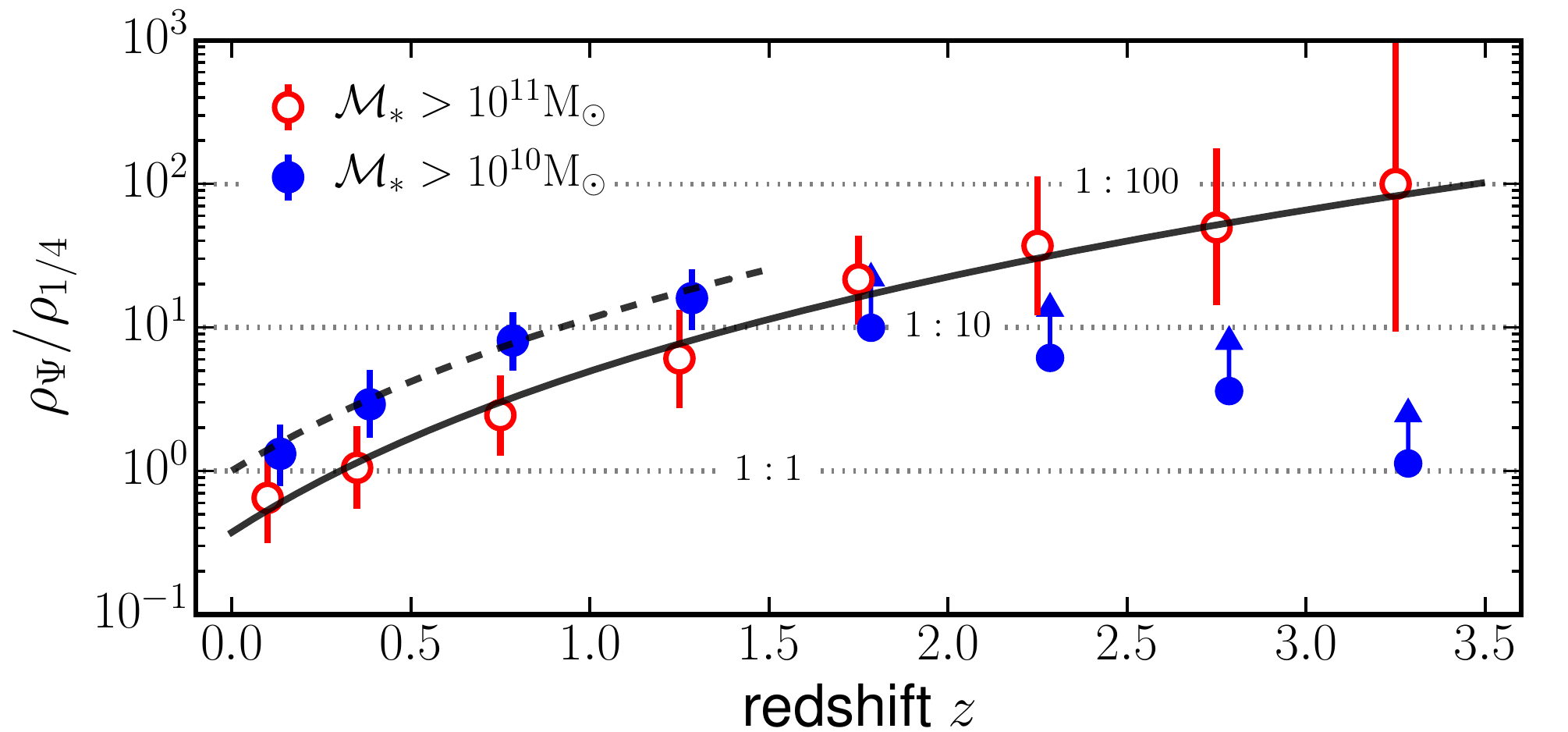}
    \label{fig:rhoratio:m}
\end{subfigure}
\begin{subfigure}[]{0.975\textwidth}  
	\includegraphics[width=0.49\textwidth]{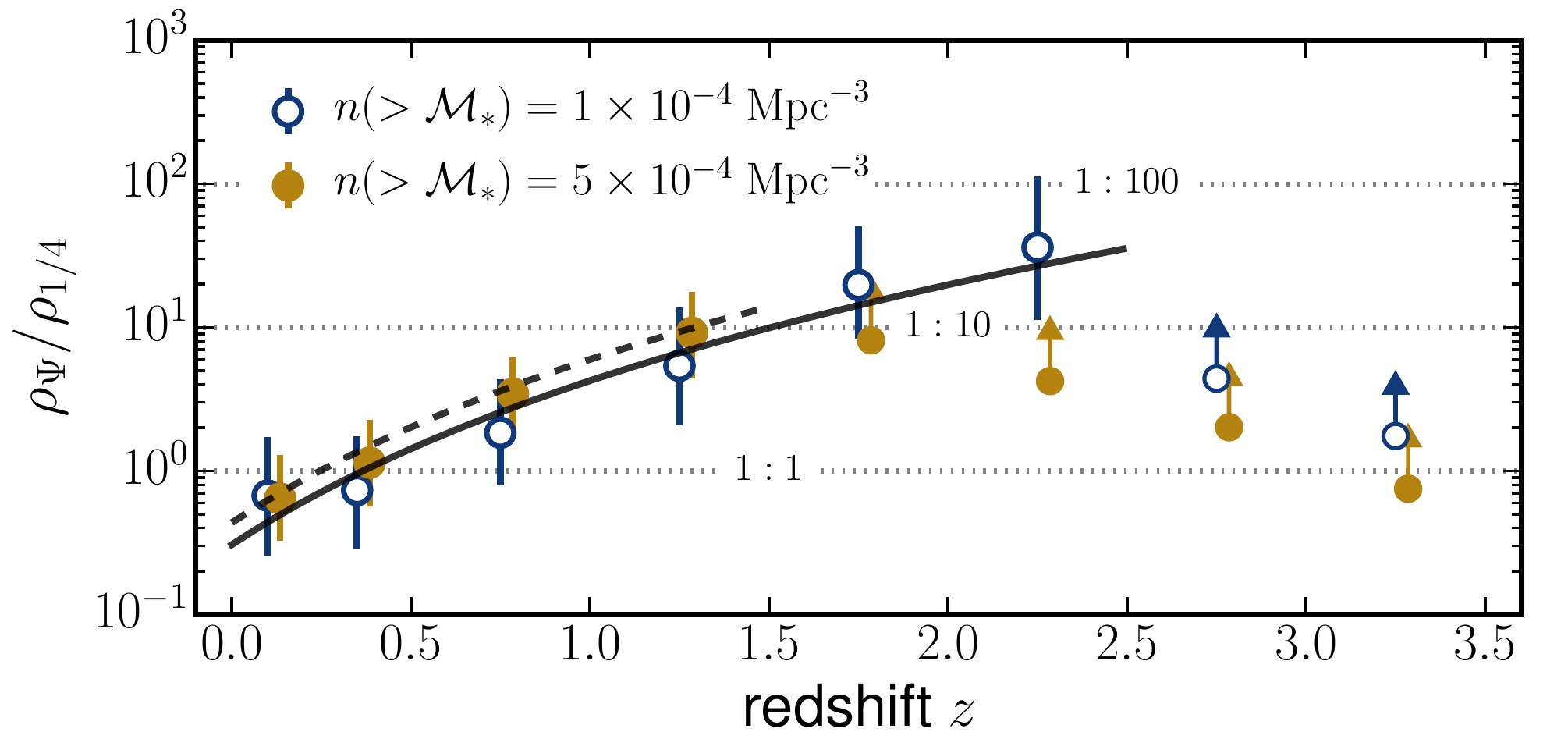}
    \label{fig:rhoratio:n}
\end{subfigure}
	\caption{Ratio of stellar mass production/accretion rates, $\rho_\Psi / \rho_{1/4}$, through the processes of star-formation and major mergers as a function of redshift. \textit{Top:} Samples of galaxies with $>10^{10}\mathrm{M}_\odot$ (blue filled circles) and $>10^{11}\mathrm{M}_\odot$ (red open circles). \textit{Bottom:} Samples of galaxies selected at constant cumulative number densities of $n = 1\times10^{-4}$ Mpc$^{-3}$ (blue open circles) and $n = 5\times10^{-4}$ Mpc$^{-3}$ (gold filled circles). Uncertainties include contributions from the GSMF. Where stellar mass completeness prohibits us from integrating $\phi_\Psi$ to smaller masses, we present lower limits on $\rho_\Psi / \rho_{1/4}$ as markers with upwards pointing arrows. Power law fits to each sample are shown as dashed and dotted black lines, respectively.}
    \label{fig:rhoratio}
\end{figure}

As the star-formation rate distributions we use are constructed from a flux-limited catalogue, we take care to only use $\rho_\Psi$ estimates at redshifts where $\mathcal{M}_{*,1} > \mathcal{M}_*^{90}(z)$, i.e. where we are not sampling distributions of $\log \Psi$ which may be incomplete. Not accounting for this would result in values of $\rho_\Psi$ which are likely to be \textit{over}-estimated, as it would assume that lower mass galaxies have the same star-formation rate distribution as higher mass systems. At these redshifts where this is the case we estimate lower limits on $\rho_\Psi$ by integrating down to the appropriate stellar mass completeness limit. These lower limits are denoted as arrows in Figure \ref{fig:rhoratio}. Nonetheless, when we incrementally integrate down to stellar masses of $10^8\ \mathrm{M}_\odot$, we achieve estimations of $\rho_\Psi$ in excellent agreement with the cosmic star-formation rate density given in Equation 15 of \citet{Madau2014}. Although taking SFR distributions from deeper survey data \citep[e.g.,][]{Laigle2016} would allow us to extend comparisons to higher redshifts, these SFR estimates would be based on SED fitting techniques which can provide substantially inaccurate SFR estimates if the assumed star-formation history is not correct \citep[e.g.,][]{Maraston2010, Papovich2011}. For completeness, we note that if the SFR and stellar mass estimates of \citeauthor{Laigle2016} are used we recover a qualitatively similar result, however the slope of the fitted ratios is approximately a factor of $\sim1.5$ smaller at all sample selections.

Presented in Figure \ref{fig:rhoratio} is the ratio between the star-formation rate density, $\rho_\Psi$, and major merger accretion rate density, $\rho_{1/4}$, as a function of redshift. This quantity, $\rho_\Psi / \rho_{1/4}$ encodes the relative significance of the two channels in the build-up of stellar mass in massive galaxies. We plot this for samples selected at a constant stellar mass (top panel), and at a constant cumulative number density (bottom panel). Massive galaxies ($>10^{11}\mathrm{M}_\odot$) exhibit a steep decline in $\rho_\Psi / \rho_{1/4}$ towards low redshift, which suggests the increasing relative significance of major mergers in recent times. We also find that $\rho_\Psi / \rho_{1/4}$ is a factor of $\sim3$ larger for intermediate mass ($>10^{10}\mathrm{M}_\odot$) galaxies compared to the most massive ($>10^{11}\mathrm{M}_\odot$). From the top panel of Figure \ref{fig:rhoratio} it is evident that the two channels of stellar mass growth are approximately equivalent at $z \lesssim 0.75$. Put another way, star-formation and major mergers contribute similar amounts of stellar mass to a `typical' massive galaxy at $z \lesssim 0.75$. This can be attributed to two main factors: the increasing rate of major mergers at $z < 1.5$ for these samples (see Figures \ref{fig:mr-11-30kpc} and \ref{fig:mr-10-30kpc}), and the decreasing average star-formation rate of galaxies since the peak of cosmic star-formation at $z \sim 2$ \citep[e.g.,][]{Madau2014}. Similar trends are seen when tracing the progenitors of $z\sim0$ massive galaxies using a constant number density selection (see the bottom panel of Figure \ref{fig:rhoratio}). However, unlike the constant stellar mass selections we find no significant difference between the two constant number density selections probed in this work.

Ratios for all selections are found to be well fit by a simple power law of the form $p_0(1+z)^m$ at the redshifts observed. These are shown in Figure \ref{fig:rhoratio} as solid and dashed black curves while best fitting parameters, uncertainties and the redshift ranges over which they are valid are given in Table \ref{tab:rhoratio}.

\begin{table}
\caption{Best fitting parameters and uncertainties, derived from a bootstrap analysis, of the ratio of star-formation rate density and major merger accretion rate density, $\rho_\Psi / \rho_{1/4}$. This quantity is parametrised as $\rho_\Psi / \rho_{1/4} = p_0(1+z)^m$. Parameters shown for the two constant stellar mass selections and two constant number density selections probed in this work. Fits are shown in Figure \ref{fig:rhoratio} as dashed and solid black curves.}
\label{tab:rhoratio}
\centering


\begin{tabular}{ccc}
\hline
$p_0$ & $m$ & redshift range \\
\hline
\multicolumn{3}{c}{$\mathcal{M}_* > 10^{10} \mathrm{M}_\odot$} \\
\hline
1.03$^{+0.53}_{-0.44}$ & 3.51$^{+0.87}_{-0.94}$ & $z < 1.5$ \\ 
\hline
\multicolumn{3}{c}{$\mathcal{M}_* > 10^{11} \mathrm{M}_\odot$} \\
\hline
0.39$^{+0.40}_{-0.23}$ & 3.65$^{+1.11}_{-1.49}$ & $z < 3.5$ \\ 
\hline
\multicolumn{3}{c}{$n(>\mathcal{M}_*) = 1 \times 10^{-4}\ \text{Mpc}^{-3}$} \\
\hline
 0.30$^{+0.89}_{-0.24}$ & 3.75$^{+1.96}_{-4.37}$ & $z < 2.5$ \\ 
\hline
\multicolumn{3}{c}{$n(>\mathcal{M}_*) = 5 \times 10^{-4}\ \text{Mpc}^{-3}$} \\
\hline
 0.44$^{+0.42}_{-0.46}$ & 3.71$^{+1.49}_{-1.69}$ & $z < 1.5$ \\ 
\hline
\end{tabular}
\end{table}


\section{Discussion}
\label{sec:discussion}
Here we discuss the implications of our results with respect to galaxy evolution at $z < 3.5$, and explore various issues with the methods we have employed. In Section \S\ref{sec:illustris} we compare our pair fractions with the \citetalias{Henriques2014} semi-analytic model, and our merger rates with those of the Illustris cosmological hydrodynamical simulation. In Section \S\ref{sec:cosmos-offset} we explore possible explanations for the systematic difference between pair fractions measured in the COSMOS region, and those measured in the other regions. Finally, in Section \S\ref{sec:fm-tests} we subject our data and measurements of the pair fraction to multiple tests which demonstrates their robustness.

Firstly, a caveat of our work and indeed any close-pair study of merger histories is the inherent uncertainty surrounding the fraction, $C_\text{merg}$, of close-pair systems that will eventually merge. Throughout this work we have explicitly assumed this fraction is constant with redshift, stellar mass and physical separation. Although numerical simulations and empirical measurements of close-pairs at $r < 30$ kpc have determined $C_\text{merg} \approx 0.4-1.0$ \citep{Kitzbichler2008, Patton2008, Bundy2009}, its dependence on stellar mass and redshift is as yet unexplored in detail. Furthermore the timescale, $\left<T_\text{obs}\right>$, over which we can observe a merger event (as defined in this paper) has been explored only at $z < 1.5$ \citep{Lotz2011}, and its constancy beyond this is unknown. In this work we assume that this timescale is fixed at earlier times. If any of these assumptions prove incorrect, the results presented here will be in doubt. Further investigation of these parameters is needed.

\subsection{Comparison with semi-analytic models and hydrodynamical simulations}
\label{sec:illustris}

Figure \ref{fig:fm-11-30kpc} and Figure \ref{fig:fm-10-30kpc}  in Section \S\ref{sec:merger-fractions} present a comparison between $f_\text{pair}$ measured observationally and from lightcones extracted from the \citetalias{Henriques2014} semi-analytic model, illustrated as the grey shaded area in these figures. We find that the model predicts pair fractions in excellent agreement with those found in this work, especially when the higher redshift CANDELS data are considered. Additionally, the (cosmic) variance seen between the lightcones appear to reflect the variance between the observational measurements in different survey regions. The measured pair fractions depend mainly on the clustering of galaxies (i.e. the cosmology) and the stellar mass of galaxies. As \citetalias{Henriques2014} uses the most current cosmological model and is able to reproduce the (total) GSMFs out to at least $z\sim3$, this is welcome agreement. This agreement also extends the argument that we are in fact measuring close-pairs with $\Delta v < 500$ km/s, as seen at $z \sim 0$ using GAMA in Section \S\ref{sec:comp-spec-photo-fm}.

Figures \ref{fig:mr-11-30kpc} and \ref{fig:mr-10-30kpc} illustrate the derived fractional merger rates of galaxies at $>10^{11}\mathrm{M}_\odot$ and $>10^{10}\mathrm{M}_\odot$, respectively. Shown as dashed lines, we also plot the fractional merger rates of galaxies within the Illustris cosmological hydrodynamical simulation using the equation given in Table 1 of \citet{Rodriguez-Gomez2015}. This equation estimates the galaxy-galaxy merger rate as a function of stellar mass, stellar mass merger ratio, and redshift. Integrating this equation with respect to stellar mass merger ratio at $0.25 < \mu < 1$, we arrive at the cumulative merger rate comparable to our observations. Predictions from Illustris are found to be inconsistent with observational estimates of the fractional merger rates at high redshift. The predictions made by the simulation evolve strongly with redshift and do not reproduce the observed values of $\mathcal{R}$ at $\mathcal{M}_* < 10^{12} \mathrm{M}_\odot$. This may well be due to the overproduction of both high ($\mathcal{M}_* > 10^{10.5}\mathrm{M}_\odot$) and low ($\mathcal{M}_* < 10^{10}\mathrm{M}_\odot$) stellar mass galaxies within Illustris \citep{Arthur2017,Genel2014, Schaye2014} compared to observed number densities.

We also calculate the major merger accretion rate density, $\rho_{1/4}$, within the Illustris simulation. This is achieved by using fits to the GSMF within the simulation (Equation 1 of \citet{Torrey2015}) combined with the fitting function of the specific merger accretion rate, $\dot{m}_\text{acc}(\mathcal{M}_*, \mu, z)$, in Table 1 of \citet{Rodriguez-Gomez2016}. It is then trivial to estimate $\rho$ within Illustris and can be written as 
\begin{equation}
\label{eqn:illustris_rho}
\rho^\text{sim} = \int_{\mathcal{M}_{l}}^{\mathcal{M}_{h}} \phi(\mathcal{M}_*, z) \mathcal{M}_*\  \int_{\mu_l}^{\mu_h} \dot{m}_\text{acc}(\mathcal{M}_*, \mu, z)\ \mathrm{d}\mu\ \mathrm{d}\mathcal{M}_*,
\end{equation}
where $\mathcal{M}_*$ is stellar mass, $\phi(\mathcal{M}_*, z)$ is the GSMF evaluated at $\mathcal{M}_*$ and redshift $z$, and the specific merger accretion rate is defined as \[\dot{m}_\text{acc}(\mathcal{M}_*, \mu, z) = \frac{1}{\mathcal{M}_*}\frac{\mathrm{d}\mathcal{M}_\text{acc}}{\mathrm{d}t\ \mathrm{d}\mu}.\] For the purposes of this work we integrate over stellar mass merger ratio between $0.25 < \mu < 1.0$ in order to attain the accretion mass from major mergers only, and choose the upper stellar mass integration limit to be $\mathcal{M}_h = 10^{12} \mathrm{M}_\odot$. Figures \ref{fig:arate:m} and \ref{fig:arate:n} show the result of integrating Equation \ref{eqn:illustris_rho} at fixed redshifts for constant stellar mass and constant number density selected samples, respectively.

Overall, we find that the major merger accretion rate densities within Illustris are qualitatively similar to that found observationally. Values for galaxies at $\mathcal{M}_* > 10^{10}\ \mathrm{M}_\odot$ and $>10^{11}\ \mathrm{M}_\odot$ within Illustris agree tenuously at $z > 1.5$ with our observational estimates. At lower redshifts $\rho_{1/4}$ tends to be smaller within the simulation by a factor of $\sim2$. Similarly, for samples of galaxies selected at constant number densities we find that the evolution of $\rho_{1/4}$ agrees well with observational estimates, however Illustris values are consistently a factor of $\sim2$ smaller across the entire redshift range probed. A puzzling observation is that Illustris predicts \textit{larger} merger rates than observed, but generally \textit{under}-predicts the stellar mass accretion rate compared to our derived values. As Illustris is found to predict larger abundances of massive galaxies at $z<4$ \citep{Genel2014,Somerville2015}, it suggests that companion galaxies inside the simulation that eventually merge possess smaller stellar masses than predicted by simply taking the average stellar mass of a potential companion sample.

\subsection{Field-to-field variation}
\label{sec:cosmos-offset}
Evident in the measured pair fractions (see Figure \ref{fig:fm-11-30kpc}) is an apparent systematic offset in the measured merger fractions between the COSMOS region and the UDS and VIDEO regions. At $z > 1$ the pair fractions measured in the COSMOS region are found to be a factor of $\sim 2$ lower than those in either the UDS or VIDEO regions. Such a consistent systematic difference over such a large redshift range cannot in all likelihood be attributed to cosmic variance alone. Here we describe the efforts undertaken to determine the cause of this systematic difference.

During the course of this work an issue with the IRAC photometry in the UDS catalogue was identified whereby fluxes were found to be underestimated by approximately $20\%$. As these filters aid in constraining the photometric redshifts and stellar masses of galaxies, the effect of such an underestimate in the flux on photometric redshifts and stellar population parameters is not trivial to predict. To probe this issue we increased our sample's IRAC fluxes by a factor of 1.2 whilst conserving the signal-to-noise ratio, then recalculated the photometric redshifts and stellar masses of galaxies in the UDS region and performed the pair fraction measurement on the adjusted photometry. No significant differences are found between the recalculated $f_\text{pair}$ and those tabulated in Table \ref{tab:fm}. A similar issue with IRAC photometry was discovered within the COSMOS catalogue as well. Spatially dependent systematic shifts in IRAC fluxes of up to one magnitude exist which essentially renders the IRAC photometry in this catalogue unusable. We thus removed the IRAC photometry and recalculated photometric redshifts and stellar masses of all galaxies and performed measurements of the pair fraction once more. We find a systematic increase of $\sim10\%$ in the pair fraction in all redshift bins. This can be attributed to a slight rise of $\sim 0.1$ dex in the estimated stellar massed calculated without IRAC photometry. While this goes part of the way to reducing the observed offset between COSMOS and the other regions, erroneous IRAC photometry cannot be the primary source of the observed offset and its absence does not significantly affect the results of this work. Further work is needed to pinpoint the cause of this difference.

Another suspected source of the discrepancy is the different pixel scales of the images from which photometry is extracted. Companion galaxies could be missed by our analysis if it was close enough to a primary galaxy to have its photometry blended in with the host galaxy's light. We doubled the minimum separation for two galaxies to be considered a close-pair to 10 kpc and reran the pair fraction measurement. Comparing the remeasured fractions revealed the discrepancy remained and thus is not predominantly due to source extraction/blending issues.

\subsection{Tests on the merger fraction}
\label{sec:fm-tests}

We perform several tests and consistency checks on the data and the method to ensure the robustness of the results presented in this work. Firstly, we test for any spatial dependence of the merger fractions within each survey region by splitting each dataset into four contiguous sub-fields and performing the merger fraction measurement once more. No significant differences are found except in the UDS region. We find a slight excess in the pair fraction, with $f_\text{pair}$ found to be a factor of $\sim 1.5$ higher at $1.5 < z < 2.0$, in one sub-field. This quadrant contains a known galaxy cluster at $z = 1.6$ \citep{Papovich2011}, to which we attribute the observed excess. Averaged over the entire region, this excess signal is not found to significantly impact the measured pair fractions. Where possible we remeasure fractions using redshift PDFs produced by independent works (e.g. \citet{Hartley2013} in the UDS region and \citet{Muzzin2013b} in the COSMOS region). No significant difference is found when these data are used. Additionally, we perform a measurement of the contribution to the measured pair fraction by the random projected positions of galaxies on the sky. Given these conditions, one would expect a negligible pair fraction extremely close to zero. Pair fractions of $\sim 10^{-4}$, approximately two orders of magnitude lower than those tabulated in Table \ref{tab:fm}, are found.

\subsubsection{Galaxy stellar mass function choice}
\label{sec:gsmf}

Various parts of this work make use of the galaxy stellar mass function (GSMF). For example, in Section \S\ref{sec:weights} we employ the GSMF to calculate statistical weightings for primary and secondary galaxies if a search in $\mathcal{M}_*$-space falls below the completeness limit of a survey. Additionally, in Section \S\ref{sec:rates:n} we use the GSMF to calculate stellar mass limits for a constant cumulative comoving number density selected sample. Finally, in Section \S\ref{sec:sfrd} we integrate the GSMF to estimate the star-formation rate density of stellar mass selected samples. 

GSMF parametrisations are sourced from various literature works for this purpose. At $z<0.2$ we use the GSMF of \citet{Baldry2012}, at $0.2 < z < 3$ we use those presented in \citet{Mortlock2015, Mortlock2016}, and at $3.0 < z < 3.5$ we use the results of \citet{Santini2012}. The numerical results presented in this paper are based on these GSMF parametrisations, making appropriate conversions to a \citet{Chabrier2003} IMF. To ensure the results presented herein are not dependent on the choice of GSMF, we perform all measurements that depend on the GSMF with other literature parametrisations. At low redshift ($z<0.2$) we substitute GSMF parameters from \citet{Pozzetti2010} and \citet{Kelvin2014}. At higher redshifts we check measurements against those using GSMFs from \citet{Muzzin2013a} and \citet{Duncan2014}. No significant change to the results presented in this work are observed using any of these GSMF parametrisations and thus our results are robust to the choice of GSMF. Summarising all aforementioned tests, we conclude that the results presented in this work are robust and not significantly influenced by any of the factors discussed.
\section{Summary}
\label{sec:summary}

In this paper we have presented the best constraints yet on the major merger fraction at $z < 3.5$. This is achieved using $\sim$350,000 galaxies drawn from the UKIDSS UDS, VIDEO/CFHT-LS, UltraVISTA/COSMOS and GAMA survey regions using a method adapted from that presented in \citet{Lopez-Sanjuan2014}. These regions provided 144 sq. deg. at $z < 0.2$, and 3.25 sq. deg at $z < 3.5$ in which to perform this measurement. In addition to merger fractions we have derived major merger rates for galaxies at $>10^{10}\ \mathrm{M}_\odot$ selected above a constant stellar mass (with redshift), and samples selected at a constant cumulative number density of $n(>\mathcal{M}_*) > 1 \times 10^{-4}$ Mpc$^{-3}$ in order to trace directly the merger histories of low redshift massive galaxies. Additionally, we compared the relative roles of major mergers and star-formation in the build-up of stellar mass in massive galaxies over this redshift range by computing the major merger accretion rate  of our samples. Finally, we compared our results with predictions made by the \citet{Henriques2014} semi-analytic model and the Illustris cosmological hydrodynamical simulation. A series of follow up papers will explore the role of minor mergers in the evolution of the most massive galaxies and determine the relative significance of (major and minor) mergers and star-formation in the evolution of massive galaxies at $z < 3.5$. Our main results can be summarised as follows:

\begin{itemize}
	\item Measured pair fractions detailed in Section \S\ref{sec:merger-fractions} are found to be approximately constant over the redshift range probed, and we find no significant difference between the normalisation or evolution of the pair fraction for galaxies selected above stellar masses of $10^{10} \mathrm{M}_\odot$ and $10^{11} \mathrm{M}_\odot$ at $z < 3.5$. Pair fractions measured photometrically and spectroscopically ($\Delta v < 500$ kms/s) using the second data release of GAMA are found to be consistent with each other at both constant stellar mass selection limits.

	\item In Section \S\ref{sec:rates:m} we find volume-averaged merger rates, $\Gamma(z)$, of galaxies selected above stellar masses of $10^{10} \mathrm{M}_\odot$ and $10^{11} \mathrm{M}_\odot$ to be a factor of 2--3 smaller than many previous works. These rates exhibit a strong evolution with redshift and are well fit by a combined power law plus exponential.
    
	\item As we find lower major merger rates, galaxies are expected to undergo less major mergers than previously found. Galaxies with $\mathcal{M}_* > 10^{10} \mathrm{M}_\odot$ undergo $0.5^{+0.3}_{-0.1}$ major mergers while galaxies with $\mathcal{M}_* > 10^{11} \mathrm{M}_\odot$ undergo $0.5^{+0.3}_{-0.1}$ major merger events. However, these precise numbers strongly depend on the assumed values of $C_\text{merg}$ and $\left<T_\text{obs}\right>$.
    
    \item Galaxies with stellar masses $> 10^{11} \mathrm{M}_\odot$ ($> 10^{10} \mathrm{M}_\odot$) at $z \approx 3.25$ accumulate additional stellar mass of $\log(\mathcal{M}_* / \mathrm{M}_\odot) = 10.6\pm0.2$ ($10.1^{+0.2}_{-0.1}$) at $z < 3.5$ solely via major mergers (Section \S\ref{sec:merger-rates:mass-added}). Tracing the direct progenitors of local massive galaxies by sampling at a constant cumulative number density of $n = 1\times10^{-4}$ ($5\times10^{-4}$) Mpc$^{-3}$, representing $z=0$ selections of $\mathcal{M}_* > 10^{11.2}\mathrm{M}_\odot$ ($> 10^{11.0}\mathrm{M}_\odot$) galaxies, we find that a stellar mass of $\log(\mathcal{M}_* / \mathrm{M}_\odot) = 10.4\pm0.2$ ($10.6\pm0.3$) is accrued via major mergers over the same redshift range.

	\item The \citet{Henriques2014} semi-analytic model predicts pair fractions (measured spectroscopically with $\Delta v < 500$ km/s) in excellent agreement with observations. Furthermore, the model variance between 1 deg$^2$ fields-of-view similar to that seen between the observed fields. Fractional merger rates, $\mathcal{R}(z)$, predicted within Illustris are qualitatively and quantitatively inconsistent with our derived rates at $z > 0.5$. This may be due to the inability of the simulation to reproduce the correct number density of galaxies over a wide range of stellar masses at most redshifts. Illustris predictions of the major merger accretion rate density, $\rho_{1/4}$, are qualitatively similar to those estimated for galaxies at a constant number density and constant stellar mass. However, the normalisation is typically smaller than that observed by a factor of $\sim$2--3.
   
   \item Finally, we compared the typical stellar mass accretion rates from major mergers to that through the process of in-situ star-formation. We find that major mergers become a comparable source of stellar mass growth compared to star-formation only at $z<1$. 
\end{itemize}
\section*{Acknowledgements}
The authors thank Gabriel Brammer and Vicente Rodrigues-Gomez for their rapid response in request for information, Adam Muzzin for supplying details on the creation of the area correction mask in the UltraVISTA field, Matt Jarvis for access to VIDEO data, and Jake Arthur and Frazer Pearce for useful discussions which contributed towards the content of this paper. CJM gratefully acknowledges support from the Science and Technology Facilities Council (STFC) and thanks Charlotte Marshall for the encouragement to finish this publication. CJC acknowledges support from the Leverhulme Trust. KJD acknowledges support from the ERC Advanced Investigator programme NewClusters 321271

This research made use of {\tt{Astropy}} (v1.0), a community-developed core Python package for Astronomy \citep{Robitaille2013}. Furthermore, we are grateful for access to the University of Nottingham High Performance Computing (HPC) Facility on which {\tt{Pyrus}} analysis was performed. The Millennium Simulation databases used in this paper and the web application providing online access to them were constructed as part of the activities of the German Astrophysical Virtual Observatory (GAVO). The CFHT data used in this paper were based on observations obtained with MegaPrime/MegaCam, a joint project of CFHT and CEA/DAPNIA, at the Canada-France-Hawaii Telescope (CFHT) which is operated by the National Research Council (NRC) of Canada, the Institut National des Science de l'Univers of the Centre National de la Recherche Scientifique (CNRS) of France and the University of Hawaii. This work is based in part on data products produced at TERAPIX and the Canadian Astronomy Data Centre as part of the Canada-France-Hawaii Telescope Legacy Survey, a collaborative project of NRC and CNRS. This paper uses data from the VIMOS Public Extragalactic Redshift Survey (VIPERS). VIPERS has been performed using the ESO Very Large Telescope, under the ``Large Programme'' 182.A-0886. GAMA is a joint European-Australasian project based around a spectroscopic campaign using the Anglo-Australian Telescope. The GAMA input catalogue is based on data taken from the Sloan Digital Sky Survey and the UKIRT Infrared Deep Sky Survey. Complementary imaging of the GAMA regions is being obtained by a number of independent survey programmes including GALEX MIS, VST KiDS, VISTA VIKING, WISE, Herschel-ATLAS, GMRT and ASKAP providing UV to radio coverage. GAMA is funded by the STFC (UK), the ARC (Australia), the AAO, and the participating institutions. The GAMA website is \url{http://www.gama-survey.org/}. Based on observations made with ESO Telescopes at the La Silla Paranal Observatory under programme ID 179.A-2004.

\bibliography{paper2refs}

\begin{thebibliography}{}
\makeatletter
\relax
\def\mn@urlcharsother{\let\do\@makeother \do\$\do\&\do\#\do\^\do\_\do\%\do\~}
\def\mn@doi{\begingroup\mn@urlcharsother \@ifnextchar [ {\mn@doi@}
  {\mn@doi@[]}}
\def\mn@doi@[#1]#2{\def\@tempa{#1}\ifx\@tempa\@empty \href
  {http://dx.doi.org/#2} {doi:#2}\else \href {http://dx.doi.org/#2} {#1}\fi
  \endgroup}
\def\mn@eprint#1#2{\mn@eprint@#1:#2::\@nil}
\def\mn@eprint@arXiv#1{\href {http://arxiv.org/abs/#1} {{\tt arXiv:#1}}}
\def\mn@eprint@dblp#1{\href {http://dblp.uni-trier.de/rec/bibtex/#1.xml}
  {dblp:#1}}
\def\mn@eprint@#1:#2:#3:#4\@nil{\def\@tempa {#1}\def\@tempb {#2}\def\@tempc
  {#3}\ifx \@tempc \@empty \let \@tempc \@tempb \let \@tempb \@tempa \fi \ifx
  \@tempb \@empty \def\@tempb {arXiv}\fi \@ifundefined
  {mn@eprint@\@tempb}{\@tempb:\@tempc}{\expandafter \expandafter \csname
  mn@eprint@\@tempb\endcsname \expandafter{\@tempc}}}

\bibitem[\protect\citeauthoryear{Arthur et~al.,}{Arthur
  et~al.}{2017}]{Arthur2017}
Arthur J.,  et~al., 2017, \mn@doi [Monthly Notices of the Royal Astronomical
  Society] {10.1093/mnras/stw2424}, 464, 2027

\bibitem[\protect\citeauthoryear{Baldry et~al.,}{Baldry
  et~al.}{2010}]{Baldry2010}
Baldry I.~K.,  et~al., 2010, \mn@doi [Monthly Notices of the Royal Astronomical
  Society] {10.1111/j.1365-2966.2010.16282.x}, 404, 86

\bibitem[\protect\citeauthoryear{Baldry et~al.,}{Baldry
  et~al.}{2012}]{Baldry2012}
Baldry I.~K.,  et~al., 2012, \mn@doi [Monthly Notices of the Royal Astronomical
  Society] {10.1111/j.1365-2966.2012.20340.x}, 421, no

\bibitem[\protect\citeauthoryear{Baldry et~al.,}{Baldry
  et~al.}{2014}]{Baldry2014}
Baldry I.~K.,  et~al., 2014, \mn@doi [Monthly Notices of the Royal Astronomical
  Society] {10.1093/mnras/stu727}, 441, 2440

\bibitem[\protect\citeauthoryear{Barnes \& Hernquist}{Barnes \&
  Hernquist}{1996}]{Barnes1996}
Barnes J.~E.,  Hernquist L.,  1996, \mn@doi [The Astrophysical Journal]
  {10.1086/177957}, 471, 115

\bibitem[\protect\citeauthoryear{Behroozi, Marchesini, Wechsler, Muzzin,
  Papovich  \& Stefanon}{Behroozi et~al.}{2013}]{Behroozi2013}
Behroozi P.~S.,  Marchesini D.,  Wechsler R.~H.,  Muzzin A.,  Papovich C.,
  Stefanon M.,  2013, \mn@doi [The Astrophysical Journal]
  {10.1088/2041-8205/777/1/L10}, 777, L10

\bibitem[\protect\citeauthoryear{Bell et~al.,}{Bell et~al.}{2005}]{Bell2005}
Bell E.~F.,  et~al., 2005, \mn@doi [The Astrophysical Journal]
  {10.1086/429552}, 625, 23

\bibitem[\protect\citeauthoryear{Bell et~al.,}{Bell et~al.}{2006}]{Bell2006}
Bell E.~F.,  et~al., 2006, \mn@doi [The Astrophysical Journal]
  {10.1086/499931}, 640, 241

\bibitem[\protect\citeauthoryear{Benitez}{Benitez}{2000}]{Benitez2000}
Benitez N.,  2000, \mn@doi [The Astrophysical Journal] {10.1086/308947}, 536,
  571

\bibitem[\protect\citeauthoryear{Bertone \& Conselice}{Bertone \&
  Conselice}{2009}]{Bertone2009}
Bertone S.,  Conselice C.~J.,  2009, \mn@doi [Monthly Notices of the Royal
  Astronomical Society] {10.1111/j.1365-2966.2009.14916.x}, 396, 2345

\bibitem[\protect\citeauthoryear{Bluck, Conselice, Bouwens, Daddi, Dickinson,
  Papovich  \& Yan}{Bluck et~al.}{2009}]{Bluck2009}
Bluck A. F.~L.,  Conselice C.~J.,  Bouwens R.~J.,  Daddi E.,  Dickinson M.,
  Papovich C.,   Yan H.,  2009, \mn@doi [Monthly Notices of the Royal
  Astronomical Society: Letters] {10.1111/j.1745-3933.2008.00608.x}, 394, L51

\bibitem[\protect\citeauthoryear{Bluck, Conselice, Buitrago, Gr{\"{u}}tzbauch,
  Hoyos, Mortlock  \& Bauer}{Bluck et~al.}{2012}]{Bluck2012}
Bluck A. F.~L.,  Conselice C.~J.,  Buitrago F.,  Gr{\"{u}}tzbauch R.,  Hoyos
  C.,  Mortlock A.,   Bauer A.~E.,  2012, \mn@doi [The Astrophysical Journal]
  {10.1088/0004-637X/747/1/34}, 747, 34

\bibitem[\protect\citeauthoryear{Bradshaw et~al.,}{Bradshaw
  et~al.}{2013}]{Bradshaw2013}
Bradshaw E.~J.,  et~al., 2013, \mn@doi [Monthly Notices of the Royal
  Astronomical Society] {10.1093/mnras/stt715}, 433, 194

\bibitem[\protect\citeauthoryear{Brammer, van Dokkum  \& Coppi}{Brammer
  et~al.}{2008}]{Brammer2008a}
Brammer G.~B.,  van Dokkum P.~G.,   Coppi P.,  2008, \mn@doi [The Astrophysical
  Journal] {10.1086/591786}, 686, 1503

\bibitem[\protect\citeauthoryear{Bridge et~al.,}{Bridge
  et~al.}{2007}]{Bridge2007}
Bridge C.~R.,  et~al., 2007, \mn@doi [The Astrophysical Journal]
  {10.1086/512029}, 659, 931

\bibitem[\protect\citeauthoryear{Bruzual \& Charlot}{Bruzual \&
  Charlot}{2003}]{Bruzual2003}
Bruzual G.,  Charlot S.,  2003, \mn@doi [Monthly Notices of the Royal
  Astronomical Society] {10.1046/j.1365-8711.2003.06897.x}, 344, 1000

\bibitem[\protect\citeauthoryear{Buitrago, Trujillo, Conselice, Bouwens,
  Dickinson  \& Yan}{Buitrago et~al.}{2008}]{Buitrago2008}
Buitrago F.,  Trujillo I.,  Conselice C.~J.,  Bouwens R.~J.,  Dickinson M.,
  Yan H.,  2008, \mn@doi [The Astrophysical Journal] {10.1086/592836}, 687, L61

\bibitem[\protect\citeauthoryear{Bundy, Fukugita, Ellis, Targett, Belli  \&
  Kodama}{Bundy et~al.}{2009}]{Bundy2009}
Bundy K.,  Fukugita M.,  Ellis R.~S.,  Targett T.~A.,  Belli S.,   Kodama T.,
  2009, \mn@doi [The Astrophysical Journal] {10.1088/0004-637X/697/2/1369},
  697, 1369

\bibitem[\protect\citeauthoryear{Calzetti, Armus, Bohlin, Kinney, Koornneef  \&
  Storchi-Bergmann}{Calzetti et~al.}{2000}]{Calzetti2000}
Calzetti D.,  Armus L.,  Bohlin R.~C.,  Kinney A.~L.,  Koornneef J.,
  Storchi-Bergmann T.,  2000, \mn@doi [The Astrophysical Journal]
  {10.1086/308692}, 533, 682

\bibitem[\protect\citeauthoryear{Capak et~al.,}{Capak et~al.}{2007}]{Capak2007}
Capak P.,  et~al., 2007, \mn@doi [The Astrophysical Journal Supplement Series]
  {10.1086/519081}, 172, 99

\bibitem[\protect\citeauthoryear{Caputi et~al.,}{Caputi
  et~al.}{2015}]{Caputi2015}
Caputi K.~I.,  et~al., 2015, \mn@doi [The Astrophysical Journal]
  {10.1088/0004-637X/810/1/73}, 810, 73

\bibitem[\protect\citeauthoryear{Carlberg, Pritchet  \& Infante}{Carlberg
  et~al.}{1994}]{Carlberg1994}
Carlberg R.~G.,  Pritchet C.~J.,   Infante L.,  1994, \mn@doi [The
  Astrophysical Journal] {10.1086/174835}, 435, 540

\bibitem[\protect\citeauthoryear{Chabrier}{Chabrier}{2003}]{Chabrier2003}
Chabrier G.,  2003, \mn@doi [Publications of the Astronomical Society of the
  Pacific] {10.1086/376392}, 115, 763

\bibitem[\protect\citeauthoryear{Chary \& Elbaz}{Chary \&
  Elbaz}{2001}]{Chary2001}
Chary R.,  Elbaz D.,  2001, \mn@doi [The Astrophysical Journal]
  {10.1086/321609}, 556, 562

\bibitem[\protect\citeauthoryear{Conselice}{Conselice}{2006}]{Conselice2006}
Conselice C.~J.,  2006, \mn@doi [The Astrophysical Journal] {10.1086/499067},
  638, 686

\bibitem[\protect\citeauthoryear{Conselice}{Conselice}{2009}]{Conselice2009a}
Conselice C.~J.,  2009, \mn@doi [Monthly Notices of the Royal Astronomical
  Society: Letters] {10.1111/j.1745-3933.2009.00708.x}, 399, L16

\bibitem[\protect\citeauthoryear{Conselice, Bershady, Dickinson  \&
  Papovich}{Conselice et~al.}{2003}]{Conselice2003}
Conselice C.~J.,  Bershady M.~A.,  Dickinson M.,   Papovich C.,  2003, \mn@doi
  [The Astronomical Journal] {10.1086/377318}, 126, 1183

\bibitem[\protect\citeauthoryear{Conselice et~al.,}{Conselice
  et~al.}{2007}]{Conselice2007}
Conselice C.~J.,  et~al., 2007, \mn@doi [Monthly Notices of the Royal
  Astronomical Society] {10.1111/j.1365-2966.2007.12316.x}, 381, 962

\bibitem[\protect\citeauthoryear{Conselice, Rajgor  \& Myers}{Conselice
  et~al.}{2008}]{Conselice2008}
Conselice C.~J.,  Rajgor S.,   Myers R.,  2008, \mn@doi [Monthly Notices of the
  Royal Astronomical Society] {10.1111/j.1365-2966.2008.13069.x}, 386, 909

\bibitem[\protect\citeauthoryear{Conselice, Mortlock, Bluck, Grutzbauch  \&
  Duncan}{Conselice et~al.}{2013}]{Conselice2013}
Conselice C.~J.,  Mortlock A.,  Bluck A. F.~L.,  Grutzbauch R.,   Duncan K.,
  2013, \mn@doi [Monthly Notices of the Royal Astronomical Society]
  {10.1093/mnras/sts682}, 430, 1051

\bibitem[\protect\citeauthoryear{Conselice, Bluck, Mortlock, Palamara  \&
  Benson}{Conselice et~al.}{2014}]{Conselice2014b}
Conselice C.~J.,  Bluck A. F.~L.,  Mortlock A.,  Palamara D.,   Benson A.~J.,
  2014, \mn@doi [Monthly Notices of the Royal Astronomical Society]
  {10.1093/mnras/stu1385}, 444, 1125

\bibitem[\protect\citeauthoryear{Croton et~al.,}{Croton
  et~al.}{2006}]{Croton2006}
Croton D.~J.,  et~al., 2006, \mn@doi [Monthly Notices of the Royal Astronomical
  Society] {10.1111/j.1365-2966.2005.09675.x}, 365, 11

\bibitem[\protect\citeauthoryear{Curtis-Lake et~al.,}{Curtis-Lake
  et~al.}{2012}]{Curtis-Lake2012}
Curtis-Lake E.,  et~al., 2012, \mn@doi [Monthly Notices of the Royal
  Astronomical Society] {10.1111/j.1365-2966.2012.20720.x}, 422, 1425

\bibitem[\protect\citeauthoryear{Daddi et~al.,}{Daddi et~al.}{2005}]{Daddi2005}
Daddi E.,  et~al., 2005, \mn@doi [The Astrophysical Journal] {10.1086/430104},
  626, 680

\bibitem[\protect\citeauthoryear{Dahlen et~al.,}{Dahlen
  et~al.}{2013}]{Dahlen2013}
Dahlen T.,  et~al., 2013, \mn@doi [The Astrophysical Journal]
  {10.1088/0004-637X/775/2/93}, 775, 93

\bibitem[\protect\citeauthoryear{De~Lucia \& Blaizot}{De~Lucia \&
  Blaizot}{2006}]{DeLucia2006}
De~Lucia G.,  Blaizot J.,  2006, \mn@doi [Monthly Notices of the Royal
  Astronomical Society] {10.1111/j.1365-2966.2006.11287.x}, 375, 2

\bibitem[\protect\citeauthoryear{De~Propris, Conselice, Liske, Driver, Patton,
  Graham  \& Allen}{De~Propris et~al.}{2007}]{DePropris2007}
De~Propris R.,  Conselice C.~J.,  Liske J.,  Driver S.~P.,  Patton D.~R.,
  Graham A.~W.,   Allen P.~D.,  2007, \mn@doi [The Astrophysical Journal]
  {10.1086/520488}, 666, 212

\bibitem[\protect\citeauthoryear{Driver \& Robotham}{Driver \&
  Robotham}{2010}]{Driver2010}
Driver S.~P.,  Robotham A. S.~G.,  2010, \mn@doi [Monthly Notices of the Royal
  Astronomical Society] {10.1111/j.1365-2966.2010.17028.x}, 407, 2131

\bibitem[\protect\citeauthoryear{Driver, Norberg, Baldry, Bamford, Hopkins,
  Liske, Loveday  \& Peacock}{Driver et~al.}{2009}]{Driver2009}
Driver S.~P.,  Norberg P.,  Baldry I.~K.,  Bamford S.~P.,  Hopkins A.~M.,
  Liske J.,  Loveday J.,   Peacock J.~A.,  2009, \mn@doi [Astronomy {\&}
  Geophysics] {10.1111/j.1468-4004.2009.50512.x}, 50, 12

\bibitem[\protect\citeauthoryear{Duncan et~al.,}{Duncan
  et~al.}{2014}]{Duncan2014}
Duncan K.,  et~al., 2014, \mn@doi [Monthly Notices of the Royal Astronomical
  Society] {10.1093/mnras/stu1622}, 444, 2960

\bibitem[\protect\citeauthoryear{Fevre et~al.,}{Fevre
  et~al.}{2004}]{LeFevre2005}
Fevre O.~L.,  et~al., 2004, \mn@doi [Astronomy and Astrophysics]
  {10.1051/0004-6361:20041960}, 439, 30

\bibitem[\protect\citeauthoryear{Fioc \& Rocca-Volmerange}{Fioc \&
  Rocca-Volmerange}{1999}]{Fioc1999}
Fioc M.,  Rocca-Volmerange B.,  1999, eprint arXiv:astro-ph/9912179

\bibitem[\protect\citeauthoryear{Furusawa et~al.,}{Furusawa
  et~al.}{2008}]{Furusawa2008}
Furusawa H.,  et~al., 2008, \mn@doi [The Astrophysical Journal Supplement
  Series] {10.1086/527321}, 176, 1

\bibitem[\protect\citeauthoryear{Garilli et~al.,}{Garilli
  et~al.}{2014}]{Garilli2014}
Garilli B.,  et~al., 2014, \mn@doi [Astronomy {\&} Astrophysics]
  {10.1051/0004-6361/201322790}, 562, A23

\bibitem[\protect\citeauthoryear{Genel et~al.,}{Genel et~al.}{2014}]{Genel2014}
Genel S.,  et~al., 2014, \mn@doi [Monthly Notices of the Royal Astronomical
  Society] {10.1093/mnras/stu1654}, 445, 175

\bibitem[\protect\citeauthoryear{Guo et~al.,}{Guo et~al.}{2011}]{Guo2011}
Guo Q.,  et~al., 2011, \mn@doi [Monthly Notices of the Royal Astronomical
  Society] {10.1111/j.1365-2966.2010.18114.x}, 413, 101

\bibitem[\protect\citeauthoryear{Hartley et~al.,}{Hartley
  et~al.}{2013}]{Hartley2013}
Hartley W.~G.,  et~al., 2013, \mn@doi [Monthly Notices of the Royal
  Astronomical Society] {10.1093/mnras/stt383}, 431, 3045

\bibitem[\protect\citeauthoryear{Henriques, White, Thomas, Angulo, Guo, Lemson,
  Springel  \& Overzier}{Henriques et~al.}{2015}]{Henriques2014}
Henriques B. M.~B.,  White S. D.~M.,  Thomas P.~A.,  Angulo R.,  Guo Q.,
  Lemson G.,  Springel V.,   Overzier R.,  2015, \mn@doi [Monthly Notices of
  the Royal Astronomical Society] {10.1093/mnras/stv705}, 451, 2663

\bibitem[\protect\citeauthoryear{Hildebrandt, Wolf  \&
  Ben{\'{i}}tez}{Hildebrandt et~al.}{2008}]{Hildebrandt2008}
Hildebrandt H.,  Wolf C.,   Ben{\'{i}}tez N.,  2008, \mn@doi [Astronomy and
  Astrophysics] {10.1051/0004-6361:20077107}, 480, 703

\bibitem[\protect\citeauthoryear{Hopkins et~al.,}{Hopkins
  et~al.}{2013}]{Hopkins2013}
Hopkins A.~M.,  et~al., 2013, \mn@doi [Monthly Notices of the Royal
  Astronomical Society] {10.1093/mnras/stt030}, 430, 2047

\bibitem[\protect\citeauthoryear{Ilbert et~al.,}{Ilbert
  et~al.}{2009}]{Ilbert2009}
Ilbert O.,  et~al., 2009, \mn@doi [The Astrophysical Journal]
  {10.1088/0004-637X/690/2/1236}, 690, 1236

\bibitem[\protect\citeauthoryear{Ilbert et~al.,}{Ilbert
  et~al.}{2010}]{Ilbert2010}
Ilbert O.,  et~al., 2010, \mn@doi [The Astrophysical Journal]
  {10.1088/0004-637X/709/2/644}, 709, 644

\bibitem[\protect\citeauthoryear{Ilbert et~al.,}{Ilbert
  et~al.}{2013}]{Ilbert2013}
Ilbert O.,  et~al., 2013, \mn@doi [Astronomy {\&} Astrophysics]
  {10.1051/0004-6361/201321100}, 556, A55

\bibitem[\protect\citeauthoryear{Jaacks, Finkelstein  \& Nagamine}{Jaacks
  et~al.}{2016}]{Jaacks2016}
Jaacks J.,  Finkelstein S.~L.,   Nagamine K.,  2016, \mn@doi [The Astrophysical
  Journal] {10.3847/0004-637X/817/2/174}, 817, 174

\bibitem[\protect\citeauthoryear{Jarvis et~al.,}{Jarvis
  et~al.}{2012}]{Jarvis2012}
Jarvis M.~J.,  et~al., 2012, \mn@doi [Monthly Notices of the Royal Astronomical
  Society] {10.1093/mnras/sts118}, 428, 1281

\bibitem[\protect\citeauthoryear{Jenkins et~al.,}{Jenkins
  et~al.}{1997}]{Jenkins1998}
Jenkins A.,  et~al., 1997, \mn@doi [The Astrophysical Journal]
  {10.1086/305615}, 499, 20

\bibitem[\protect\citeauthoryear{Jogee et~al.,}{Jogee et~al.}{2009}]{Jogee2009}
Jogee S.,  et~al., 2009, \mn@doi [The Astrophysical Journal]
  {10.1088/0004-637X/697/2/1971}, 697, 1971

\bibitem[\protect\citeauthoryear{Kartaltepe et~al.,}{Kartaltepe
  et~al.}{2007}]{Kartaltepe2007}
Kartaltepe J.~S.,  et~al., 2007, \mn@doi [The Astrophysical Journal Supplement
  Series] {10.1086/519953}, 172, 320

\bibitem[\protect\citeauthoryear{Kelvin et~al.,}{Kelvin
  et~al.}{2014}]{Kelvin2014}
Kelvin L.~S.,  et~al., 2014, \mn@doi [Monthly Notices of the Royal Astronomical
  Society] {10.1093/mnras/stu1507}, 444, 1647

\bibitem[\protect\citeauthoryear{Kennicutt}{Kennicutt}{1998}]{Kennicutt1998}
Kennicutt R.~C.,  1998, \mn@doi [Annual Review of Astronomy and Astrophysics]
  {10.1146/annurev.astro.36.1.189}, 36, 189

\bibitem[\protect\citeauthoryear{Kitzbichler \& White}{Kitzbichler \&
  White}{2008}]{Kitzbichler2008}
Kitzbichler M.~G.,  White S. D.~M.,  2008, \mn@doi [Monthly Notices of the
  Royal Astronomical Society] {10.1111/j.1365-2966.2008.13873.x}, 391, 1489

\bibitem[\protect\citeauthoryear{Laigle et~al.,}{Laigle
  et~al.}{2016}]{Laigle2016}
Laigle C.,  et~al., 2016, \mn@doi [The Astrophysical Journal Supplement Series]
  {10.3847/0067-0049/224/2/24}, 224, 24

\bibitem[\protect\citeauthoryear{Lawrence et~al.,}{Lawrence
  et~al.}{2007}]{Lawrence2007}
Lawrence A.,  et~al., 2007, \mn@doi [Monthly Notices of the Royal Astronomical
  Society] {10.1111/j.1365-2966.2007.12040.x}, 379, 1599

\bibitem[\protect\citeauthoryear{Le~Fevre et~al.,}{Le~Fevre
  et~al.}{2000}]{LeFevre2000}
Le~Fevre O.,  et~al., 2000, \mn@doi [Monthly Notices of the Royal Astronomical
  Society] {10.1046/j.1365-8711.2000.03083.x}, 311, 565

\bibitem[\protect\citeauthoryear{Leja, van Dokkum  \& Franx}{Leja
  et~al.}{2013}]{Leja2013}
Leja J.,  van Dokkum P.,   Franx M.,  2013, \mn@doi [The Astrophysical Journal]
  {10.1088/0004-637X/766/1/33}, 766, 33

\bibitem[\protect\citeauthoryear{Lemson \& Consortium}{Lemson \&
  Consortium}{2006}]{Lemson2006}
Lemson G.,  Consortium t.~V.,  2006, eprint arXiv:astro-ph/0608019, 1, 1

\bibitem[\protect\citeauthoryear{Lilly et~al.,}{Lilly
  et~al.}{2007}]{Lilly2007a}
Lilly S.~J.,  et~al., 2007, \mn@doi [The Astrophysical Journal Supplement
  Series] {10.1086/516589}, 172, 70

\bibitem[\protect\citeauthoryear{Lin et~al.,}{Lin et~al.}{2004}]{Lin2004}
Lin L.,  et~al., 2004, \mn@doi [The Astrophysical Journal] {10.1086/427183},
  617, L9

\bibitem[\protect\citeauthoryear{Lin et~al.,}{Lin et~al.}{2008}]{Lin2008}
Lin L.,  et~al., 2008, \mn@doi [The Astrophysical Journal] {10.1086/587928},
  681, 232

\bibitem[\protect\citeauthoryear{Liske et~al.,}{Liske et~al.}{2015}]{Liske2015}
Liske J.,  et~al., 2015, \mn@doi [Monthly Notices of the Royal Astronomical
  Society] {10.1093/mnras/stv1436}, 452, 2087

\bibitem[\protect\citeauthoryear{L{\'{o}}pez-Sanjuan, Balcells,
  P{\'{e}}rez-Gonz{\'{a}}lez, Barro, Garc{\'{i}}a-Dab{\'{o}}, Gallego  \&
  Zamorano}{L{\'{o}}pez-Sanjuan et~al.}{2009}]{Lopez-Sanjuan2009}
L{\'{o}}pez-Sanjuan C.,  Balcells M.,  P{\'{e}}rez-Gonz{\'{a}}lez P.~G.,  Barro
  G.,  Garc{\'{i}}a-Dab{\'{o}} C.~E.,  Gallego J.,   Zamorano J.,  2009,
  \mn@doi [Astronomy and Astrophysics] {10.1051/0004-6361/200911923}, 501, 505

\bibitem[\protect\citeauthoryear{L{\'{o}}pez-Sanjuan
  et~al.,}{L{\'{o}}pez-Sanjuan et~al.}{2011}]{Lopez-Sanjuan2011}
L{\'{o}}pez-Sanjuan C.,  et~al., 2011, \mn@doi [Astronomy {\&} Astrophysics]
  {10.1051/0004-6361/201015839}, 530, A20

\bibitem[\protect\citeauthoryear{L{\'{o}}pez-Sanjuan
  et~al.,}{L{\'{o}}pez-Sanjuan et~al.}{2012}]{Lopez-Sanjuan2012}
L{\'{o}}pez-Sanjuan C.,  et~al., 2012, \mn@doi [Astronomy {\&} Astrophysics]
  {10.1051/0004-6361/201219085}, 548, A7

\bibitem[\protect\citeauthoryear{L{\'{o}}pez-Sanjuan
  et~al.,}{L{\'{o}}pez-Sanjuan et~al.}{2013}]{Lopez-Sanjuan2013}
L{\'{o}}pez-Sanjuan C.,  et~al., 2013, \mn@doi [Astronomy {\&} Astrophysics]
  {10.1051/0004-6361/201220286}, 553, A78

\bibitem[\protect\citeauthoryear{L{\'{o}}pez-Sanjuan
  et~al.,}{L{\'{o}}pez-Sanjuan et~al.}{2015}]{Lopez-Sanjuan2014}
L{\'{o}}pez-Sanjuan C.,  et~al., 2015, \mn@doi [Astronomy {\&} Astrophysics]
  {10.1051/0004-6361/201424913}, 576, A53

\bibitem[\protect\citeauthoryear{Lotz, Primack  \& Madau}{Lotz
  et~al.}{2004}]{Lotz2004}
Lotz J.~M.,  Primack J.,   Madau P.,  2004, \mn@doi [The Astronomical Journal]
  {10.1086/421849}, 128, 163

\bibitem[\protect\citeauthoryear{Lotz et~al.,}{Lotz et~al.}{2008}]{Lotz2008}
Lotz J.~M.,  et~al., 2008, \mn@doi [The Astrophysical Journal]
  {10.1086/523659}, 672, 177

\bibitem[\protect\citeauthoryear{Lotz, Jonsson, Cox, Croton, Primack,
  Somerville  \& Stewart}{Lotz et~al.}{2011}]{Lotz2011}
Lotz J.~M.,  Jonsson P.,  Cox T.~J.,  Croton D.,  Primack J.~R.,  Somerville
  R.~S.,   Stewart K.,  2011, \mn@doi [The Astrophysical Journal]
  {10.1088/0004-637X/742/2/103}, 742, 103

\bibitem[\protect\citeauthoryear{Madau \& Dickinson}{Madau \&
  Dickinson}{2014}]{Madau2014}
Madau P.,  Dickinson M.,  2014, \mn@doi [Annual Review of Astronomy and
  Astrophysics] {10.1146/annurev-astro-081811-125615}, 52, 415

\bibitem[\protect\citeauthoryear{Maller, Katz, Kere{\v{s}}, Dave  \&
  Weinberg}{Maller et~al.}{2006}]{Maller2006}
Maller A.~H.,  Katz N.,  Kere{\v{s}} D.,  Dave R.,   Weinberg D.~H.,  2006,
  \mn@doi [The Astrophysical Journal] {10.1086/503319}, 647, 763

\bibitem[\protect\citeauthoryear{Man, Toft, Zirm, Wuyts  \& van~der Wel}{Man
  et~al.}{2012}]{Man2012}
Man A. W.~S.,  Toft S.,  Zirm A.~W.,  Wuyts S.,   van~der Wel A.,  2012,
  \mn@doi [The Astrophysical Journal] {10.1088/0004-637X/744/2/85}, 744, 85

\bibitem[\protect\citeauthoryear{Man, Zirm  \& Toft}{Man
  et~al.}{2016}]{Man2014}
Man A. W.~S.,  Zirm A.~W.,   Toft S.,  2016, \mn@doi [The Astrophysical
  Journal] {10.3847/0004-637X/830/2/89}, 830, 89

\bibitem[\protect\citeauthoryear{Maraston}{Maraston}{2005}]{Maraston2005}
Maraston C.,  2005, \mn@doi [Monthly Notices of the Royal Astronomical Society]
  {10.1111/j.1365-2966.2005.09270.x}, 362, 799

\bibitem[\protect\citeauthoryear{Maraston, Pforr, Renzini, Daddi, Dickinson,
  Cimatti  \& Tonini}{Maraston et~al.}{2010}]{Maraston2010}
Maraston C.,  Pforr J.,  Renzini A.,  Daddi E.,  Dickinson M.,  Cimatti A.,
  Tonini C.,  2010, \mn@doi [Monthly Notices of the Royal Astronomical Society]
  {10.1111/j.1365-2966.2010.16973.x}, 407, 830

\bibitem[\protect\citeauthoryear{Martin et~al.,}{Martin
  et~al.}{2005}]{Martin2005}
Martin D.~C.,  et~al., 2005, \mn@doi [The Astrophysical Journal]
  {10.1086/426387}, 619, L1

\bibitem[\protect\citeauthoryear{McCracken et~al.,}{McCracken
  et~al.}{2012}]{McCracken2012}
McCracken H.~J.,  et~al., 2012, \mn@doi [Astronomy {\&} Astrophysics]
  {10.1051/0004-6361/201219507}, 544, A156

\bibitem[\protect\citeauthoryear{McLure et~al.,}{McLure
  et~al.}{2013}]{McLure2013}
McLure R.~J.,  et~al., 2013, \mn@doi [Monthly Notices of the Royal Astronomical
  Society] {10.1093/mnras/sts092}, 428, 1088

\bibitem[\protect\citeauthoryear{Mihos}{Mihos}{1995}]{Mihos1995}
Mihos J.~C.,  1995, \mn@doi [The Astrophysical Journal] {10.1086/187719}, 438,
  L75

\bibitem[\protect\citeauthoryear{Molino et~al.,}{Molino
  et~al.}{2013}]{Molino2013}
Molino A.,  et~al., 2013, eprint arXiv:1306.4968

\bibitem[\protect\citeauthoryear{Mortlock et~al.,}{Mortlock
  et~al.}{2013}]{Mortlock2013}
Mortlock A.,  et~al., 2013, \mn@doi [Monthly Notices of the Royal Astronomical
  Society] {10.1093/mnras/stt793}, 433, 1185

\bibitem[\protect\citeauthoryear{Mortlock et~al.,}{Mortlock
  et~al.}{2015}]{Mortlock2015}
Mortlock A.,  et~al., 2015, \mn@doi [Monthly Notices of the Royal Astronomical
  Society] {10.1093/mnras/stu2403}, 447, 2

\bibitem[\protect\citeauthoryear{Mortlock et~al.,}{Mortlock
  et~al.}{2016}]{Mortlock2016}
Mortlock A.,  et~al., 2016, \mn@doi [Monthly Notices of the Royal Astronomical
  Society] {10.1093/mnras/stw390}, 458, 3478

\bibitem[\protect\citeauthoryear{Moster, Somerville, Newman  \& Rix}{Moster
  et~al.}{2011}]{Moster2011}
Moster B.~P.,  Somerville R.~S.,  Newman J.~A.,   Rix H.-W.,  2011, \mn@doi
  [The Astrophysical Journal] {10.1088/0004-637X/731/2/113}, 731, 113

\bibitem[\protect\citeauthoryear{Mundy, Conselice  \& Ownsworth}{Mundy
  et~al.}{2015}]{Mundy2015}
Mundy C.~J.,  Conselice C.~J.,   Ownsworth J.~R.,  2015, \mn@doi [Monthly
  Notices of the Royal Astronomical Society] {10.1093/mnras/stv860}, 450, 3696

\bibitem[\protect\citeauthoryear{Muzzin et~al.,}{Muzzin
  et~al.}{2013a}]{Muzzin2013a}
Muzzin A.,  et~al., 2013a, \mn@doi [The Astrophysical Journal Supplement
  Series] {10.1088/0067-0049/206/1/8}, 206, 8

\bibitem[\protect\citeauthoryear{Muzzin et~al.,}{Muzzin
  et~al.}{2013b}]{Muzzin2013b}
Muzzin A.,  et~al., 2013b, \mn@doi [The Astrophysical Journal]
  {10.1088/0004-637X/777/1/18}, 777, 18

\bibitem[\protect\citeauthoryear{Newman, Ellis, Bundy  \& Treu}{Newman
  et~al.}{2012}]{Newman2012}
Newman A.~B.,  Ellis R.~S.,  Bundy K.,   Treu T.,  2012, \mn@doi [The
  Astrophysical Journal] {10.1088/0004-637X/746/2/162}, 746, 162

\bibitem[\protect\citeauthoryear{Oke \& Gunn}{Oke \& Gunn}{1983}]{Oke1983}
Oke J.~B.,  Gunn J.~E.,  1983, \mn@doi [The Astrophysical Journal]
  {10.1086/160817}, 266, 713

\bibitem[\protect\citeauthoryear{Onodera et~al.,}{Onodera
  et~al.}{2012}]{Onodera2012}
Onodera M.,  et~al., 2012, \mn@doi [The Astrophysical Journal]
  {10.1088/0004-637X/755/1/26}, 755, 26

\bibitem[\protect\citeauthoryear{Ownsworth, Conselice, Mortlock, Hartley,
  Almaini, Duncan  \& Mundy}{Ownsworth et~al.}{2014}]{Ownsworth2014}
Ownsworth J.~R.,  Conselice C.~J.,  Mortlock A.,  Hartley W.~G.,  Almaini O.,
  Duncan K.,   Mundy C.~J.,  2014, \mn@doi [Monthly Notices of the Royal
  Astronomical Society] {10.1093/mnras/stu1802}, 445, 2198

\bibitem[\protect\citeauthoryear{Ownsworth, Conselice, Mundy, Mortlock,
  Hartley, Duncan  \& Almaini}{Ownsworth et~al.}{2016}]{Ownsworth2016}
Ownsworth J.~R.,  Conselice C.~J.,  Mundy C.~J.,  Mortlock A.,  Hartley W.~G.,
  Duncan K.,   Almaini O.,  2016, \mn@doi [Monthly Notices of the Royal
  Astronomical Society] {10.1093/mnras/stw1207}, 461, 1112

\bibitem[\protect\citeauthoryear{Papovich, Finkelstein, Ferguson, Lotz  \&
  Giavalisco}{Papovich et~al.}{2011}]{Papovich2011}
Papovich C.,  Finkelstein S.~L.,  Ferguson H.~C.,  Lotz J.~M.,   Giavalisco M.,
   2011, \mn@doi [Monthly Notices of the Royal Astronomical Society]
  {10.1111/j.1365-2966.2010.17965.x}, 412, 1123

\bibitem[\protect\citeauthoryear{Patton \& Atfield}{Patton \&
  Atfield}{2008}]{Patton2008}
Patton D.~R.,  Atfield J.~E.,  2008, \mn@doi [The Astrophysical Journal]
  {10.1086/590542}, 685, 235

\bibitem[\protect\citeauthoryear{Patton, Pritchet, Yee, Ellingson  \&
  Carlberg}{Patton et~al.}{1997}]{Patton1997}
Patton D.~R.,  Pritchet C.~J.,  Yee H. K.~C.,  Ellingson E.,   Carlberg R.~G.,
  1997, The Astrophysical Journal, 475, 29

\bibitem[\protect\citeauthoryear{Patton, Carlberg, Marzke, Pritchet, da Costa
  \& Pellegrini}{Patton et~al.}{2000}]{Patton2000}
Patton D.~R.,  Carlberg R.~G.,  Marzke R.~O.,  Pritchet C.~J.,  da Costa L.~N.,
    Pellegrini P.~S.,  2000, \mn@doi [The Astrophysical Journal]
  {10.1086/308907}, 536, 153

\bibitem[\protect\citeauthoryear{Patton et~al.,}{Patton
  et~al.}{2002}]{Patton2002}
Patton D.~R.,  et~al., 2002, \mn@doi [The Astrophysical Journal]
  {10.1086/324543}, 565, 208

\bibitem[\protect\citeauthoryear{Pawlik, Wild, Walcher, Johansson, Villforth,
  Rowlands, Mendez-Abreu  \& Hewlett}{Pawlik et~al.}{2016}]{Pawlik2016}
Pawlik M.~M.,  Wild V.,  Walcher C.~J.,  Johansson P.~H.,  Villforth C.,
  Rowlands K.,  Mendez-Abreu J.,   Hewlett T.,  2016, \mn@doi [Monthly Notices
  of the Royal Astronomical Society] {10.1093/mnras/stv2878}, 456, 3032

\bibitem[\protect\citeauthoryear{{Planck Collaboration (XVI)}}{{Planck
  Collaboration (XVI)}}{2014}]{Planck2014params}
{Planck Collaboration (XVI)} 2014, \mn@doi [Astronomy {\&} Astrophysics]
  {10.1051/0004-6361/201321591}, 571, A16

\bibitem[\protect\citeauthoryear{Pozzetti et~al.,}{Pozzetti
  et~al.}{2010}]{Pozzetti2010}
Pozzetti L.,  et~al., 2010, \mn@doi [Astronomy {\&} Astrophysics]
  {10.1051/0004-6361/200913020}, 523, A13

\bibitem[\protect\citeauthoryear{Robitaille et~al.,}{Robitaille
  et~al.}{2013}]{Robitaille2013}
Robitaille T.~P.,  et~al., 2013, \mn@doi [Astronomy {\&} Astrophysics]
  {10.1051/0004-6361/201322068}, 558, A33

\bibitem[\protect\citeauthoryear{Robotham et~al.,}{Robotham
  et~al.}{2010}]{Robotham2010}
Robotham A.,  et~al., 2010, \mn@doi [Publications of the Astronomical Society
  of Australia] {10.1071/AS09053}, 27, 76

\bibitem[\protect\citeauthoryear{Rodriguez-Gomez et~al.,}{Rodriguez-Gomez
  et~al.}{2015}]{Rodriguez-Gomez2015}
Rodriguez-Gomez V.,  et~al., 2015, \mn@doi [Monthly Notices of the Royal
  Astronomical Society] {10.1093/mnras/stv264}, 449, 49

\bibitem[\protect\citeauthoryear{Rodriguez-Gomez et~al.,}{Rodriguez-Gomez
  et~al.}{2016}]{Rodriguez-Gomez2016}
Rodriguez-Gomez V.,  et~al., 2016, \mn@doi [Monthly Notices of the Royal
  Astronomical Society] {10.1093/mnras/stw456}, 458, 2371

\bibitem[\protect\citeauthoryear{Ruiz, Trujillo  \& Marmol-Queralto}{Ruiz
  et~al.}{2014}]{Ruiz2014}
Ruiz P.,  Trujillo I.,   Marmol-Queralto E.,  2014, \mn@doi [Monthly Notices of
  the Royal Astronomical Society] {10.1093/mnras/stu821}, 442, 347

\bibitem[\protect\citeauthoryear{Sanders et~al.,}{Sanders
  et~al.}{2007}]{Sanders2007}
Sanders D.~B.,  et~al., 2007, \mn@doi [The Astrophysical Journal Supplement
  Series] {10.1086/517885}, 172, 86

\bibitem[\protect\citeauthoryear{Santini et~al.,}{Santini
  et~al.}{2012}]{Santini2012}
Santini P.,  et~al., 2012, \mn@doi [Astronomy {\&} Astrophysics]
  {10.1051/0004-6361/201117513}, 538, A33

\bibitem[\protect\citeauthoryear{Schaye et~al.,}{Schaye
  et~al.}{2014}]{Schaye2014}
Schaye J.,  et~al., 2014, \mn@doi [Monthly Notices of the Royal Astronomical
  Society] {10.1093/mnras/stu2058}, 446, 521

\bibitem[\protect\citeauthoryear{Scoville et~al.,}{Scoville
  et~al.}{2007}]{Scoville2007b}
Scoville N.,  et~al., 2007, \mn@doi [The Astrophysical Journal Supplement
  Series] {10.1086/516585}, 172, 1

\bibitem[\protect\citeauthoryear{S{\'{e}}rsic}{S{\'{e}}rsic}{1963}]{Sersic1963}
S{\'{e}}rsic J.~L.,  1963, Boletin de la Asociacion Argentina de Astronomia, 6

\bibitem[\protect\citeauthoryear{Somerville \& Dav{\'{e}}}{Somerville \&
  Dav{\'{e}}}{2015}]{Somerville2015}
Somerville R.~S.,  Dav{\'{e}} R.,  2015, \mn@doi [Annual Review of Astronomy
  and Astrophysics] {10.1146/annurev-astro-082812-140951}, 53, 51

\bibitem[\protect\citeauthoryear{Somerville, Lee, Ferguson, Gardner, Moustakas
  \& Giavalisco}{Somerville et~al.}{2004}]{Somerville2004}
Somerville R.~S.,  Lee K.,  Ferguson H.~C.,  Gardner J.~P.,  Moustakas L.~A.,
  Giavalisco M.,  2004, \mn@doi [The Astrophysical Journal] {10.1086/378628},
  600, L171

\bibitem[\protect\citeauthoryear{Springel et~al.,}{Springel
  et~al.}{2005}]{Springel2005}
Springel V.,  et~al., 2005, \mn@doi [Nature] {10.1038/nature03597}, 435, 629

\bibitem[\protect\citeauthoryear{Taylor et~al.,}{Taylor
  et~al.}{2011}]{Taylor2011}
Taylor E.~N.,  et~al., 2011, \mn@doi [Monthly Notices of the Royal Astronomical
  Society] {10.1111/j.1365-2966.2011.19536.x}, 418, 1587

\bibitem[\protect\citeauthoryear{Toomre \& Toomre}{Toomre \&
  Toomre}{1972}]{Toomre1972}
Toomre A.,  Toomre J.,  1972, \mn@doi [The Astrophysical Journal]
  {10.1086/151823}, 178, 623

\bibitem[\protect\citeauthoryear{Torrey et~al.,}{Torrey
  et~al.}{2015}]{Torrey2015}
Torrey P.,  et~al., 2015, \mn@doi [Monthly Notices of the Royal Astronomical
  Society] {10.1093/mnras/stv1986}, 454, 2770

\bibitem[\protect\citeauthoryear{Trujillo, Conselice, Bundy, Cooper, Eisenhardt
   \& Ellis}{Trujillo et~al.}{2007}]{Trujillo2007}
Trujillo I.,  Conselice C.~J.,  Bundy K.,  Cooper M.~C.,  Eisenhardt P.,
  Ellis R.~S.,  2007, \mn@doi [Monthly Notices of the Royal Astronomical
  Society] {10.1111/j.1365-2966.2007.12388.x}, 382, 109

\bibitem[\protect\citeauthoryear{Vogelsberger et~al.,}{Vogelsberger
  et~al.}{2014a}]{Vogelsberger2014}
Vogelsberger M.,  et~al., 2014a, \mn@doi [Monthly Notices of the Royal
  Astronomical Society] {10.1093/mnras/stu1536}, 444, 1518

\bibitem[\protect\citeauthoryear{Vogelsberger et~al.,}{Vogelsberger
  et~al.}{2014b}]{Vogelsberger2014a}
Vogelsberger M.,  et~al., 2014b, \mn@doi [Nature] {10.1038/nature13316}, 509,
  177

\bibitem[\protect\citeauthoryear{Williams, Quadri  \& Franx}{Williams
  et~al.}{2011}]{Williams2011}
Williams R.~J.,  Quadri R.~F.,   Franx M.,  2011, \mn@doi [The Astrophysical
  Journal] {10.1088/2041-8205/738/2/L25}, 738, L25

\bibitem[\protect\citeauthoryear{Xu, Zhao, Scoville, Capak, Drory  \& Gao}{Xu
  et~al.}{2012}]{Xu2012}
Xu C.~K.,  Zhao Y.,  Scoville N.,  Capak P.,  Drory N.,   Gao Y.,  2012,
  \mn@doi [The Astrophysical Journal] {10.1088/0004-637X/747/2/85}, 747, 85

\bibitem[\protect\citeauthoryear{de Ravel et~al.,}{de~Ravel
  et~al.}{2009}]{DeRavel2009}
de Ravel L.,  et~al., 2009, \mn@doi [Astronomy and Astrophysics]
  {10.1051/0004-6361/200810569}, 498, 379

\bibitem[\protect\citeauthoryear{van~der Wel et~al.,}{van~der Wel
  et~al.}{2014}]{vanderwel2014}
van~der Wel A.,  et~al., 2014, \mn@doi [The Astrophysical Journal]
  {10.1088/0004-637X/788/1/28}, 788, 28

\makeatother
\end{thebibliography}
\bibliographystyle{mnras}

\appendix
\section{VIDEO Completeness Simulations}
\label{sec:appendix-video-completeness}

\begin{figure}
  \includegraphics[width=0.475\textwidth]{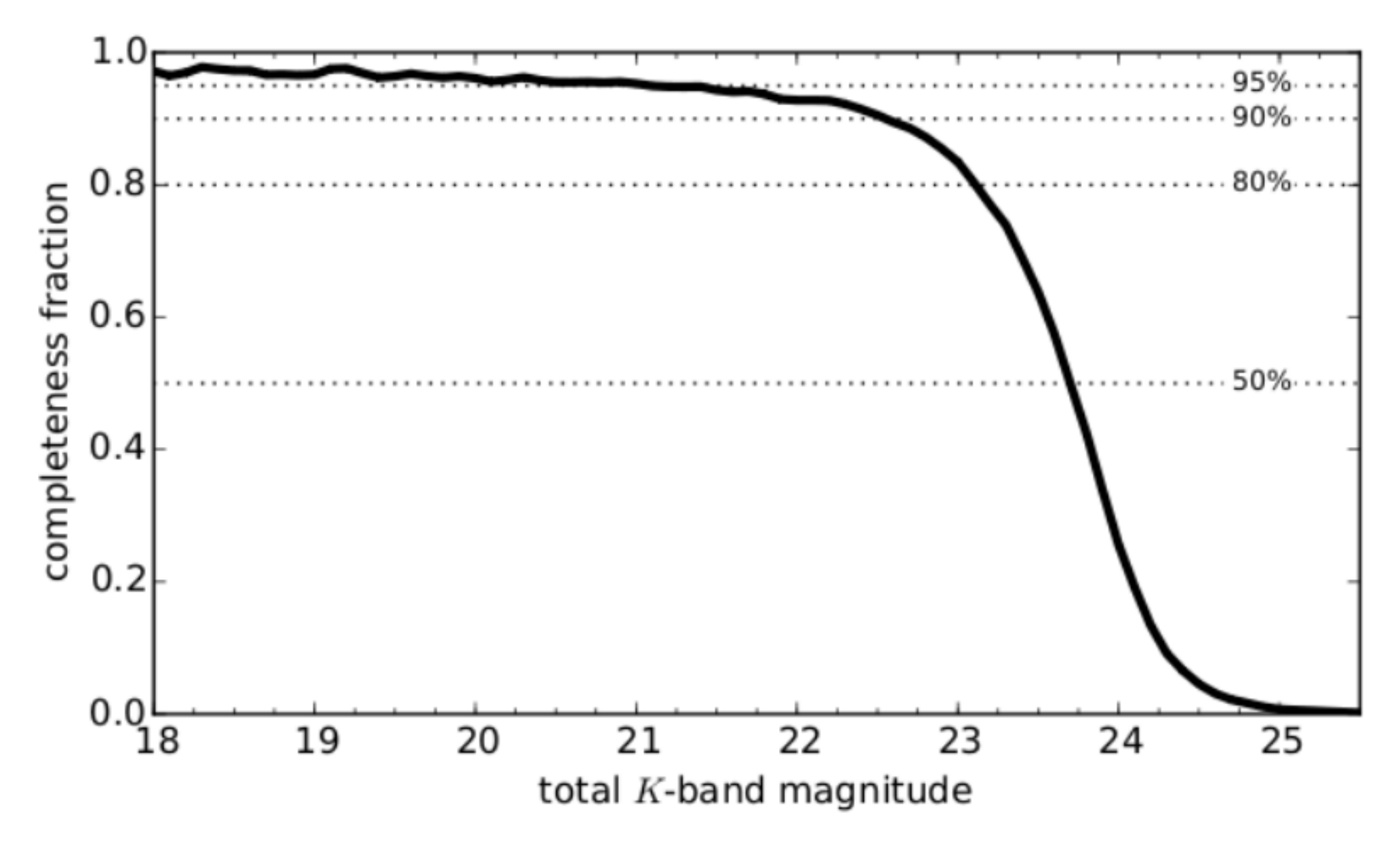}
  \caption{Fraction of artificial sources recovered from the VIDEO sky background as a function of total $K$-band magnitude.}
  \label{fig:video-comp}
\end{figure}

We perform comprehensive completeness simulations on a pixel scale matched VIDEO $K$-band image in the CFHT-LS D1 deep field. This field provides exactly 1 square degree with matched photometry from the VIDEO near-IR filters and optical data from the CFHT-LS. The total $K$-band completeness fraction, shown in Figure \ref{fig:video-comp}, is calculated in the following way. Firstly, areas of background sky are identified in the $K$-band image and patched together to create an image the size of the original image. This image is then populated with realistic galaxy light profiles with a range of magnitudes ($17 < m_K < 27$), S\'{e}rsic \citep{Sersic1963} indices ($0.5 < n < 8$), and sizes taken from observed distributions within the UDS DR8. These profiles are added randomly to the image using the \textit{IRAF} \textit{mkobjects} routine. SExtractor is then run on these modified images using exactly the same configuration as used to create the catalogue used in this work. Comparing the input parameters with those extracted, the completeness fraction as a function of total $K$-band magnitude is calculated. An artificial source is considered to be recovered if it is found to within 1" of its input location, and 1 mag of its input magnitude. We find that the data is 95, 90, 80 and 50 per cent complete in $m_{K,\textrm{tot}}$ at 21.5, 22.5, 23.1 and 23.7 AB mags, respectively.

\section{Odds parameter selection}
\label{appendix:odds}
\begin{figure}
\includegraphics[width=0.475\textwidth]{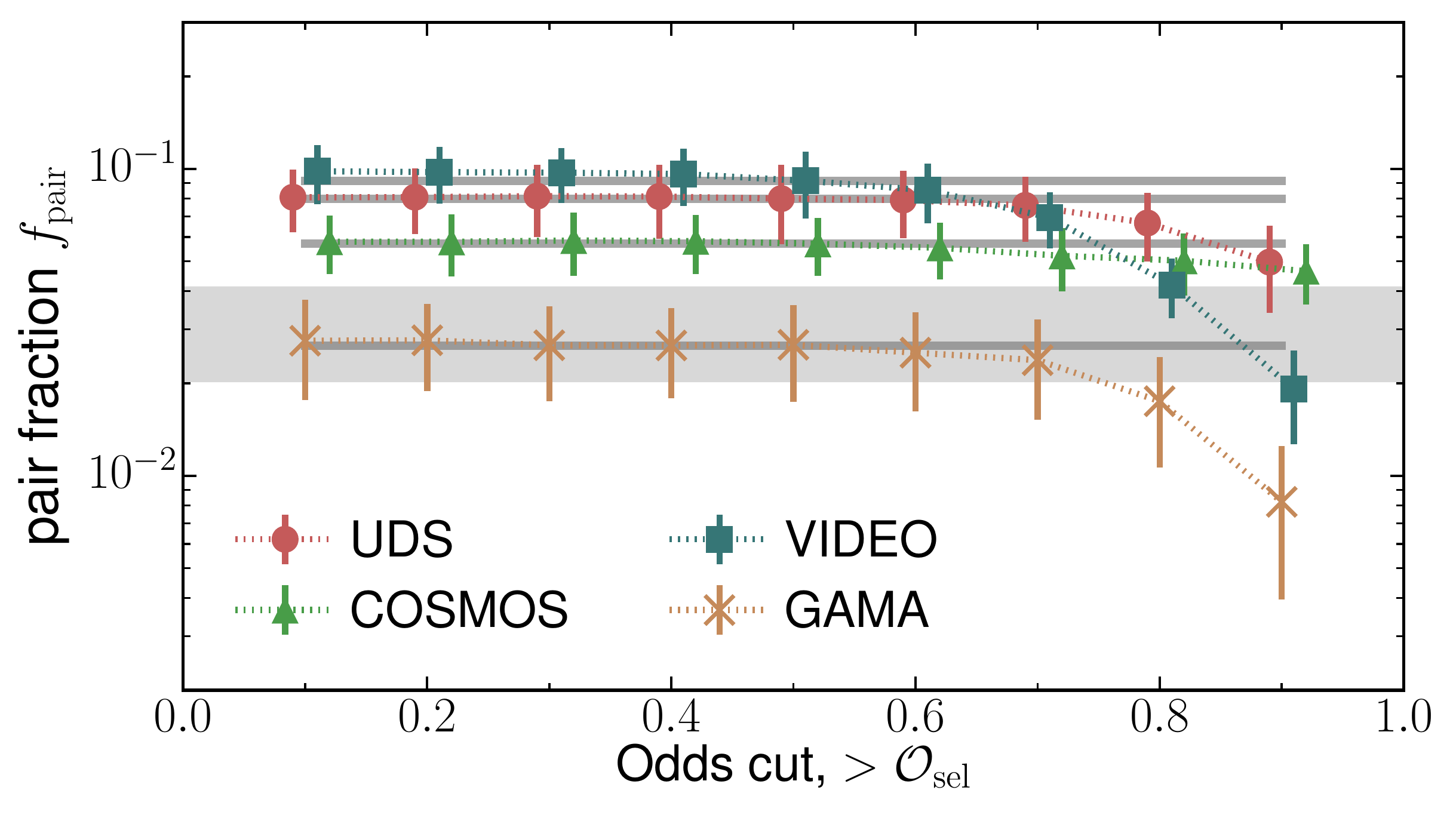}
\caption{Measured photometric pair fractions as a function of cut adopted in the Odds parameter, $\mathcal{O}$ for the GAMA (gold crosses), UDS (red circles), VIDEO (blue squares) and COSMOS (green triangles) regions. Pair fractions are measured at $0.01 < z < 0.15$ in the GAMA region and at $0.2 < z < 1.0$ in the remaining three survey regions. The grey shaded region represents the measured spectroscopic pair fraction in the GAMA region over the same redshift range. Horizontal solid grey lines represent the median pair fraction over all $\mathcal{O}$ in each survey region.}
\label{fig:odds_selection}
\end{figure}

\begin{figure}
  \includegraphics[width=0.475\textwidth]{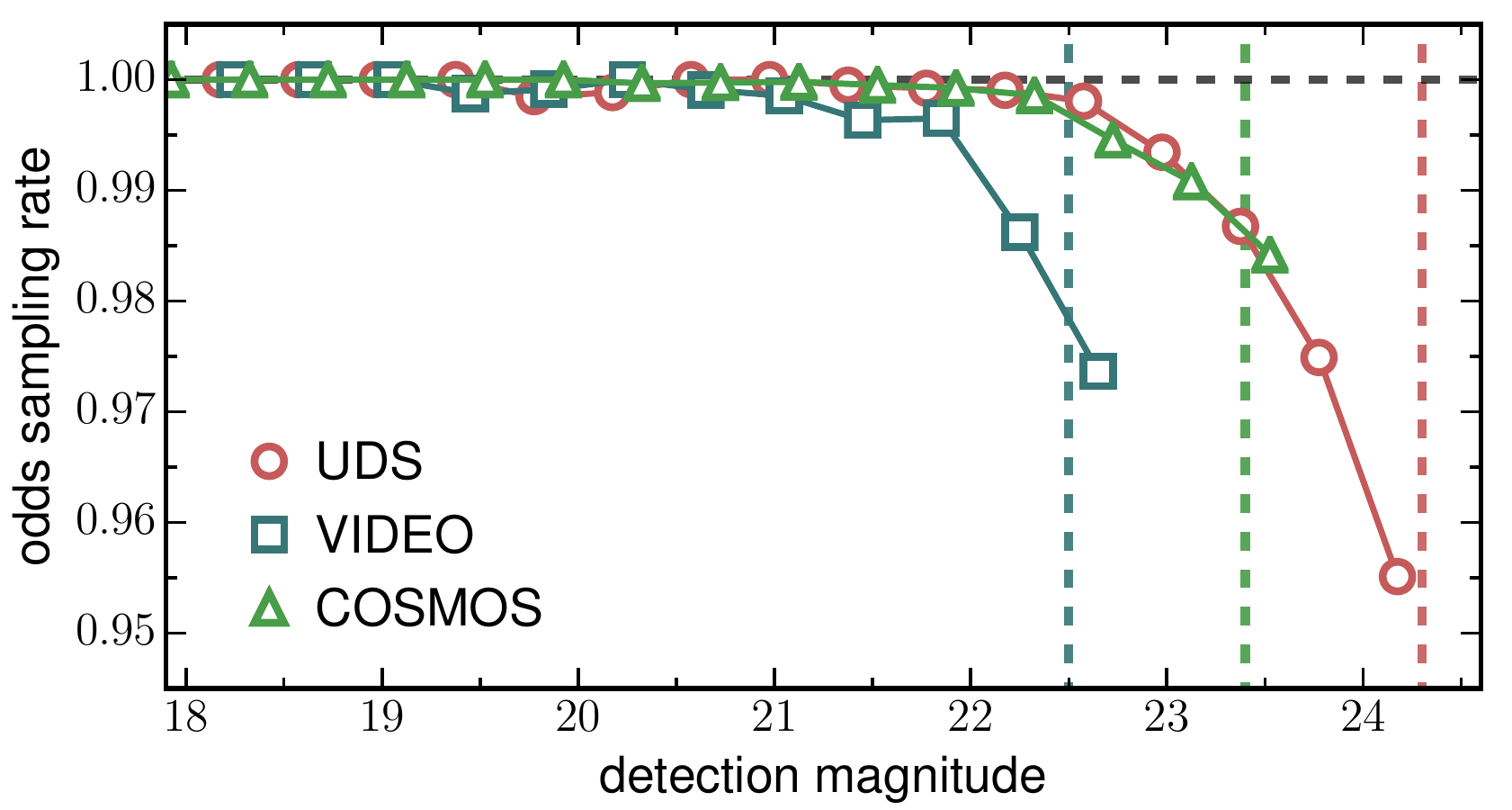}
  \caption{Measured odds sampling rate (OSR), defined in Equation \ref{eqn:osr}, as a function of detection band apparent magnitude for the UDS (red circles), VIDEO (blue squares) and COSMOS (green triangles) regions. Vertical dashed lines represent the various completeness magnitudes for the regions: $m_K = 24.3, 23.4, 22.5$ for the UDS, VIDEO and COSMOS, respectively.}
  \label{fig:osr}
\end{figure}

We follow \citet{Lopez-Sanjuan2014} in determining the most practical choice of cut in the Odds parameter, $\mathcal{O}$. This is achieved by measuring the pair fraction with samples of galaxies selected at an increasing cut in $\mathcal{O}$. The appropriate cut in $\mathcal{O}$ should ideally be where the measured pair fractions are stable and agree with the pair fraction measured spectroscopically. Figure \ref{fig:odds_selection} displays the measured pair fraction for galaxies at $M_* > 10^{10} M_\odot$ over the redshift range $0.01 < z < 0.15$ in the GAMA region, and over the redshift range $0.2 < z < 1.0$ in the UDS, VIDEO and COSMOS regions as a function of the chosen cut in the Odds parameter. It is found that the measured pair fractions are stable to the choice of cut up until large values of $\mathcal{O} > 0.6$ where measured pair fractions begin to sharply decrease. The grey shaded area in Figure \ref{fig:odds_selection} represents the measured pair fraction using spectroscopic data in the GAMA region. It is found that all cuts at $\mathcal{O} < 0.6$ produce photometrically measured pair fractions that agree with the spectroscopic measurement. It is therefore chosen that a cut of $\mathcal{O} > 0.3$ is the chosen cut adopted in this paper. Figure \ref{fig:osr} displays the Odds sampling rate (OSR) as a function of apparent magnitude in the UDS, VIDEO and COSMOS fields. Even at the faintest magnitudes, only a few percent of sources are removed by the Odds cut. Similar results are found for GAMA.

\label{lastpage}

\end{document}